\documentclass{article}

\usepackage{arxiv}

\usepackage{amsmath}

\newcommand{\bfalpha}{\mbox{\boldmath $\alpha$}}		
\newcommand{\bfbeta}{\mbox{\boldmath $\beta$}}      	
\newcommand{\bfgamma}{\mbox{\boldmath $\gamma$}}	
		
\newcommand{\bfepsilon}{\mbox{\boldmath $\epsilon$}}	
		
\newcommand{\bfeta}{\mbox{\boldmath $\eta$}}		
\newcommand{\bftheta}{\mbox{\boldmath $\theta$}}		\newcommand{\bfTheta}{\mbox{\boldmath $\Theta$}}

\newcommand{\bflambda}{\mbox{\boldmath $\lambda$}}	\newcommand{\bfLambda}{\mbox{\boldmath $\Lambda$}}
\newcommand{\bfmu}{\mbox{\boldmath $\mu$}}

			\newcommand{\bfPi}{\mbox{\boldmath $\Pi$}}
	        
	\newcommand{\bfSigma}{\mbox{\boldmath $\Sigma$}}

	\newcommand{\bfOmega}{\mbox{\boldmath $\Omega$}}

	\newcommand{\bfA}{{\bf A}}	\newcommand{\bfb}{{\bf b}}	\newcommand{\bfB}{{\bf B}}
	\newcommand{\bfC}{{\bf C}}		\newcommand{\bfD}{{\bf D}}
\newcommand{\bfe}{{\bf e}}		\newcommand{\bff}{{\bf f}}		
	\newcommand{\bfG}{{\bf G}}	\newcommand{\bfh}{{\bf h}}	
		\newcommand{\bfI}{{\bf I}}				\newcommand{\bfJ}{{\bf J}}
				\newcommand{\bfL}{{\bf L}}
			
\newcommand{\bfo}{{\bf o}}			\newcommand{\bfP}{{\bf P}}
\newcommand{\bfq}{{\bf q}}	\newcommand{\bfQ}{{\bf Q}}	\newcommand{\bfr}{{\bf r}}		\newcommand{\bfR}{{\bf R}}
\newcommand{\bfs}{{\bf s}}	\newcommand{\bfS}{{\bf S}}			
\newcommand{\bfu}{{\bf u}}		\newcommand{\bfv}{{\bf v}}	\newcommand{\bfV}{{\bf V}}
\newcommand{\bfw}{{\bf w}}	\newcommand{\bfW}{{\bf W}}	\newcommand{\bfx}{{\bf x}}	\newcommand{\bfX}{{\bf X}}
\newcommand{\bfy}{{\bf y}}		\newcommand{\bfz}{{\bf z}}	\newcommand{\bfZ}{{\bf Z}}

\newcommand{\bfzero}{{\bf 0}}

\usepackage[utf8]{inputenc} % allow utf-8 input
\usepackage[T1]{fontenc}    % use 8-bit T1 fonts
\usepackage{hyperref}       % hyperlinks
\usepackage{url}            % simple URL typesetting
\usepackage{booktabs,array,graphicx,longtable}       % professional-quality tables
\usepackage{amsfonts}       % blackboard math symbols
\usepackage{nicefrac}       % compact symbols for 1/2, etc.
\usepackage{microtype}      % microtypography
\usepackage{cleveref}       % smart cross-referencing
\usepackage{lipsum}         % Can be removed after putting your text content
\usepackage{graphicx}
\usepackage{natbib}
\usepackage{doi}
\usepackage{xcolor}

\title{Robust and Sparse Group Dynamic Causal Modeling via Student-$t$ Parametric Empirical Bayes and Nonlocal Priors}

\newif\ifuniqueAffiliation
\uniqueAffiliationtrue

\ifuniqueAffiliation % Standard variant of author block
\author{ \href{https://orcid.org/0009-0007-5426-5409}{\includegraphics[scale=0.06]{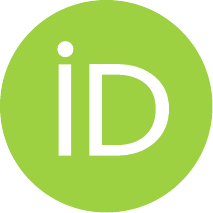}\hspace{1mm}Godfred~Arhin}\\
	Department of Mathematical Sciences\\
	The University of Texas at El Paso\\
	El Paso, TX 79968 \\
	\texttt{garhin@miners.utep.edu} \\
	\And
	\href{https://orcid.org/0000-0003-4814-7602}{\includegraphics[scale=0.06]{orcid.pdf}\hspace{1mm}Nilotpal~Sanyal\thanks{Corresponding author}} \\
	Department of Mathematical Sciences\\
	The University of Texas at El Paso\\
	El Paso, TX 79968\\
	\texttt{nsanyal@utep.edu} \\
}
\else
\usepackage{authblk}

\newbox{\orcid}\sbox{\orcid}{\includegraphics[scale=0.06]{orcid.pdf}} 
\author[1]{%
	\href{https://orcid.org/0000-0000-0000-0000}{\usebox{\orcid}\hspace{1mm}David S.~Hippocampus\thanks{\texttt{hippo@cs.cranberry-lemon.edu}}}%
}
\author[1,2]{%
	\href{https://orcid.org/0000-0000-0000-0000}{\usebox{\orcid}\hspace{1mm}Elias D.~Striatum\thanks{\texttt{stariate@ee.mount-sheikh.edu}}}%
}
\affil[1]{Department of Computer Science, Cranberry-Lemon University, Pittsburgh, PA 15213}
\affil[2]{Department of Electrical Engineering, Mount-Sheikh University, Santa Narimana, Levand}
\fi

\renewcommand{\headeright}{} %Technical Report
\renewcommand{\undertitle}{} %Technical Report
\renewcommand{\shorttitle}{Robust and Sparse Group Dynamic Causal Modeling}

\hypersetup{
pdftitle={Robust and Sparse Group Dynamic Causal Modeling via Student-$t$ Parametric Empirical Bayes and Nonlocal Priors},
pdfsubject={q-bio.NC, q-bio.QM},
pdfauthor={Godfred~Arhin, Nilotpal~Sanyal},
pdfkeywords={dynamic causal modeling, effective connectivity, parametric empirical Bayes, robust Bayesian inference, nonlocal prior, spike-and-slab prior, Student-$t$ distribution, fMRI},
}

\begin{document}
\maketitle

\begin{abstract}
Dynamic causal modeling (DCM) estimates directed effective connectivity, while parametric empirical Bayes (PEB) supports group inference using subject-specific posterior summaries. Standard PEB relies on Gaussian models and continuous shrinkage, making it sensitive to atypical estimates and unable to distinguish negligible from nonzero effects. We develop a robust and sparse group-DCM extension combining a Student-$t$ likelihood with a spike-and-slab prior using a nonlocal product-moment (pMOM) slab. A normal--gamma representation of the Student-$t$ distribution yields weights that downweight atypical {whitened residual coordinates}. The pMOM slab vanishes at zero, sharpening coefficient selection and yielding inclusion probabilities. {We propagate first-level posterior uncertainty through block-covariance pre-whitening and use a multistart generalized expectation--maximization (EM) algorithm with conditional maximization for mode fitting, followed by parallel-tempering posterior integration for inclusion probabilities. A simulation study showed that Student-$t$ weighting supplied the main protection against contamination, while spike-and-slab regularization reduced fixed-threshold false selections. In an openly shared mixed-gambles fMRI application, we found mild heterogeneity, concentrated inclusion support on departures of all four diagonal self-inhibition scaling parameters from the SPM default, and close directional agreement with standard PEB implemented in Statistical Parametric Mapping.} Our framework therefore adds interpretable {whitened-coordinate} robustness diagnostics and sparse coefficient selection to hierarchical DCM while retaining first-level uncertainty.
\end{abstract}

% keywords can be removed
\keywords{dynamic causal modeling \and effective connectivity \and parametric empirical Bayes \and robust Bayesian inference \and nonlocal prior \and spike-and-slab prior \and Student-$t$ distribution \and fMRI}

\addtocontents{toc}{\protect\setcounter{tocdepth}{-5}}

%%%
\section{Introduction}
\label{sec:intro}

Effective connectivity concerns the directed influence that one neuronal population exerts on another. Dynamic causal modeling (DCM) estimates such influences by combining a neuronal state equation with a biophysical observation model for functional magnetic resonance imaging (fMRI) data \cite{friston2003dcm,stephan2010rules,zeidman2019part1}. In a group study, inversion of each participant's DCM yields a posterior mean and covariance for the connectivity parameters. Parametric empirical Bayes (PEB) carries these posterior summaries into a hierarchical model for population-average effects, covariate effects, and between-subject variability while accounting for differences in first-level precision \cite{friston2016peb,zeidman2019part2}. More broadly, Bayesian hierarchical models have been used to borrow information across subjects and spatial scales in fMRI analysis \cite{sanyal2012fmri,bowman2008bayesian,woolrich2004multilevel}. We propose a robust and sparse extension of group-level PEB that combines Student-$t$ weighting for robustness to atypical subject--parameter estimates with nonlocal spike-and-slab selection of group effects.

Although conventional PEB is computationally efficient and can be combined with Bayesian model reduction to compare reduced parameter sets, it has two group-level limitations. First, its Gaussian sampling model can give disproportionate influence to atypical subject--parameter estimates arising from heterogeneous participants, noisy acquisitions, imperfect first-level inversion, or model misspecification. A participant may be atypical for only selected connections, making complete-case exclusion unnecessarily coarse. Second, the Gaussian priors used for retained coefficients are local: they assign positive density at zero and can regularize weak effects without sharply separating negligible coefficients from effects credibly different from zero. Robustness to atypical posterior summaries and sparse identification of group effects are related but distinct inferential goals, and addressing only one leaves the other unresolved.

{Related Bayesian regression methods have combined robust error distributions with sparsity-inducing priors. Robust variable selection has been studied under general spherically symmetric errors \citep{maruyama2014robust}, under asymmetric and thick-tailed likelihoods with nonlocal priors \citep{rossell2018tractable}, through Dirichlet-process scale mixtures paired with spike-and-slab or horseshoe priors \citep{cabezas2024robust}, and through spike-and-slab model averaging with error families that include Student-$t$ tails \citep{de2026robust}. These developments establish a broader precedent for treating robustness and sparsity jointly, but they do not address the subject-specific posterior covariances, whitened-coordinate residuals, and hierarchical effective-connectivity structure of group DCM.}

Our formulation preserves the PEB covariance architecture while modifying its second-level likelihood and coefficient prior. After propagating the subject-specific posterior covariances through block pre-whitening, we replace the Gaussian second-level likelihood with a Student-$t$ model represented as a normal--gamma scale mixture. The resulting latent precision weights adapt to individual whitened residuals, allowing a participant to be down-weighted for one whitened coordinate without being discarded from the analysis. For the group coefficients, we use a spike-and-slab prior whose slab is a nonlocal product-moment (pMOM) density. Related nonlocal-prior constructions have previously been used in Bayesian wavelet methods for fMRI and other noisy imaging and audio data \cite{sanyal2017wavelet,sanyal2025nonlocal}. Because the pMOM slab has zero density at the null value, it discourages small coefficients from being classified as non-null and yields coefficient-level posterior inclusion probabilities (PIPs) \cite{johnson2010nonlocal,rossell2017nonlocal}. {A multistart generalized expectation--maximization (EM) algorithm with conditional maximization (i.e., ECM) estimates the empirical-Bayes mode while monitoring the observed and expected-complete-data objectives. After convergence, parallel-tempering posterior sampling integrates the conditional slab responsibilities over the joint coefficient posterior to obtain the reported PIPs.}

Because robustness and sparsity address distinct inferential goals, we assess their separate and joint contributions in a simulation study. We compare the proposed robust--sparse estimator, \emph{Proposed}, with a Student-$t$ ridge model, a Gaussian pMOM model, {a Student-$t$ Gaussian spike-and-slab model,} and standard \emph{SPM-PEB} across settings varying in contamination fraction and geometry, sample size, network dimension, signal strength, covariate structure, sparsity, and first-level covariance. {The results show that Student-$t$ weighting provides the primary protection against contamination, while sparse mixture regularization reduces false selections at fixed score thresholds. The proposed estimator combines these features and achieves the lowest average RMSE and highest MCC in the principal comparison, although the pMOM slab does not uniformly outperform the local Gaussian slab.} The combined estimator is consequently most advantageous when contamination and sparsity occur together, although it is not uniformly superior under every loss function or data-generating regime.

We additionally demonstrate the method using DCM summaries from an openly shared mixed-gambles fMRI data set. In the analyzed sample, the robustness weights indicated mild heterogeneity without supporting participant exclusion, and the nonlocal prior concentrated inclusion support on {departures of all four diagonal self-inhibition scaling parameters from the SPM default}. The \emph{Proposed}-model estimates agreed in direction with standard \emph{SPM-PEB} for nearly all group-average parameters while applying stronger shrinkage to weak modulatory effects. {Adjustment for task version left the group-average coefficient estimates stable but showed that the inclusion probabilities were more sensitive to the second-level design.} This application assesses interpretability, diagnostics, and compatibility with an established neuroimaging workflow rather than establishing new neuroscientific or demographic conclusions.

The paper makes four main contributions. First, it formulates a whitened-coordinate robust group DCM while preserving first-level posterior uncertainty and the conventional PEB group design. Second, it introduces nonlocal coefficient selection within the same hierarchical model. {Third, it provides a multistart generalized EM/ECM procedure with post-fit parallel-tempering integration that returns group estimates, integrated PIPs, robustness weights, variance-component estimates, and numerical diagnostics.} Fourth, it evaluates the separate and combined roles of robustness and sparsity through replicated simulations and a reproducible application to open neuroimaging data. The rest of the paper is organized as follows: Section~\ref{sec:background} reviews the DCM and PEB framework, Section~\ref{sec:method} develops the proposed estimator, Section~\ref{sec:simulation} presents the simulation study, Section~\ref{sec:application} presents the fMRI data application, Section~\ref{sec:discussion} discusses the implications, and finally, Section~\ref{sec:conclusion} concludes.

%%%
\section{Background}\label{sec:background}

\subsection{Effective connectivity and dynamic causal modeling}

Functional connectivity summarizes statistical dependence between measured brain signals, whereas effective connectivity concerns directed influences within a specified generative model. Dynamic causal modeling (DCM) formalizes the latter by treating a network of neuronal populations as a dynamical system whose parameters describe how activity in one region changes the rate of activity in another. The model also specifies how experimental inputs enter the network and how latent neuronal activity is transformed into the observed fMRI BOLD signal. Inferences are therefore conditional on the chosen regions, inputs, connectivity architecture, and hemodynamic model \cite{friston2003dcm,stephan2010rules,zeidman2019part1}.
%; they should be interpreted as model-based directed influences rather than intervention-free causal effects in the broadest sense.

DCM is particularly useful for group studies because fitting the first-level model to each subject’s data yields both a point estimate and an uncertainty estimate for every parameter. These posterior summaries can be carried forward into a hierarchical model, allowing population-level connectivity and covariate modulation to be estimated while accounting for differences in first-level precision. The following subsections review the parts of DCM that are relevant to the proposed robust group-level extension.

\subsection{Neuronal state equation and connectivity parameters}

Let $\bfz(t)\in\mathbb{R}^{l}$ denote the neuronal states of $l$ regions and let $\bfu(t)\in\mathbb{R}^{m}$ denote the experimental inputs. {Let $\bftheta_c$ denote the neuronal coupling parameters, and let $F(\bfz,\bfu;\bftheta_c)$ denote the neuronal vector field, that is, the function governing the instantaneous rate of change of the neuronal states.} DCM starts from the nonlinear state equation
\begin{equation}\label{eq:dcm-general}
\dot{\bfz}(t)=F\!\left(\bfz(t),\bfu(t);\bftheta_c\right),
\end{equation}
and commonly uses a bilinear approximation around an operating point,
\begin{equation}\label{eq:neuronal}
\dot{\bfz}(t)=\left(\bfA+\sum_{j=1}^{m}u_j(t)\bfB^{(j)}\right)\bfz(t)+\bfC\bfu(t).
\end{equation}
The matrix $\bfA\in\mathbb{R}^{l\times l}$ contains intrinsic coupling in the absence of experimental input, $\bfB^{(j)}\in\mathbb{R}^{l\times l}$ contains the change in coupling induced by input $u_j$, and $\bfC\in\mathbb{R}^{l\times m}$ maps external inputs directly to neuronal states. In terms of the underlying vector field,
\begin{equation}\label{eq:dcm-derivatives}
\bfA=\left.\frac{\partial F}{\partial\bfz}\right|_{\mathrm{op}},\qquad \bfB^{(j)}=\left.\frac{\partial^2F}{\partial\bfz\,\partial u_j}\right|_{\mathrm{op}},\qquad \bfC=\left.\frac{\partial F}{\partial\bfu}\right|_{\mathrm{op}},
\end{equation}
where the derivatives are evaluated at the operating point. Thus, $a_{ik}$ quantifies the instantaneous influence of region $k$ on the rate of change in region $i$, $b_{ik}^{(j)}$ quantifies how that influence changes under input $u_j$, and $c_{ik}$ quantifies the direct effect of input $k$ on region $i$. The neuronal coupling parameters are collected in $\bftheta_c=\{\bfA,\bfB^{(1)},\ldots,\bfB^{(m)},\bfC\}$.

{In the one-state fMRI DCM considered here, the off-diagonal elements of $\bfA$ and the elements of $\bfB^{(j)}$ and $\bfC$ are coupling parameters in Hz. SPM instead stores each diagonal $A$-matrix parameter as an unconstrained dimensionless log-scaling parameter, denoted by $\widetilde a_{ii}$, and maps it to the physical self-connection
\begin{equation}\label{eq:dcm-stability}
a_{ii}^{\mathrm{Hz}}=-\frac{1}{2}\exp(\widetilde a_{ii}).
\end{equation}
Thus, $\widetilde a_{ii}=0$ corresponds to the default self-inhibition of $-0.5$ Hz. The exponential transformation ensures that every physical self-connection is negative and promotes stable neuronal dynamics. Group-level inference for a diagonal $A$ parameter is therefore performed on the internal log-scaling scale.}

\subsection{Hemodynamic observation model for fMRI}

Neuronal states are not observed directly in fMRI. DCM therefore couples \labelcref{eq:neuronal} to a Balloon--Windkessel model that represents neurovascular coupling through four regional states: vasodilatory signal $s_i$, blood inflow $f_i$, venous blood volume $v_i$, and deoxyhemoglobin content $q_i$. A standard form is
\begin{equation}\label{eq:hemo}
\begin{aligned}
\dot s_i&=z_i-\kappa_i s_i-\gamma_i(f_i-1),&\dot f_i&=s_i,\\
\tau_i\dot v_i&=f_i-v_i^{1/\alpha_i},&
\tau_i\dot q_i&=\frac{f_i\{1-(1-\rho_i)^{1/f_i}\}}{\rho_i}-q_i v_i^{1/\alpha_i-1}.
\end{aligned}
\end{equation}
The predicted regional BOLD response is
\begin{equation}\label{eq:bold}
y_i=g(q_i,v_i)=V_0\left[k_1(1-q_i)+k_2\left(1-\frac{q_i}{v_i}\right)+k_3(1-v_i)\right],
\end{equation}
where $V_0$ is the resting venous blood-volume fraction and $k_1,k_2,k_3$ depend on acquisition and physiological parameters. The hemodynamic parameter vector, denoted by $\bftheta_h$, includes signal-decay and autoregulation parameters $(\kappa_i,\gamma_i)$, transit time $\tau_i$, Grubb's exponent $\alpha_i$, and resting oxygen-extraction fraction $\rho_i$. Equations \labelcref{eq:hemo,eq:bold} make explicit why a change in neuronal coupling does not translate linearly or instantaneously into a change in BOLD amplitude.

\subsection{Forward and observation models}
{Collect the neuronal and hemodynamic states in $\bfx(t)=\{\bfz(t),\bfs(t),\bff(t),\allowbreak\bfv(t),\bfq(t)\}$. Define the full subject-level parameter vector as $\bftheta=(\bftheta_c^\top,\bftheta_h^\top)^\top\in\mathbb{R}^{p}$, where $\bftheta_c$ and $\bftheta_h$ contain the neuronal and hemodynamic parameters, respectively. The complete forward model can then be written compactly as}
\begin{equation}\label{eq:forward}
\dot{\bfx}(t)=f\!\left(\bfx(t),\bfu(t),\bftheta\right),\qquad \bfy(t)=\psi\!\left(\bfx(t)\right),
\end{equation}
where $\psi$ applies the regional BOLD output map to the hemodynamic states. Numerical integration of the state equation followed by the observation map gives the predicted time series
\begin{equation}\label{eq:predictedbold}
h(\bfu,\bftheta)=\psi\!\left(\bfx(t;\bfu,\bftheta)\right)\in\mathbb{R}^{T\times l}.
\end{equation}
The selected regions of interest and the nonzero entries of $\bfA$, $\bfB^{(j)}$, and $\bfC$ define the model architecture. In practice, regions are often selected using anatomical hypotheses and task-related activity from a first-level general linear model. Restricting the network is important because the number of possible directed parameters grows rapidly with the number of regions \cite{stephan2010rules}.

The measured BOLD data additionally contain nuisance structure and measurement error. The considered observation model is therefore
\begin{equation}\label{eq:observation}
\bfy=h(\bfu,\bftheta)+\bfX\bfbeta+\bfepsilon,\qquad \bfepsilon\sim N(\bfzero,\bfC_\epsilon),
\end{equation}
where $\bfX$ contains a constant and nuisance regressors such as low-frequency drift bases, and $\bfbeta$ contains their coefficients. Temporal and regional error structure can be represented through covariance components,
\begin{equation}\label{eq:firstlevelcov}
\bfC_\epsilon=\sum_{r=1}^{R}\lambda_r\bfQ_r,
\end{equation}
with known component matrices $\bfQ_r$ and nonnegative hyperparameters $\lambda_r$. Around the current posterior mean $\bfeta_{\theta\mid y}$, local linearization gives
\begin{equation}\label{eq:linearized}
\bfy\approx h(\bfu,\bfeta_{\theta\mid y})+\bfJ(\bftheta-\bfeta_{\theta\mid y})+\bfX\bfbeta+\bfepsilon,\qquad \bfJ=\left.\frac{\partial h(\bfu,\bftheta)}{\partial\bftheta}\right|_{\bfeta_{\theta\mid y}}.
\end{equation}

\subsection{First-level Bayesian inversion}

{DCM combines the likelihood from \labelcref{eq:observation} with informed priors. Gaussian priors are typically assigned to the free neuronal parameters, including the allowed off-diagonal elements of $\bfA$, the internal diagonal parameters $\widetilde a_{ii}$, and the allowed elements of $\bfB^{(j)}$ and $\bfC$. Connections fixed as absent by the model architecture are assigned zero prior variance. The transformation in Equation~\eqref{eq:dcm-stability} constrains the physical self-connections to be negative and regularizes the neuronal dynamics. Hemodynamic parameters receive Gaussian priors centered at empirically plausible physiological values, with variances reflecting uncertainty in the neurovascular response \cite{friston2003dcm,stephan2007hemodynamic,penny2010families}.} If $\bftheta_0$ and $\bfC_\theta$ denote the resulting prior mean and covariance, the first-level prior is
\begin{equation}\label{eq:firstlevelprior}
\bftheta\sim N(\bftheta_0,\bfC_\theta).
\end{equation}
These priors stabilize inversion of the nonlinear state-space model and provide a principled way to encode the fact that some connections are not estimable from the available inputs and time series.

The first-level algorithm alternates a variational E-step for the parameters with an M-step for the noise components. To make the E-step explicit, let $\bfr=\bfy-h(\bfu,\bfeta_{\theta\mid y})$ and $\Delta\bftheta=\bftheta-\bfeta_{\theta\mid y}$. Under the local approximation in \labelcref{eq:linearized}, the posterior mode solves the augmented generalized least-squares system
\begin{equation}\label{eq:augmented}
\underbrace{\begin{bmatrix}\bfr\\\bftheta_0-\bfeta_{\theta\mid y}\end{bmatrix}}_{\overline{\bfy}}\approx
\underbrace{\begin{bmatrix}\bfJ&\bfX\\\bfI&\bfzero\end{bmatrix}}_{\overline{\bfJ}}
\begin{bmatrix}\Delta\bftheta\\\bfbeta\end{bmatrix}+\overline{\bfepsilon},\qquad
\operatorname{Cov}(\overline{\bfepsilon})=\operatorname{blkdiag}(\bfC_\epsilon,\bfC_\theta).
\end{equation}
Consequently,
\begin{equation}\label{eq:firstlevelposterior}
\begin{bmatrix}\Delta\bfeta_{\theta\mid y}\\\bfeta_{\beta\mid y}\end{bmatrix}=
\left(\overline{\bfJ}^{\top}\overline{\bfC}^{-1}\overline{\bfJ}\right)^{-1}\overline{\bfJ}^{\top}\overline{\bfC}^{-1}\overline{\bfy},\qquad
\bfC_{\theta\mid y}=\left(\overline{\bfJ}^{\top}\overline{\bfC}^{-1}\overline{\bfJ}\right)^{-1}_{\theta\theta},
\end{equation}
where $\overline{\bfC}=\operatorname{blkdiag}(\bfC_\epsilon,\bfC_\theta)$ and the subscript $\theta\theta$ denotes the parameter block. The posterior mean is updated as $\bfeta_{\theta\mid y}\leftarrow\bfeta_{\theta\mid y}+\Delta\bfeta_{\theta\mid y}$, and the Jacobian is recomputed after the nonlinear forward model is updated.

In the M-step, the nuisance covariance parameters $\bflambda=(\lambda_1,\ldots,\lambda_R)^{\top}$ are updated by maximizing the restricted likelihood or variational free energy. Integrating out the local Gaussian parameter approximation gives the effective precision
\begin{equation}\label{eq:firstlevelprecision}
\bfP=\overline{\bfC}^{-1}-\overline{\bfC}^{-1}\overline{\bfJ}\bfC_{\theta\mid y}\overline{\bfJ}^{\top}\overline{\bfC}^{-1}.
\end{equation}
For $\bfG_r=\partial\overline{\bfC}/\partial\lambda_r$, the score and expected information have the familiar covariance-component form
\begin{align}
s_r&=-\frac12\operatorname{tr}(\bfP\bfG_r)+\frac12\overline{\bfy}^{\top}\bfP\bfG_r\bfP\overline{\bfy},\label{eq:firstlevelscore}\\
\mathcal{I}_{rs}&=\frac12\operatorname{tr}(\bfP\bfG_r\bfP\bfG_s).\label{eq:firstlevelinfo}
\end{align}
Fisher scoring, nonnegative line searches, and iteration of the E- and M-steps produce the first-level posterior summaries $\bfeta_{\theta\mid y}^{n}$ and $\bfC_{\theta\mid y}^{n}$ used below. 
%Thus, the proposed method changes the group-level likelihood and coefficient prior; it does not replace the DCM forward model or the first-level Bayesian inversion.

%%
\subsection{Conventional group-level PEB}

For clarity, let $\bfeta_n=\bfeta_{\theta\mid y}^{n}$ and $\bfC_n=\bfC_{\theta\mid y}^{n}$ denote the outputs of the first-level inversions. Stack the posterior means as
\begin{equation}\label{eq:peb-stack}
\bfeta=\begin{bmatrix}\bfeta_1^{\top}&\cdots&\bfeta_N^{\top}\end{bmatrix}^{\top}\in\mathbb{R}^{Np}.
\end{equation}
Let $\bfX_G\in\mathbb{R}^{N\times r}$ contain an intercept and centered group-level covariates, and define the Kronecker-expanded design
\begin{equation}\label{eq:peb-design}
\widetilde{\bfX}_G=\bfX_G\otimes\bfI_p\in\mathbb{R}^{Np\times pr},\qquad \bfbeta_G\in\mathbb{R}^{pr}.
\end{equation}
The conventional PEB mean model is
\begin{equation}\label{eq:peb-mean}
E(\bfeta\mid\bfbeta_G)=\widetilde{\bfX}_G\bfbeta_G.
\end{equation}
Its covariance propagates first-level uncertainty and between-subject variation:
\begin{equation}\label{eq:peb-cov}
\bfOmega=\operatorname{blkdiag}(\bfC_1,\ldots,\bfC_N)+\bfI_N\otimes\bfSigma_b,\qquad \bfSigma_b=\sum_{k=1}^{K}\alpha_k\bfV_k,
\end{equation}
where $\bfV_k$ are prespecified positive-semidefinite covariance bases and $\alpha_k\geq0$ are variance components. Standard PEB uses a Gaussian second-level likelihood and Gaussian or empirically estimated shrinkage priors for $\bfbeta_G$ \cite{zeidman2019part2}. It is efficient and often effective, but a Gaussian likelihood can be sensitive to heterogeneous subjects and local shrinkage does not explicitly separate small nonzero effects from null effects. Section~\ref{sec:method} retains the first-level summaries, the design in \labelcref{eq:peb-design}, and the covariance representation in \labelcref{eq:peb-cov}, while replacing the second-level error model and coefficient prior.

%%% 
\section{Robust and Sparse Group Dynamic Causal Modeling}
\label{sec:method}

Building on the first-level DCM and conventional PEB framework established in Section~\ref{sec:background}, we develop a robust and sparse group-level model that combines a Student-$t$ likelihood with a nonlocal pMOM spike-and-slab prior, moving beyond conventional Gaussian error modeling and local shrinkage. We develop an empirical-Bayes (posterior-mode) procedure using a { generalized expectation--maximization algorithm with conditional maximization (EM/ECM)}, which returns group effects, {integrated} posterior inclusion probabilities (PIPs), Student-$t$ robustness weights, {a fixed or estimated residual scale}, optional between-subject variance components{, and conditional Laplace standard deviations}. All vectors and matrices are written in bold throughout this section.

\subsection{Robust group-level formulation}

Section~\ref{sec:background} introduced the first-level posterior summaries, Kronecker-expanded group design, and covariance structure underlying conventional PEB. The proposed formulation preserves these components while introducing a robust second-level likelihood and a nonlocal prior for the group-level coefficients. For notational continuity, let $\bfPi=\bfOmega^{-1}$ denote the precision associated with \labelcref{eq:peb-cov}.

The stacked posterior means, covariance, and group design are defined in \labelcref{eq:peb-stack,eq:peb-design,eq:peb-cov}. The robust analysis uses the same $\bfeta$, $\bfOmega$, $\bfSigma_b$, $\bfX_G$, and $\bfbeta_G$. Only their second-level likelihood and coefficient prior are changed. Let $q=Np$ denote the length of the stacked group response and $d=pr$ the number of group-level coefficients. Pre-whitening, described below, preserves these dimensions.

\subsection{Pre-whitening}
The components of $\bfeta$ are correlated and heteroscedastic because the matrices $\bfC_n$ differ across subjects and because $\bfSigma_b$ can be non-diagonal. We therefore pre-whiten the response and the expanded design before applying the Student-$t$ likelihood. For each subject, choose a Cholesky or symmetric square-root factor $\bfL_n$ satisfying

\begin{equation}\label{eq:whitenfactor}
\bfL_n^{\top}\bfL_n=(\bfC_n+\bfSigma_b)^{-1}.
\end{equation}

Define the block-diagonal whitening matrix and the transformed quantities by

\begin{equation}\label{eq:white}
\bfB=\operatorname{blkdiag}(\bfL_1,\ldots,\bfL_N),\qquad \bfy^{*}=\bfB\bfeta,\qquad \bfX^{*}=\bfB\widetilde{\bfX}_G.
\end{equation}

When the same $\bfSigma_b$ is used for all subjects, $\bfB^{\top}\bfB=\bfOmega^{-1}$ under the block representation in \labelcref{eq:whitenfactor}. {Unlike a Gaussian likelihood, however, a product of univariate Student-$t$ densities is not invariant to orthogonal rotations. The whitening rule and input parameter ordering therefore form part of the coordinate-wise robust model and must be held fixed and reported.} Consequently, the transformed mean is $E(\bfy^*\mid\bfbeta_G)=\bfX^*\bfbeta_G$ and the {reference Gaussian scale matrix is the identity}. If the variance components $\alpha_k$ are updated, the whitening factors are recomputed before the next outer iteration{, and the likelihood includes the corresponding transformation Jacobian}.

The index $i=1,\ldots,q$ refers to {a coordinate} of the whitened response and $j=1,\ldots,d$ refers to a group coefficient. {The robustness weights therefore apply to whitened coordinates. Under diagonal covariance, a whitened coordinate corresponds to one original parameter, whereas under correlated covariance it can combine residual information from several parameters. The coordinate-wise formulation permits different transformed residual directions from the same participant to receive different weights without discarding the entire participant.} A multivariate subject-level Student-$t$ block can be substituted when an entire scan is suspected to be atypical. The present implementation uses the {whitened-coordinate} formulation.

\subsection{Robust Student-$t$ group-level likelihood}
For each whitened element, introduce a latent precision $\lambda_i$. Conditional on $\lambda_i$, we consider a Gaussian model:
\begin{align}
y_i^{*}\mid\bfbeta_G,\lambda_i,\sigma^2&\sim N\!\left({\bfx_i^*}^{\top}\bfbeta_G,\frac{\sigma^2}{\lambda_i}\right),\label{eq:normalmix}\\
\lambda_i&\sim\operatorname{Gamma}\!\left(\frac{\nu}{2},\frac{\nu}{2}\right),\label{eq:gamma}
\end{align}
where the gamma distribution is parameterized by shape and rate, $\nu>0$ is the degrees of freedom, and $\sigma^2$ is a global residual scale. Marginalizing $\lambda_i$ yields
\begin{equation}\label{eq:marginalt}
y_i^{*}\mid\bfbeta_G,\sigma^2\sim t_{\nu}\!\left({\bfx_i^*}^{\top}\bfbeta_G,\sigma^2\right).
\end{equation}
The Student-$t$ distribution approaches the Gaussian model as $\nu\to\infty$ and becomes increasingly heavy-tailed as $\nu$ decreases. {When $\sigma^2$ is estimated,} we assign a weak Jeffreys prior to the scale,
\begin{equation}\label{eq:jeffreys}
p(\sigma^2)\propto(\sigma^2)^{-1}.
\end{equation}

{Because a free global residual scale can be confounded with the scale of $\bfSigma_b(\bfalpha)$, the default implementation fixes $\sigma^2=1$. The optional update of $\sigma^2$ described below is used only when scale identification has been established for the specified covariance basis.}

Let $r_i=y_i^{*}-{\bfx_i^*}^{\top}\bfbeta_G$. The conditional distribution of the latent precision is then

\begin{equation}\label{eq:lambdacond}
\lambda_i\mid\bfy^*,\bfbeta_G,\sigma^2\sim\operatorname{Gamma}\!\left(\frac{\nu+1}{2},\frac{\nu+r_i^2/\sigma^2}{2}\right).
\end{equation}

Thus, our E-step uses

\begin{equation}\label{eq:weight}
w_i=E(\lambda_i\mid\cdot)=\frac{\nu+1}{\nu+r_i^2/\sigma^2}.
\end{equation}

The weight decreases smoothly as the standardized residual increases. It is therefore an influence diagnostic, not an observation-deletion rule. A low $w_i$ identifies a subject--parameter combination that is atypical under the current group model, but it does not by itself establish poor data quality.

In matrix notation, with $\bfLambda=\operatorname{diag}(\lambda_1,\ldots,\lambda_q)$,

\begin{equation}\label{eq:matrixlik}
\bfy^*\mid\bfbeta_G,\bfLambda,\sigma^2\sim N\!\left(\bfX^*\bfbeta_G,\sigma^2\bfLambda^{-1}\right).
\end{equation}

\subsection{Nonlocal spike-and-slab prior}
For each group coefficient $\beta_{G,j}$, we introduce an inclusion indicator $\gamma_j\in\{0,1\}$ such that, conditional on the indicator, the coefficient follows a Gaussian spike or a first-order product-moment slab:

\begin{equation}\label{eq:mixprior}
p(\beta_{G,j}\mid\gamma_j,\tau_0^2,\tau_1^2)=f_0(\beta_{G,j};\tau_0^2)^{1-\gamma_j}f_1(\beta_{G,j};\tau_1^2)^{\gamma_j},
\end{equation}

where

\begin{align}
f_0(\beta;\tau_0^2)&=\phi(\beta;0,\tau_0^2)=(2\pi\tau_0^2)^{-1/2}\exp\!\left(-\frac{\beta^2}{2\tau_0^2}\right),\label{eq:spike}\\
f_1(\beta;\tau_1^2)&=\frac{\beta^2}{\tau_1^2}\phi(\beta;0,\tau_1^2),\label{eq:pmom}\\
\gamma_j&\sim\operatorname{Bernoulli}(\pi),\qquad j=1,\ldots,d,\quad 0<\pi<1.\label{eq:indicator}
\end{align}

Here $\phi(\cdot;0,v)$ denotes the $N(0,v)$ density. The factor $\beta^2/\tau_1^2$ normalizes the first-order pMOM slab because $E_{N(0,\tau_1^2)}(\beta^2/\tau_1^2)=1$. Crucially, $f_1(0;\tau_1^2)=0$, whereas a local Gaussian slab has positive density at zero. The pMOM slab therefore discourages tiny nonzero coefficients that are neither clearly null nor meaningfully separated from zero.

For regularization of the scale parameters, we use independent inverse-gamma priors,

\begin{equation}\label{eq:igprior}
\tau_0^2\sim\operatorname{Inverse\text{-}Gamma}(a_0,b_0),\qquad \tau_1^2\sim\operatorname{Inverse\text{-}Gamma}(a_1,b_1),
\end{equation}

with density

\begin{equation}\label{eq:igdensity}
p(\tau_\ell^2)=\frac{b_\ell^{a_\ell}}{\Gamma(a_\ell)}(\tau_\ell^2)^{-(a_\ell+1)}\exp\!\left(-\frac{b_\ell}{\tau_\ell^2}\right),\qquad \ell\in\{0,1\}.
\end{equation}

Small positive $(a_\ell,b_\ell)$ give proper but diffuse priors. {The implementation enforces $\tau_1\geq c\tau_0$, with a prespecified $c\geq1$, after each scale update.} This avoids an empirical-Bayes solution whose component labels are difficult to interpret.

Integrating out $\gamma_j$ gives the marginal prior

\begin{equation}\label{eq:margprior}
m_j(\beta_{G,j})=(1-\pi)f_0(\beta_{G,j};\tau_0^2)+\pi f_1(\beta_{G,j};\tau_1^2),
\end{equation}

and the coefficient-wise factorization

\begin{equation}\label{eq:priorfactor}
p(\bfbeta_G)=\prod_{j=1}^{d}m_j(\beta_{G,j}),\qquad p({\bfgamma})=\prod_{j=1}^{d}\pi^{\gamma_j}(1-\pi)^{1-\gamma_j}.
\end{equation}

\subsection{Posterior inference}

\subsubsection{Joint posterior and conditional distributions}
Combining \labelcref{eq:matrixlik,eq:gamma,eq:jeffreys,eq:mixprior,eq:indicator,eq:igprior}, the joint posterior is, up to a constant independent of the unknowns,

\begin{equation}\label{eq:posterior}
\begin{aligned}
p(\bfbeta_G,\bfLambda,\sigma^2,{\bfgamma},\tau_0^2,\tau_1^2{,\bfalpha}\mid{\bfeta})
&\propto {|\bfB(\bfalpha)|}\,\phi\!\left({\bfy^*(\bfalpha)};{\bfX^*(\bfalpha)}\bfbeta_G,\sigma^2\bfLambda^{-1}\right)p(\sigma^2)\prod_{i=1}^{q}p(\lambda_i)\\
&\quad\times\prod_{j=1}^{d}p(\beta_{G,j}\mid\gamma_j,\tau_0^2,\tau_1^2)p(\gamma_j) \times p(\tau_0^2)p(\tau_1^2).
\end{aligned}
\end{equation}

{Here $|\bfB(\bfalpha)|$ is the absolute Jacobian determinant of the block whitening transformation. With $r_i^*(\bfalpha)=y_i^*(\bfalpha)-{\bfx_i^*(\bfalpha)}^{\top}\bfbeta_G$ and $t_\nu(\cdot)$ denoting the standard Student-$t$ density, all empirical-Bayes updates are tied to the normalized observed-data log posterior}

\begin{equation}\label{eq:observedobjective}
{
\begin{aligned}
\ell(\bfTheta)
&=\log|\bfB(\bfalpha)|+\sum_{i=1}^{q}\left[\log t_\nu\!\left(\frac{r_i^*(\bfalpha)}{\sigma}\right)-\log\sigma\right]
+\sum_{j=1}^{d}\log m_j(\beta_{G,j})\\
&\quad+\log p(\tau_0^2)+\log p(\tau_1^2)+\log p(\sigma^2),
\end{aligned}
}
\end{equation}

{where the final term is constant and may be omitted when $\sigma^2$ is fixed. The Jacobian is essential when $\bfalpha$ changes because whitening then depends on an estimated parameter.}

The latent variables $\bfLambda$ and {$\bfgamma$} make direct maximization of the marginal posterior inconvenient. We therefore alternate conditional-expectation updates with {conditional} maximization of the expected complete-data log posterior.

{If $\sigma^2$ is estimated,} the scale conditional distribution follows from conjugacy. With $S=\sum_{i=1}^{q}\lambda_i r_i^2$,

\begin{equation}\label{eq:sigmacond}
\sigma^2\mid\bfy^*,\bfbeta_G,\bfLambda\sim\operatorname{Inverse\text{-}Gamma}\!\left(\frac{q}{2},\frac{S}{2}\right),
\end{equation}

The second argument is the scale under the convention in \labelcref{eq:igdensity}. The conditional MAP update is therefore

\begin{equation}\label{eq:sigma}
\widehat{\sigma}^{2}=\frac{S}{q+2}.
\end{equation}

If the scale prior is omitted or replaced by a different proper prior, the corresponding posterior-mode denominator should be changed accordingly.

Given a current coefficient value and current scale parameters, the E-step responsibility for the nonlocal component is

\begin{equation}\label{eq:pip}
q_j=E(\gamma_j\mid\beta_{G,j},\tau_0^2,\tau_1^2)=\frac{\pi f_1(\beta_{G,j};\tau_1^2)}{(1-\pi)f_0(\beta_{G,j};\tau_0^2)+\pi f_1(\beta_{G,j};\tau_1^2)}.
\end{equation}

{At convergence, $q_j$ is the conditional slab responsibility evaluated at the empirical-Bayes mode. Because it conditions on a single value of $\beta_{G,j}$, it is an optimization quantity rather than the reported PIP.}

{Holding the E-step weights $w_i^{(t)}$ and responsibilities $q_j^{(t)}$ fixed, the common expected complete-data objective used by the conditional-maximization steps is, up to terms independent of the fitted parameters,}

\begin{equation}\label{eq:qcomplete}
{
\begin{aligned}
Q(\bfTheta\mid\bfTheta^{(t)})
&=\log|\bfB(\bfalpha)|-\left(\frac{q}{2}+1\right)\log\sigma^2
-\frac{1}{2\sigma^2}\sum_{i=1}^{q}w_i^{(t)}\{r_i^*(\bfalpha)\}^2\\
&\quad+\sum_{j=1}^{d}\left[(1-q_j^{(t)})\log\{(1-\pi)f_0(\beta_{G,j};\tau_0^2)\}
+q_j^{(t)}\log\{\pi f_1(\beta_{G,j};\tau_1^2)\}\right]\\
&\quad+\log p(\tau_0^2)+\log p(\tau_1^2).
\end{aligned}
}
\end{equation}

\subsubsection{Coefficient update}
Conditional on $\bfW=\operatorname{diag}(w_1,\ldots,w_q)$ and the current variance parameters, the coefficient-dependent part of the expected complete log posterior is

\begin{equation}\label{eq:qbeta}
\begin{aligned}
Q(\bfbeta_G)&=-\frac{1}{2\sigma^2}(\bfy^*-\bfX^*\bfbeta_G)^{\top}\bfW(\bfy^*-\bfX^*\bfbeta_G)\\
&\quad+\sum_{j=1}^{d}\left[(1-q_j)\log f_0(\beta_{G,j})+q_j\log f_1(\beta_{G,j})\right].
\end{aligned}
\end{equation}

The pMOM term contains $2\log|\beta_{G,j}|$, so $Q$ is not defined at zero for an included coefficient and need not be globally concave. In numerical work, coefficients are initialized away from zero, a small numerical floor is used in evaluating $\log|\beta_{G,j}|$, and Newton steps are damped by step halving whenever they decrease $Q$ or cross the numerical floor.

The likelihood gradient and Hessian are

\begin{align}
\nabla_{\bfbeta_G}\ell_L&=\frac{1}{\sigma^2}{\bfX^*}^{\top}\bfW(\bfy^*-\bfX^*\bfbeta_G),\label{eq:likgrad}\\
\nabla^2_{\bfbeta_G}\ell_L&=-\frac{1}{\sigma^2}{\bfX^*}^{\top}\bfW\bfX^*.\label{eq:likhess}
\end{align}

For a single coefficient, the component-specific derivatives are

\begin{align}
\frac{\partial}{\partial\beta}\log f_0(\beta;\tau_0^2)&=-\frac{\beta}{\tau_0^2},&
\frac{\partial^2}{\partial\beta^2}\log f_0(\beta;\tau_0^2)&=-\frac{1}{\tau_0^2},\label{eq:spikederiv} \\
\frac{\partial}{\partial\beta}\log f_1(\beta;\tau_1^2)&=\frac{2}{\beta}-\frac{\beta}{\tau_1^2},&
\frac{\partial^2}{\partial\beta^2}\log f_1(\beta;\tau_1^2)&=-\frac{2}{\beta^2}-\frac{1}{\tau_1^2}.\label{eq:slabderiv}
\end{align}

Writing ${\bf x}_j^*$ for the $j$th column of $\bfX^*$ and $\beta_{G,j}$ for the current coordinate, the expected-prior contribution is

\begin{align}
g_j^{(P)}&=-(1-q_j)\frac{\beta_{G,j}}{\tau_0^2}+q_j\left(\frac{2}{\beta_{G,j}}-\frac{\beta_{G,j}}{\tau_1^2}\right),\label{eq:priorgrad}\\
h_j^{(P)}&=-(1-q_j)\frac{1}{\tau_0^2}+q_j\left(-\frac{2}{\beta_{G,j}^2}-\frac{1}{\tau_1^2}\right).\label{eq:priorhess}
\end{align}

The likelihood contributions for the same coordinate are

\begin{equation}\label{eq:coordlik}
g_j^{(L)}=\frac{{\bf x}_j^{*\top}\bfW(\bfy^*-\bfX^*\bfbeta_G)}{\sigma^2},\qquad h_j^{(L)}=-\frac{{\bf x}_j^{*\top}\bfW\bf x_j^*}{\sigma^2}.
\end{equation}

The damped coordinate-Newton update is

\begin{equation}\label{eq:newton}
\beta_{G,j}^{\mathrm{new}}=\beta_{G,j}^{\mathrm{old}}-\rho\frac{g_j^{(L)}+g_j^{(P)}}{h_j^{(L)}+h_j^{(P)}},\qquad 0<\rho\leq1,
\end{equation}

where $\rho$ is reduced until the objective increases and the proposed coefficient remains numerically admissible. Residuals can be updated incrementally after each coordinate, reducing the cost of a sweep from repeated full matrix multiplication.

\subsubsection{Scale and sparsity-hyperparameter updates}
With the inverse-gamma priors in \labelcref{eq:igprior}, maximization of the expected complete log posterior gives closed-form updates for the two variance scales. Let $q_j$ denote the current responsibility in \labelcref{eq:pip}. Then

\begin{align}
\tau_0^{2\,\mathrm{new}}&=\frac{b_0+\frac12\sum_{j=1}^{d}(1-q_j)\beta_{G,j}^2}{a_0+1+\frac12\sum_{j=1}^{d}(1-q_j)},\label{eq:tau0}\\
\tau_1^{2\,\mathrm{new}}&=\frac{b_1+\frac12\sum_{j=1}^{d}q_j\beta_{G,j}^2}{a_1+1+\frac32\sum_{j=1}^{d}q_j}.\label{eq:tau1}
\end{align}

%The $3/2$ term in the denominator of the pMOM update arises because $f_1(\beta;\tau_1^2)$ contains $(\tau_1^2)^{-3/2}$: one factor $(\tau_1^2)^{-1}$ comes from the moment multiplier and one factor $(\tau_1^2)^{-1/2}$ from the Gaussian density. 
{The prior inclusion probability $\pi$ is fixed at a prespecified sparsity level and is not updated within the current GEM/ECM implementation.} In applications with limited $N$ and large $d$, fixing $\pi$ or regularizing the scale ratio can prevent an overly conservative empirical-Bayes solution.

\subsubsection{Variance-component update}
{The between-subject covariance is parameterized as $\bfSigma_b(\bfalpha)=\sum_{k=1}^{K}\alpha_k\bfV_k$. Rather than applying a separate Gaussian ReML or Fisher-scoring step, the current implementation conditionally maximizes the same expected complete-data objective in \labelcref{eq:qcomplete} over the log variance components:}

\begin{equation}\label{eq:alphacm}
{
\bfalpha^{\mathrm{new}}
=\underset{\alpha_{\min}\leq\alpha_k\leq\alpha_{\max}}{\arg\max}\;
Q\!\left(\bfbeta_G^{\mathrm{new}},\sigma^{2\,\mathrm{new}},\tau_0^{2\,\mathrm{new}},\tau_1^{2\,\mathrm{new}},\bfalpha\mid\bfTheta^{(t)}\right).
}
\end{equation}

{This objective includes both $\log|\bfB(\bfalpha)|$ and the $\bfalpha$-dependent whitened residuals. Bounded numerical optimization on $\log\bfalpha$ preserves positivity, and the covariance basis matrices are required to be symmetric and linearly independent. After an accepted update, $\bfSigma_b$, the whitening factors, $\bfy^*$, and $\bfX^*$ are recomputed.}

\subsection{{Generalized EM/ECM algorithm and post-fit PIP integration}}
The complete estimation procedure has {a GEM/ECM loop followed by posterior integration for the PIPs}.

\begin{enumerate}

\item \textbf{Initialization.} Set $\bfSigma_b^{(0)}$ using a scaled identity or diagonal summary of the first-level posterior covariances, for example
\begin{equation*}
\bfSigma_b^{(0)}=c\,\operatorname{diag}\!\left\{N^{-1}\sum_{n=1}^{N}\operatorname{diag}(\bfC_n)\right\},\qquad 0.1\leq c\leq1.
\end{equation*}
Initialize $\bfbeta_G$ by ordinary or robust least squares, {fix $\sigma^2=1$ by default}, initialize $\lambda_i=1$, and choose positive starting values for $\tau_0^2$ and $\tau_1^2${. Fit the model from multiple dispersed initializations}.

\item \textbf{Whitening.} Construct $\bfB$, $\bfy^*$, and $\bfX^*$ from \labelcref{eq:whitenfactor,eq:white} using the current $\bfSigma_b$ {and the prespecified whitening convention}.
\item \textbf{E-step.} Compute $r_i=y_i^*-{\bfx_i^*}^{\top}\bfbeta_G$, update $w_i$ using \labelcref{eq:weight}, and update the inclusion responsibilities $q_j$ using \labelcref{eq:pip}{. Hold both sets of expectations fixed for the subsequent conditional-maximization steps}.

\item \textbf{{Conditional maximization} for group effects.} With $\bfW=\operatorname{diag}(w_i)$, perform coordinate-Newton updates using \labelcref{eq:priorgrad,eq:priorhess,eq:coordlik,eq:newton} until the inner coefficient objective is stable{, with backtracking to prevent a decrease in $Q$}.

\item \textbf{{Conditional maximization} for scales.} Update {$\tau_0^2$ and $\tau_1^2$} using \labelcref{eq:tau0,eq:tau1}; {when scale identification has been established and $\sigma^2$ is not fixed, update it using \labelcref{eq:sigma}. The value of $\pi$ remains fixed}.

\item \textbf{Covariance-component step.} If $\bfSigma_b$ is estimated, update $\bfalpha$ using {\labelcref{eq:alphacm}}, enforce {its numerical bounds}, and repeat the whitening step.

\item \textbf{Convergence{ and start selection}.} Stop when the {maximum relative parameter change}, {relative change in the normalized observed-data objective}, and {maximum absolute coefficient score} are below their prespecified tolerances, or when the maximum iteration count is reached. {Each iteration must not decrease either $Q$ or the normalized observed-data log posterior; a numerically decreasing step is rejected and rolled back. Among the multiple initializations, retain the converged solution with the largest normalized log posterior.}

\item {\textbf{Integrated PIPs.} Conditional on the selected empirical-Bayes hyperparameters, sample $\bfbeta_G$ from its joint posterior using independent parallel-tempering ladders. Each ladder contains a cold replica targeting the posterior and heated replicas that facilitate movement between separated spike and slab modes, with coordinate-wise random-walk proposals, occasional slab-scale jumps, and adjacent-temperature swaps.}
\end{enumerate}

{Let $\widehat{\bfh}=(\widehat{\sigma}^2,\widehat{\tau}_0^2,\widehat{\tau}_1^2,\widehat{\bfalpha})$ collect the fitted hyperparameters, and let $\bfbeta_G^{(m)}$ denote retained draws from the cold replicas. The reported PIP is the Rao--Blackwellized posterior average}

\begin{equation}\label{eq:integratedpip}
{
\widehat{\operatorname{PIP}}_j
=\frac{1}{M}\sum_{m=1}^{M}
\frac{\pi f_1(\beta_{G,j}^{(m)};\widehat{\tau}_1^2)}
{(1-\pi)f_0(\beta_{G,j}^{(m)};\widehat{\tau}_0^2)+\pi f_1(\beta_{G,j}^{(m)};\widehat{\tau}_1^2)}.
}
\end{equation}

{The implementation also reports chain-specific PIPs, batch-means Monte Carlo standard errors, cold- and heated-chain acceptance rates, adjacent-temperature swap rates, and the largest between-chain PIP range. These quantities assess numerical reliability of the posterior integration but do not account for empirical-Bayes uncertainty in $\widehat{\bfh}$.}

The fitted output includes $\widehat{\bfbeta}_G$, the PIP vector {$\widehat{\bfq}=(\widehat{\operatorname{PIP}}_1,\ldots,\widehat{\operatorname{PIP}}_d)^{\top}$}, the {whitened-coordinate} robustness weights $\widehat{\bfw}$, $\widehat{\sigma}^2$, the learned prior scales, convergence information, and (when applicable) $\widehat{\bfalpha}${, together with the mode responsibilities from \labelcref{eq:pip} stored separately from the PIPs}. A coefficient can be declared selected using a prespecified PIP threshold, but reporting the continuous PIP and the corresponding coefficient estimate is preferable to a threshold-only summary.

{For coefficient uncertainty, the implementation returns a conditional Laplace approximation based on the final weighted curvature,}

\begin{equation}\label{eq:laplacecov}
{
\widehat{\bfC}_{\bfbeta}
=\left[
{\bfX^*}^{\top}\widehat{\bfW}\bfX^*/\widehat{\sigma}^2
-\operatorname{diag}\!\left\{
\left.\frac{\partial^2\log m_j(\beta)}{\partial\beta^2}\right|_{\beta=\widehat{\beta}_{G,j}}
\right\}_{j=1}^{d}
\right]^{-1}.
}
\end{equation}

{The square roots of its diagonal entries are reported as approximate posterior standard deviations conditional on the fitted hyperparameters. Any nonpositive local eigenvalue is flagged and regularized only for matrix inversion.}

\subsection{Numerical safeguards, interpretation, and computational cost}
The pMOM density vanishes at zero and contains $\log|\beta_{G,j}|$ in the objective. We therefore use a small positive numerical floor, avoid evaluating the slab exactly at zero, and use step halving {and step-size caps} for coordinate updates. All covariance matrices are symmetrized numerically before factorization. Small ridge terms may be added if a first-level covariance or a variance-component update is nearly singular. The degrees of freedom $\nu$ controls the robustness--efficiency trade-off: smaller values give stronger downweighting of large residuals, whereas large values approach Gaussian PEB. It should be fixed in advance or selected through a prespecified sensitivity analysis rather than tuned after inspecting the scientific contrasts.

Low Student-$t$ weights identify observations that are influential under the fitted model, not necessarily erroneous observations. They should be compared with motion, signal quality, first-level posterior variance, and scientifically meaningful subject heterogeneity before any data-quality decision is made. PIPs likewise describe evidence for the fitted spike--slab decomposition and are sensitive to the prior inclusion probability, scale hyperpriors, and sample size{. Because coordinate-wise Student-$t$ robustness is whitening-dependent, sensitivity to a scientifically reasonable alternative whitening convention should also be examined}.

For dense matrices, an incremental-residual coordinate sweep costs approximately $O(qd)$. Whitening and covariance-component updates can dominate when $p$ is large because they require factorization of subject-level covariance blocks. Coefficient and weighting updates are readily parallelized across independent model specifications or parameter blocks, whereas the variance-component step is updated centrally to maintain a common covariance representation{. Parallel-tempering PIP integration adds a post-fit cost proportional to the number of ladders, temperatures, retained iterations, and coefficient updates}.

%%%
\section{Simulation study}
\label{sec:simulation}

We evaluated the proposed robust and sparse group-level estimator under controlled contamination while varying sample size, network dimension, first-level uncertainty, covariate structure, signal strength, and sparsity. We compared it with {three simplified variants} and an R port of the standard SPM implementation of PEB. Unless stated otherwise, the evaluation summaries are based on {50} independent replicates per condition. {The tuning experiment used separate random-number streams and five replicates under each of four tuning conditions.}

\subsection{Data-generating process}
\label{sec:simulation-dgp}

Each replicate was generated from a controlled hierarchical model matching the group-level mean and covariance structure in \labelcref{eq:peb-mean,eq:peb-cov}. The data-generating distribution was deliberately separated from the fitted priors: clean first-level errors were Gaussian, and contamination was introduced separately to evaluate robustness. Specifically,
\begin{equation}
\bfmu_n={(\bfx_{G,n}^{\top}\otimes\bfI_p)\bfbeta_G},\qquad
\bfeta_n=\bfmu_n+\bfb_n+\bfe_n+{\bfo_n},
\label{eq:simulation-dgp}
\end{equation}
where $\bfmu_n$ is the subject-specific block of the group mean, {$\bfx_{G,n}$ is the corresponding row of the group design,} $\bfb_n\sim N(\bfzero,\sigma_b^2\bfI_p)$ represents between-subject variation, $\bfe_n\sim N(\bfzero,\bfC_n)$ represents first-level posterior uncertainty, and {$\bfo_n$} is a contamination term. We used $\sigma_b=0.15$. {Thus, heavy tails were not built into the clean data-generating model, and departures from Gaussianity arose only through the explicitly controlled contamination mechanism.} The first-level covariance was varied across
\begin{equation}
\bfC_n\in\{0.05\bfI_p,\;0.20\bfI_p,\;0.60\bfI_p,\;0.20\bfR\},
\label{eq:simulation-covariance}
\end{equation}
where $\bfR$ is a correlation matrix. These settings represent small, moderate, and large diagonal uncertainty and a correlated moderate-uncertainty regime.

For the correlated regime, a $p\times p$ matrix $\bfZ$ with independent standard normal entries was generated {and we set}
$$
\bfS=\frac{\bfZ\bfZ^\top}{p},\qquad
\bfD=\operatorname{diag}\left(S_{11}^{-1/2},\ldots,S_{pp}^{-1/2}\right),\qquad
\bfR=\bfD\bfS\bfD.
$$
Thus, $\bfR$ was a randomly generated positive-definite correlation matrix with unit diagonal, and $\bfC_n=0.20\bfR$. All subjects within a replicate shared the same first-level covariance matrix {in the principal evaluation conditions. In the correlated tuning conditions, a new correlation matrix was generated for each subject. A separate sensitivity analysis replaced the stylized matrices with subject-level covariance matrices sampled with replacement from the empirical NARPS covariance archive.}

Parameters $1,\ldots,p-6$ were treated as intrinsic-like coefficients, and the final six represented two gain-related and four loss-related modulatory effects. {Unless a separate sparsity setting was specified, each intrinsic coefficient was initially activated with probability $0.25$, subject to at least three active intrinsic coefficients whenever $p-6\geq3$.} The signs {of the active intrinsic coefficients} were sampled with equal probability, and their magnitudes were drawn from $\operatorname{Uniform}(0.3,0.6)$. At the baseline effect scale $\kappa=1$, the six modulatory coefficients were fixed at $0.30$, $0.25$, $-0.28$, $-0.10$, $-0.30$, and $-0.08$. The effect-scale parameter $\kappa$ multiplied every nonzero coefficient, so $\kappa=0$ produced the global null and larger values increased the signal-to-noise ratio. {The separate $p=40$ sparsity experiment used the simpler construction described in Section~\ref{sec:simulation-conditions}.}

The truth-generation seed depended on the replicate index rather than the simulation condition. Thus, the active support, signs, and baseline magnitudes varied across the {50} replicates but were shared across compatible conditions having the same $p$ and number of group-level regressors. The truth was first generated at $\kappa=1$ and then multiplied by the condition-specific value of $\kappa$, ensuring that the effect-scale sweep changed only signal magnitude. {Separate replicate-indexed seed blocks were used for coefficient generation, group-design generation, data generation, and posterior sampling. Tuning, evaluation, and validation used disjoint seed blocks. This design supports paired comparisons among methods and prevents changes in the true coefficient vector from being confounded with nominal factor sweeps.}

{Contamination was added after generating the clean subject-level summaries. Let $\phi$ be the contamination fraction, $\delta$ its magnitude, and $N_e$ the number of subjects eligible for contamination.} We considered three contamination geometries. Under cell-wise contamination, every eligible subject--parameter entry was independently contaminated with probability $\phi$ and shifted by either {$+\delta$ or $-\delta$}, with equal probability. Under whole-subject contamination, $m_o=\max\{1,\operatorname{round}(\phi N_e)\}$ of the $N_e$ eligible subjects were sampled without replacement. All $p$ entries for each selected subject were contaminated, with signs generated independently across parameters. Structured contamination selected the same number of subjects but generated one sign per subject, applying the resulting {$+\delta$ or $-\delta$} shift coherently to all $p$ entries. Thus, the two subject-level geometries differed only in whether signs varied across parameters. {When $\phi=0$, no contamination was added.}

\subsection{Estimators under comparison, tuning, and {computation}}
\label{sec:simulation-estimators}

{Five} estimators were fitted to each simulated dataset. \emph{Proposed} denotes the robust--sparse estimator, which combines the Student-$t$ likelihood with the nonlocal pMOM spike-and-slab prior. \emph{Robust-only} retains the Student-$t$ likelihood but replaces the spike-and-slab prior with Gaussian ridge regularization. \emph{Sparse-only} retains the pMOM prior but approximates the Gaussian limit of the likelihood by setting $\nu=10^6$. {\emph{Local-slab} combines Student-$t$ robustness with a Gaussian spike-and-slab prior. Conditional on the whitening induced by the robust-ridge covariance estimate, it uses a Gaussian spike with standard deviation $0.05$ and a Gaussian slab with standard deviation $\sqrt{3}$, the latter having the same variance as a first-order pMOM slab with scale $1$.} \emph{SPM-PEB} denotes an R port of the standard SPM-PEB procedure, with Bayesian model-reduction (BMR) inclusion probabilities used as its selection scores. {Its coefficient-prior variance was $1$.}

For the two pMOM estimators, $\tau_0=0.05$ and $\tau_1=1$ {were starting values. The two scales were then updated under the inverse-gamma hyperpriors specified in Section~\ref{sec:method}, with $(a_0,b_0)=(2,0.01)$ and $(a_1,b_1)=(2,1)$. The residual scale was fixed at $\sigma^2=1$ in the simulations so that it was not confounded with the covariance components.} \emph{Robust-only} used a ridge precision of $10^{-3}$. The principal and robustness-breadth experiments used parameter-specific diagonal between-subject covariance components, whereas the $p=40$ {sparsity and covariance-recovery experiments} used one identity-matrix component.

{The nonlocal objective can have separated modes, so the pMOM and robust-ridge estimators were fitted from three dispersed initializations and the solution with the largest normalized log posterior was retained. Generalized-EM fitting was allowed up to 2000 iterations. For the pMOM estimator, convergence required relative parameter and objective changes below $5\times10^{-4}$, a maximum absolute coefficient score below $10^{-3}$, and successful covariance-component optimization. The robust-ridge estimator required the same parameter and objective tolerances together with successful covariance-component optimization, while its coefficient score was retained as a diagnostic rather than used as a stopping rule. Objective and expected-complete-data monotonicity, the selected initialization, the between-start objective range, and any Laplace-Hessian regularization were recorded when applicable.}
%A simulation stage stopped before producing summaries if a required fit failed its convergence criteria. Such fits were written to a separate diagnostic file rather than silently discarded.

{For \emph{Proposed} and \emph{Sparse-only}, coefficient estimates were posterior means conditional on the fitted empirical-Bayes hyperparameters. Integrated PIPs were computed after generalized-EM convergence by parallel-tempering Metropolis sampling of $\bfbeta_G$. Four independent temperature ladders were run for 5000 iterations, with 1500 burn-in iterations and thinning by five. Each ladder used eight geometrically spaced inverse temperatures from $1$ to $0.05$, adjacent-temperature swaps every two iterations, and slab-scale large-jump proposals with probability $0.10$. At each retained draw from the cold chain, the conditional slab responsibility was evaluated and then averaged, giving a Rao--Blackwellized estimate of the integrated PIP. Batch-means Monte Carlo standard errors and the largest between-chain PIP range were retained as sampling diagnostics. The \emph{Local-slab} PIPs were estimated by averaging conditional Gaussian-mixture responsibilities over four post-burn-in Gibbs chains.}

For \emph{Robust-only}, selection was based on the two-sided normal-approximation confidence score
\begin{equation}
S_j=1-2\Phi\!\left(-\frac{|\widehat{\beta}_j|}{s_j}\right),
\label{eq:simulation-ridge-score}
\end{equation}
where $s_j$ is the {conditional} posterior standard deviation from the final weighted ridge fit and $\Phi$ is the standard normal cumulative distribution function. {The five selection-score series were therefore the integrated PIP for \emph{Proposed} and \emph{Sparse-only}, the local-mixture PIP for \emph{Local-slab}, the normal-approximation score for \emph{Robust-only}, and the BMR inclusion probability for \emph{SPM-PEB}.}

{The Student-$t$ degrees of freedom $\nu$ and prior inclusion probability $\pi$ were chosen in a separate experiment that differed deliberately from the evaluation design. Tuning used $N=54$, $p=18$, subject-specific covariance matrices in the correlated conditions, and the four settings
\begin{equation*}
\begin{aligned}
(\phi,\delta,\text{covariance})\in\{&
(0,3,\text{moderate diagonal}),
(0.06,3,\text{correlated}),\\
&(0.15,3,\text{moderate diagonal}),
(0.15,6,\text{correlated})\}.
\end{aligned}
\end{equation*}
Five independently generated tuning replicates were used per setting, giving 20 datasets for each candidate. The same datasets were fitted by all candidates. The candidate grids were $\nu\in\{2,3,4,6,8\}$ and $\pi\in\{0.2,0.3,0.4,0.5,0.6,0.7\}$, yielding 30 candidates for \emph{Proposed}, five for \emph{Robust-only}, and six for \emph{Sparse-only}. Within each estimator family, mean PR-AUC and RMSE were separately min--max normalized across candidates, and the selected setting maximized
\begin{equation}
0.5\,\widetilde{\mathrm{PR\mbox{-}AUC}}+0.5\{1-\widetilde{\mathrm{RMSE}}\}.
\label{eq:simulation-tuning-score}
\end{equation}
The primary ablation comparison used the pair selected for \emph{Proposed}: its selected $\nu$ was also used by \emph{Robust-only}, and its selected $\pi$ by \emph{Sparse-only} and \emph{Local-slab}. This holds the shared components fixed while removing or replacing one component of the proposed method. A supplementary fairness analysis instead used the values selected separately for \emph{Robust-only} and \emph{Sparse-only}. The selected settings and their tuning summaries are reported with the simulation results.}

\subsection{Performance measures {and numerical diagnostics}}
\label{sec:simulation-measures}

Let $A_j\in\{0,1\}$ indicate whether coefficient $j$ is active and let $S_j$ denote its method-specific selection score. The primary threshold-free selection measure was the area under the precision--recall curve (PR-AUC), computed in its non-interpolated average-precision form after ranking scores from largest to smallest. {ROC-AUC was reported as a complementary measure.} For scores exceeding $0.5$, we also recorded the true-positive rate, false-positive rate, false-discovery rate, Matthews correlation coefficient, {and number selected}. Because {integrated PIPs, local-mixture PIPs}, BMR probabilities, and normal-approximation confidence scores are not identically calibrated, these fixed-threshold quantities should be interpreted {cautiously}. We additionally computed coefficient RMSE for all, active, and null coefficients and recorded estimator fitting time, excluding data generation.

To assess the purity of highly supported selections, we calculated the replicate-level false-discovery proportion among coefficients with scores greater than $0.95$, denoted $\mathrm{FDP}_{>0.95}$. {We also recorded the number selected above this threshold.} Replicates with no selections above this threshold were excluded from the average. Thus, $\mathrm{FDP}_{>0.95}$ represents a conditional mean FDP rather than the conventional FDR, for which the FDP is defined as zero when no discoveries are made. {Its denominator was therefore reported.}

{To compare selection performance at approximately common error rates, we also used cross-fitted, method-specific thresholds. Replicates were split into two folds by replicate number. For each method and target false-positive rate in $\{0.05,0.10\}$, the threshold was the corresponding empirical quantile of null scores in the other fold and was then applied to the held-out fold. The combined held-out selections were used to report the observed false-positive rate, power, false-discovery rate, mean number selected, discovery frequency, number of evaluated coefficients, and number of replicate--condition units. This analysis compares methods at matched empirical false-positive rates rather than imposing numerical equality of their native scores.}

{For intercept-only conditions, we additionally considered two scientifically structured contrasts: the mean of the two gain-related coefficients and the mean of the four loss-related coefficients. Using 4000 draws from the conditional normal approximation defined by each fitted coefficient vector and covariance matrix, we estimated the probability that the gain contrast was positive and the probability that the loss contrast was negative.}

PR-AUC {and ROC-AUC are} undefined in the global-null condition ($\kappa=0$), because no positive class is present. For that condition, we report false-positive and selection frequencies instead.

For a condition-specific metric with $B_m$ observed replicate values $m_1,\ldots,m_{B_m}$, the Monte Carlo standard error of the replicate mean was calculated as
\begin{equation}
\widehat{\operatorname{MCSE}}(\overline m)=\frac{s_m}{\sqrt{B_m}},
\label{eq:simulation-mcse}
\end{equation}
where $s_m$ is the corresponding sample standard deviation. Here $B_m={50}$ except for the conditional $\mathrm{FDP}_{>0.95}$, for which $B_m$ is the number of replicates containing at least one selection above $0.95$. For {summaries of multiple matched conditions}, each metric was first averaged with equal weight across the included conditions within a replicate. The MCSE was then computed across the available replicate-level averages. Figures display replicate means, with MCSE bars omitted for visual clarity.

{In addition to performance summaries, we recorded convergence status, stopping reason, iteration count, objective and expected-complete-data monotonicity, the maximum coefficient score, covariance-optimizer status, between-start objective variation, conditional Laplace regularization, integrated-PIP MCSEs, and between-chain PIP ranges. These diagnostics distinguish statistical performance from numerical reliability.}

\subsection{Simulation conditions {and sensitivity analyses}}
\label{sec:simulation-conditions}

\paragraph{Principal factor sweeps.}
The baseline condition used $N=48$, $p=16$, an intercept-only design, $\kappa=1$, $\phi=0.08$, and moderate diagonal first-level uncertainty. {Contamination was cell-wise and had magnitude $\delta=6$. From this baseline, one factor was varied at a time.} The principal factor sweeps examined contamination fraction $\phi\in\{0,0.10,0.20\}$ crossed with $N\in\{24,48,80\}$; effect scale $\kappa\in\{0,0.5,1,1.5,2\}$; network size $p\in\{12,16,22\}$; the four covariance regimes in \labelcref{eq:simulation-covariance}; and an $r=3$ design comprising an intercept, one binary covariate, and one continuous covariate. Under the default truth construction, the $p=12$ condition contained three active intrinsic coefficients, six active modulatory coefficients, and three null coefficients.

In the covariate condition, the binary covariate represented gender and was generated as $\operatorname{Bernoulli}(0.5)$, with $1$ denoting male and $0$ denoting female. The continuous age-like covariate was drawn from a standard normal distribution and standardized within each simulated sample. The binary covariate had true effects $0.35\kappa$ and $0.30\kappa$ on the first gain-related and first loss-related parameters, respectively, and the continuous covariate had true effect $-0.25\kappa$ on the third loss-related parameter. All other covariate effects were zero. A new covariate design was generated for each replicate and, in the robustness-breadth experiments, was shared across matched contamination conditions within that replicate. {Performance in this condition was evaluated on the covariate coefficients, excluding the intercept.}

\paragraph{Contamination magnitude and geometry.}
To distinguish mild or moderate perturbations from the severe shift used in the baseline, we compared clean data with $10\%$ cell-wise contamination at magnitudes $\delta\in\{2,3,6\}$. The contamination-geometry experiment used $N=48$, $p=16$, and $15\%$ contamination under each of the three geometries. {The contamination magnitude was $\delta=6$. Covariate robustness was examined at $\phi\in\{0,0.10,0.20\}$ under cell-wise contamination.} The subgroup-masking experiment compared clean data with whole-subject contamination applied to $30\%$ of simulated male subjects. Signs varied independently across parameters within each contaminated subject.

\paragraph{Higher-dimensional sparsity.}
A separate higher-dimensional experiment used $p=40$, $N=48$, active fractions $s\in\{0.05,0.10,0.25\}$, and contamination fractions $\phi\in\{0,0.10\}$. For each replicate and value of $s$, $k=\max\{1,\operatorname{round}(sp)\}$ active indices were sampled from all $p$ coefficients. Their signs were sampled with equal probability, and their magnitudes were sampled from $\operatorname{Uniform}(0.4,0.7)$. Thus, $s=0.05$ yielded two active and 38 null coefficients. {This experiment used a single between-subject variance component multiplying $\bfI_p$.}

\paragraph{Covariance and whitening sensitivity.}
The empirical-covariance experiment used $N=48$, $p=22$, $10\%$ cell-wise contamination of magnitude $3$, and an intercept-only group design. It compared the stylized correlated covariance $0.20\bfR$ with subject-level first-level covariance matrices sampled with replacement from the archive of 48 corrected NARPS summaries. To assess whether element-wise Student-$t$ weights depended materially on the whitening representation, a separate $N=48$, $p=16$ experiment with correlated covariance and $10\%$ contamination of magnitude $3$ compared precision-Cholesky and symmetric whitening. The precision-Cholesky analysis was also repeated after reversing the parameter order and then transforming the estimates and PIPs back to the original order. Maximum absolute differences in coefficient estimates and PIPs were recorded.

\paragraph{Algorithmic validation and robustness checks.}
In 50 clean replicates with $N=80$, $p=12$, moderate diagonal first-level uncertainty, and true between-subject covariance $0.15^2\bfI_p$, we fitted a single variance component $\alpha\bfI_p$ using the Gaussian-limit likelihood matched to the Gaussian validation design. This experiment assessed bias and RMSE in $\widehat{\alpha}$, finite-difference gradients of the common normalized log posterior with respect to $\bfbeta_G$ and $\log\alpha$, convergence, and monotonicity of the observed and expected-complete-data objectives. A scalar test case also compared the responsibility evaluated at the posterior mode with an integrated PIP obtained by direct numerical integration. Finally, sensitivity to the Student-$t$ degrees of freedom and to shared versus arm-specific tuning was assessed on three representative conditions: the baseline, the clean $N=48$ condition, and the $N=48$ condition with $20\%$ contamination. All sensitivity fits used the same datasets as their primary counterparts.

\subsection{Results}
\label{sec:simulation-results}

\paragraph{Tuning and numerical performance.}
Tuning selected $(\nu,\pi)=(2,0.6)$ for \emph{Proposed}. The combination $(2,0.7)$ was nearly tied, with composite scores of $0.969$ and $0.968$, respectively. Separate tuning selected $\nu=2$ for \emph{Robust-only} and $\pi=0.5$ for \emph{Sparse-only}. The primary ablation comparison nevertheless used the proposed estimator's shared settings, namely $\nu=2$ for the Student-$t$ arms and $\pi=0.6$ for the spike-and-slab arms. Complete tuning results are given in Table~S1 of the Supplementary Materials. %\ref{tab:supp-tuning}

All 5000 final fits in the principal grid met their estimator-specific convergence criteria. Two proposed-model fits initially ended with a covariance-optimizer status code despite negligible parameter, objective, and coefficient-score changes. Both converged when rerun with the prespecified \texttt{nlminb} fallback and were replaced only after passing the full diagnostic checks. Across the principal grid, the maximum PIP MCSE was $0.036$ for \emph{Proposed}, $0.033$ for \emph{Sparse-only}, and $0.020$ for \emph{Local-slab}. The median largest between-chain PIP ranges were $0.058$, $0.109$, and $0.051$, respectively, while the corresponding maxima were $0.309$, $0.349$, and $0.157$. These ranges show residual sampling sensitivity for some integrated PIPs, especially under the Gaussian pMOM ablation, so the aggregate performance results are accompanied by the full numerical diagnostics in Table~S2. %\ref{tab:supp-diagnostics}

\begin{figure}[htbp]
\centering
\includegraphics[width=\textwidth]{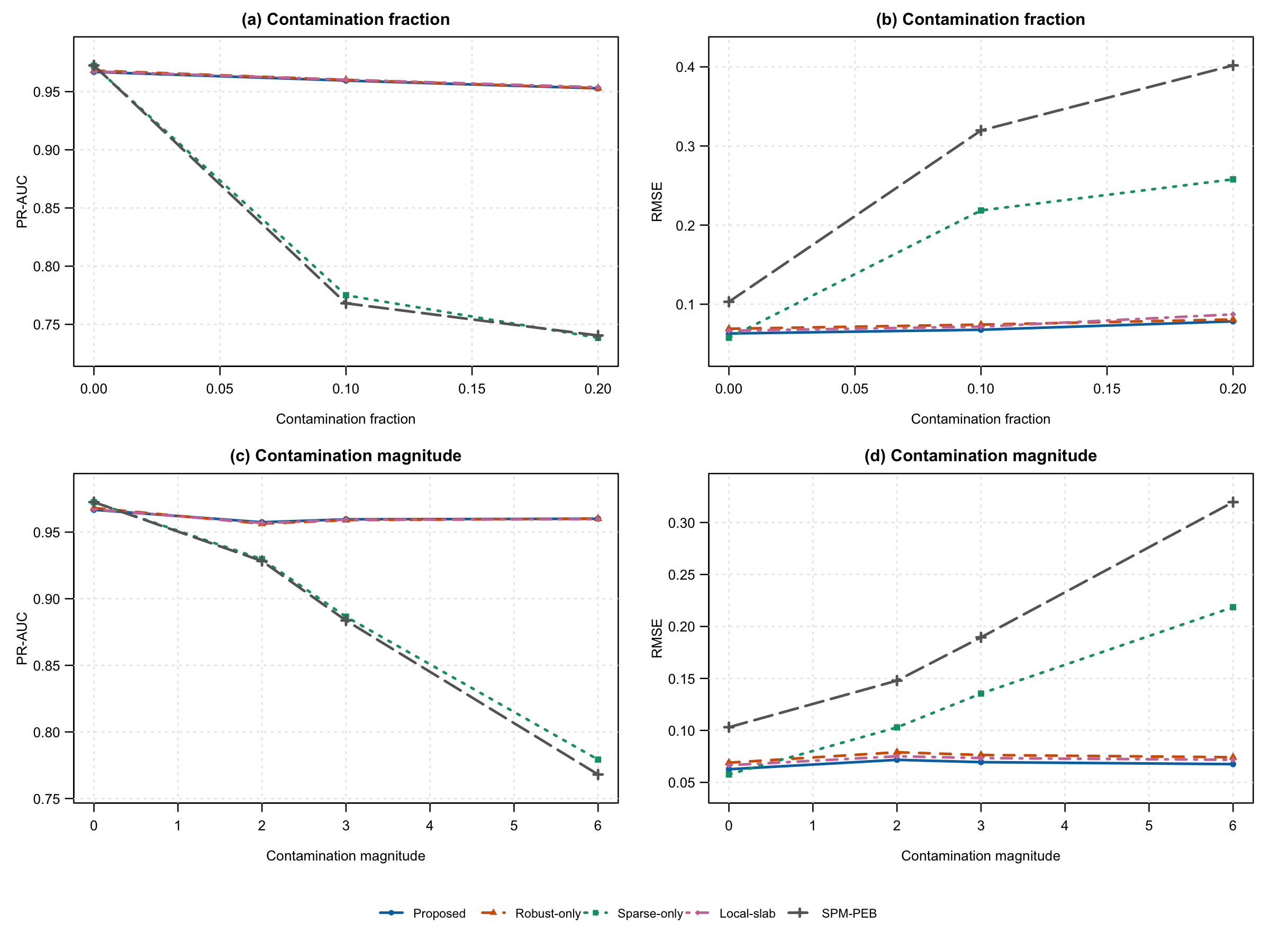}
\caption{Performance under cell-wise contamination. Panels (a) and (b) vary the contamination fraction at $N=48$ and magnitude $6$. Panels (c) and (d) compare clean data with $10\%$ contamination of magnitudes $2$, $3$, and $6$. Curves show means over 50 matched replicates.}
\label{fig:sim-main}
\end{figure}

\paragraph{Principal comparison.}
Table~\ref{tab:sim-headline} and Figure~\ref{fig:sim-main} summarize the principal baseline and contamination results. {Across the baseline and the nine contamination--sample-size conditions, the three Student-$t$ estimators had nearly identical threshold-free discrimination, with mean PR-AUC between $0.954$ and $0.956$ (Table~\ref{tab:sim-headline}). The proposed estimator combined this robustness with the lowest average RMSE, $0.075$, and the highest MCC at the common $0.5$ threshold, $0.697$. \emph{Local-slab} was close, with RMSE $0.081$ and MCC $0.663$, whereas \emph{Robust-only} had MCC $0.488$ because its normal-approximation score selected many null coefficients.} As contamination increased, the Student-$t$ estimators maintained substantially higher PR-AUC and lower RMSE than \emph{Sparse-only} and \emph{SPM-PEB}. SPM-PEB was much faster than the other estimators, while the proposed estimator was the most computationally intensive.

\begin{table}[htbp]
\centering
\small
\setlength{\tabcolsep}{4pt}
\caption{Performance averaged with equal weight over the baseline and the nine contamination--sample-size conditions. Conditions were first averaged within each matched replicate, after which means and Monte Carlo standard errors (in parentheses) were calculated across 50 replicate-level averages. Runtime is in seconds.}
\label{tab:sim-headline}
\begin{tabular}{lrrrr}
\toprule
Estimator & PR-AUC & RMSE & MCC & Runtime\\
\midrule
\emph{Proposed} & 0.956 (0.002) & 0.075 (0.001) & 0.697 (0.006) & 342.7 (20.8)\\
\emph{Robust-only} & 0.954 (0.002) & 0.082 (0.001) & 0.488 (0.012) & 225.5 (3.3)\\
\emph{Sparse-only} & 0.828 (0.005) & 0.183 (0.002) & 0.373 (0.012) & 90.3 (0.6)\\
\emph{Local-slab} & 0.956 (0.002) & 0.081 (0.001) & 0.663 (0.006) & 238.7 (3.3)\\
\emph{SPM-PEB} & 0.822 (0.005) & 0.288 (0.003) & 0.386 (0.011) & 15.1 (0.1)\\
\bottomrule
\end{tabular}
\end{table}

{The contamination trends in Figure~\ref{fig:sim-main} show where these differences arose. With $N=48$ and no contamination, \emph{Sparse-only} and \emph{SPM-PEB} had slightly higher PR-AUC than \emph{Proposed}, although the proposed estimator had lower RMSE than SPM-PEB. At $20\%$ contamination, \emph{Proposed}, \emph{Robust-only}, and \emph{Local-slab} retained PR-AUCs of $0.953$, $0.953$, and $0.954$, respectively. Their RMSEs were $0.078$, $0.081$, and $0.087$. In contrast, PR-AUC fell to approximately $0.74$ for both Gaussian-likelihood estimators, with RMSEs of $0.258$ for \emph{Sparse-only} and $0.402$ for \emph{SPM-PEB}. Increasing the contamination magnitude produced the same separation. The Student-$t$ methods remained stable, while the Gaussian methods deteriorated progressively.}

{The matched-error-rate analysis led to the same conclusion without treating the five native scores as identically calibrated. At an observed false-positive rate near $0.05$, power was $0.773$ for \emph{Proposed}, $0.790$ for \emph{Local-slab}, and $0.766$ for \emph{Robust-only}, compared with $0.389$ for \emph{Sparse-only} and $0.382$ for \emph{SPM-PEB}. At a target false-positive rate of $0.10$, the corresponding powers were $0.821$, $0.824$, $0.805$, $0.455$, and $0.460$ (Supplementary Table~S3). Thus, Student-$t$ weighting supplied the main protection against contamination, while spike-and-slab regularization improved fixed-threshold selection relative to robust ridge.} %\ref{tab:supp-matched-fpr}

\begin{figure}[htbp]
\centering
\includegraphics[width=\textwidth]{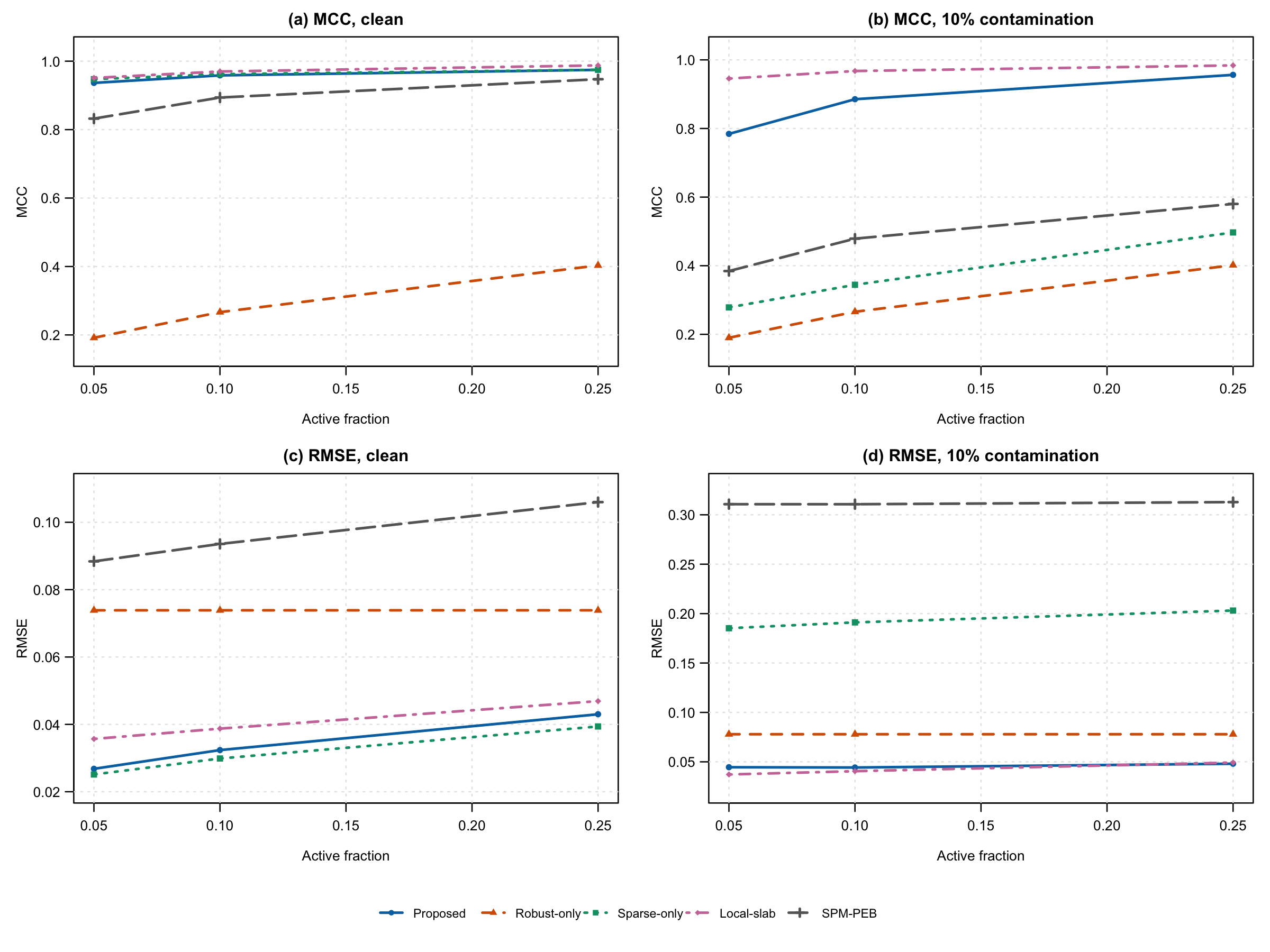}
\caption{Higher-dimensional sparse experiment with $N=48$ and $p=40$. Panels (a) and (b) show MCC at the common score threshold $0.5$, and panels (c) and (d) show coefficient RMSE. The left column contains clean data and the right column contains $10\%$ cell-wise contamination. Curves show means over 50 matched replicates.}
\label{fig:sim-sparse}
\end{figure}

\paragraph{Sparsity and broader robustness conditions.}
The $p=40$ experiment further separated robust estimation from score calibration (Figure~\ref{fig:sim-sparse}). Under clean data, the strong simulated signals yielded PR-AUC equal to $1$ for all methods. The spike-and-slab methods nevertheless achieved far higher MCC than \emph{Robust-only} because they avoided widespread null selections. Under $10\%$ contamination, \emph{Proposed} and \emph{Local-slab} retained PR-AUC equal to $1$ at every active fraction, whereas PR-AUC ranged from $0.553$ to $0.755$ for \emph{Sparse-only} and from $0.551$ to $0.754$ for \emph{SPM-PEB}. The local Gaussian slab had somewhat higher MCC than the pMOM slab in this strong-signal setting, particularly when only $5\%$ of coefficients were active. The pMOM slab therefore did not uniformly improve on the local slab. Rather, the broad gain came from combining robust weighting with a sparse mixture prior.

Under the global null, \emph{Proposed} selected an average of $0.14$ coefficients and \emph{Local-slab} selected $0.06$, compared with $8.40$ for \emph{Robust-only}, $3.44$ for \emph{Sparse-only}, and $2.26$ for \emph{SPM-PEB}. The Student-$t$ methods were also stable across cell-wise, whole-subject, and structured-subject contamination. For \emph{Proposed}, PR-AUC ranged from $0.958$ to $0.961$ and RMSE from $0.073$ to $0.074$ across these geometries. In the covariate experiment, the proposed estimator retained the highest PR-AUC among the five methods at $10\%$ and $20\%$ contamination, although \emph{Local-slab} had slightly lower RMSE and fewer false selections. Complete results for signal strength, network size, covariance regime, covariates, contamination geometry, subgroup masking, and higher-dimensional sparsity appear in Supplementary Tables~S4--S12. %\ref{tab:supp-prauc-full}--\ref{tab:supp-sparse}

\paragraph{Sensitivity and validation.}
Changing the Student-$t$ degrees of freedom from $2$ to $3$ had little effect in the baseline and clean conditions. Under $20\%$ contamination at $N=48$, however, the proposed estimator's PR-AUC decreased from $0.953$ to $0.936$ and RMSE increased from $0.078$ to $0.094$. This supports $\nu=2$ for the severe-contamination simulation benchmark. Replacing the shared $\pi=0.6$ by the separately tuned $\pi=0.5$ for \emph{Sparse-only} produced only small changes and did not alter the comparison (Supplementary Tables~S13 and~S14). %\ref{tab:supp-nu}; \ref{tab:supp-arm-tuning}

The whitening experiment yielded similar aggregate PR-AUC and RMSE across precision-Cholesky, symmetric, and reversed-order analyses. Individual estimates and PIPs were more sensitive. Relative to the primary precision-Cholesky analysis, the mean replicate-level maximum absolute PIP difference was $0.251$ under symmetric whitening and $0.341$ after reversing the parameter order, with maxima of $0.649$ and $0.672$. These findings confirm that {independent Student-$t$ weighting of the whitened coordinates} is not rotation invariant. The precision-Cholesky factor and DCM parameter ordering must therefore be treated as part of the model specification rather than as arbitrary computational choices (Supplementary Table~S16). %\ref{tab:supp-whitening}

Replacing the stylized correlated covariance with resampled NARPS covariance matrices preserved the qualitative ordering of the methods. Under the empirical covariance matrices, \emph{Proposed} achieved PR-AUC $0.998$, RMSE $0.045$, and MCC $0.799$, compared with $0.943$, $0.066$, and $0.763$ under the stylized correlated covariance. \emph{Local-slab} produced similar PR-AUC and MCC, while \emph{Robust-only} had slightly lower RMSE but selected many null coefficients, with a false-positive rate of $0.551$. Performance of \emph{Sparse-only} changed little across the two covariance regimes. All 500 fits in this sensitivity analysis converged. Complete results are reported in Supplementary Table~S15. %\ref{tab:supp-empirical-covariance}

The scalar validation gave a responsibility of $0.00004$ at the posterior mode but an integrated PIP of $0.740$, demonstrating why posterior integration was necessary. In the Gaussian-limit covariance-component validation, the mean estimate was $0.0192$ when the true value was $0.0225$, with bias $-0.0033$ and RMSE $0.0111$. Five of the 50 estimates reached the lower bound, indicating some finite-sample boundary behavior. Finite-difference coefficient and log-variance-component gradients were at most $8.3\times10^{-4}$ and $2.3\times10^{-5}$, respectively, and both objective sequences were monotone in all 50 validation replicates. Full validation results are reported in Supplementary Tables~S17 and~S18. %\ref{tab:supp-objective-validation}; \ref{tab:supp-pip-validation}

%%%
\section{Real-data analysis}
\label{sec:application}
Having established the estimator's operating characteristics under known truth in Section~\ref{sec:simulation}, we next examined its behavior on observed effective-connectivity data. This application asks a complementary methodological question---whether the proposed model yields interpretable group estimates, transparent robustness diagnostics, and conclusions that remain comparable with standard \emph{SPM-PEB}. It is not intended to establish a new substantive claim about valuation connectivity.

\subsection{Data and analysis pipeline}
The source NARPS data set (OpenNeuro ds001734, version 1.0.4) contains 108 healthy participants who completed one of two versions of a mixed-gambles task \cite{botvinik2019narps,botvinik2019fmri}. On each trial, participants accepted or rejected a prospect offering equal probabilities of a monetary gain and loss. In the equal-indifference version, the range of potential losses was half the range of potential gains, whereas the equal-range version used matched gain and loss ranges. Each of the four runs comprised 64 trials and 453 functional volumes acquired at a repetition time of 1~s. Responses were required within 4~s, and the intertrial intervals were jittered. Imaging was performed using a 3~T Siemens Prisma scanner. Functional images were acquired using T2*-weighted multiband echo-planar imaging with acceleration factor 4, repetition time 1000~ms, echo time 30~ms, 2-mm isotropic voxels, and 64 slices. T1-weighted MPRAGE images were acquired at 1-mm resolution.

Because of local storage and processing constraints associated with the subject-level DCM derivatives, the analysis was restricted to the first 48 available NARPS participants in ascending subject-ID order. Participants \texttt{sub-020} and \texttt{sub-021} were skipped because their NIfTI files were not detected owing to an error in the local processing pipeline at that stage. Selection was not based on behavioral outcomes, image quality, demographic characteristics, or estimated connectivity. The resulting sample included 24 female participants, 25 participants who completed the equal-indifference task, and 23 who completed the equal-range task. Ages ranged from 20 to 37 years, with a mean of 26.1 years and a standard deviation of 3.6 years. The identifiers of all 48 analyzed participants are provided in the public reproducibility repository.

Functional images were preprocessed using fMRIPrep. Processing included susceptibility-distortion and slice-timing correction, head-motion correction, boundary-based coregistration, and normalization to the MNI152NLin2009cAsym template. Spatial smoothing was not applied because the regional summaries described below already provide spatial averaging and additional smoothing could reduce regional specificity by mixing signals across neighboring regions \cite{alakorkko2017effects}. First-level modeling, regional time-series extraction, and DCM specification were performed in SPM25 (Wellcome Centre for Human Neuroimaging, London, UK) using MATLAB R2025a.

For each participant, a single first-level general linear model treated the four runs as separate sessions. Within each run, trials were classified using the gain-to-loss ratio $R=\text{gain}/\text{loss}$ as gain-dominant ($R\geq2$), loss-dominant ($R\leq0.5$), or intermediate otherwise. The thresholds reflect the commonly observed loss-aversion indifference ratio, under which a gain of approximately twice the potential loss is required for acceptance, together with its log-symmetric counterpart \cite{tom2007}. The corresponding event regressors were convolved with the canonical hemodynamic response function. Each run additionally included the six head-motion parameters and their temporal derivatives, framewise displacement, anatomical CompCor components, and non-steady-state outlier indicators as nuisance regressors, together with a 128-s high-pass filter. Modeling all runs within one general linear model ensured consistent temporal filtering, whitening, and scaling across sessions.

The four regions were defined using bilateral anatomical masks from the Automated Anatomical Labeling atlas. The vmPFC mask comprised the medial-orbital and superior-medial-frontal parcels. The region labeled ventral striatum comprised the caudate and putamen parcels, and the remaining masks represented the amygdala and anterior insula. The regional signal was summarized by the first eigenvariate of the voxels within each mask. {For DCM, each run's experimental inputs were taken from the SPM microtime regressors after removing the 32 leading microtime bins introduced by SPM and were retained at the microtime resolution \texttt{SPM.xBF.dt}, thereby aligning the DCM inputs with the original event onsets.} For each run, we specified a one-state, deterministic, bilinear DCM with full endogenous coupling among the four regions, the six modulatory effects listed in Table~\ref{tab:realparams}, and task inputs driving the vmPFC \cite{friston2003dcm}. Each run-level DCM was inverted using variational Laplace.

Because the two task versions provided different amounts of information about loss modulation, we examined the event counts and design ranks before conducting the group analysis. Across the four runs, each participant completing the equal-indifference version contributed only one loss-dominant trial, which occurred in run 1, whereas each participant completing the equal-range version contributed 36 loss-dominant trials, with 8, 8, 11, and 9 trials in runs 1--4. The DCM input rank was 3 for all equal-range runs and for run 1 of the equal-indifference version, and it was 2 for equal-indifference runs 2--4 because the loss-dominant input was identically zero. All DCM input matrices were full rank relative to their nonzero columns. The task-regressor portions of the first-level GLM had ranks ranging from 21 to 27. Complete subject-by-run condition counts and rank diagnostics are available in the public reproducibility repository, with compact summaries in Supplementary Table~S20. %\ref{tab:supp-first-level-inputs}

The four run-level DCMs for each participant were combined using within-subject parametric empirical Bayes with an intercept-only model across runs \cite{friston2016peb}. This produced the 22-dimensional subject-level posterior mean and covariance used in the group analysis, comprising the full $4\times4$ intrinsic connectivity matrix and six modulatory connections. The first two modulatory parameters corresponded to gain-related effects indexed by $k=1$, and the remaining four corresponded to loss-related effects indexed by $k=2$. The group-level between-subject covariance was diagonal, with a separate empirical-Bayes variance component for each of the 22 connectivity parameters.

First-level fit and information diagnostics are summarized in Supplementary Tables~S21 and~S22. %\ref{tab:supp-first-level-fit} and \ref{tab:supp-variance-reduction}
Across the 192 run-level DCMs, the median percentage of variance explained was 3.36\% (interquartile range 1.36\%--8.87\%), and all 192 posterior covariance matrices were finite and positive semidefinite to numerical tolerance. Pooling the four loss-modulatory parameters across participants, the median posterior-to-prior variance ratio at the within-subject PEB stage was 0.760 in the equal-indifference version and 0.099 in the equal-range version. Thus, the equal-indifference data reduced uncertainty in the loss-modulatory parameters substantially less than the equal-range data. These diagnostics quantify the unequal first-level information available across task versions and motivate the task-version sensitivity analysis reported in Section~5.6 and Supplementary Table~S19. %\ref{tab:supp-task-version}

Following the finite-variance choice motivated by the sensitivity analysis in Section~\ref{sec:simulation}, we used $\nu=3$ for the empirical fits. {The prior inclusion probability was fixed at $\pi=0.5$, giving equal prior probability to excluded and included states, and standard \emph{SPM-PEB} served as the reference method.} The proposed group-level analysis was implemented in R version 4.5.1. {For the \emph{Proposed} model, Tables~\ref{tab:realparams} and~\ref{tab:realcov} also report conditional Laplace standard deviations computed from the final weighted curvature, conditional on the fitted empirical-Bayes hyperparameters.}

%The source NARPS data set (OpenNeuro ds001734, version 1.0.4) contains $108$ healthy participants who completed one of two versions of a mixed-gambles task \cite{botvinik2019narps}. On each trial, participants accepted or rejected a prospect offering equal probabilities of a monetary gain and loss. The present analysis used the $N=48$ participants with the required subject-level DCM posterior summaries. For each participant, the $22$-dimensional posterior mean and covariance represented a four-region valuation network comprising ventromedial prefrontal cortex (vmPFC), ventral striatum (vStr), amygdala, and anterior insula (aIns). The parameter vector contained the full $4\times4$ intrinsic matrix ($16$ parameters) and six modulatory connections: two gain-related effects indexed by $k=1$ and four loss-related effects indexed by $k=2$, as listed in Table~\ref{tab:realparams}.
%
%Following the finite-variance choice motivated by the sensitivity analysis in Section~\ref{sec:simulation}, we used $\nu=3$ for the empirical fits. The reported analysis used prior inclusion probability $\pi=0.5$, and standard \emph{SPM-PEB} served as the reference. We first examined the group-average intercept model and its robustness weights, then compared the resulting estimates with \emph{SPM-PEB}, and finally extended the design to gender and age covariates.
%

%%
\subsection{Group-average intercept model}
\begin{figure}[tbp]
\centering
\includegraphics[width=\linewidth]{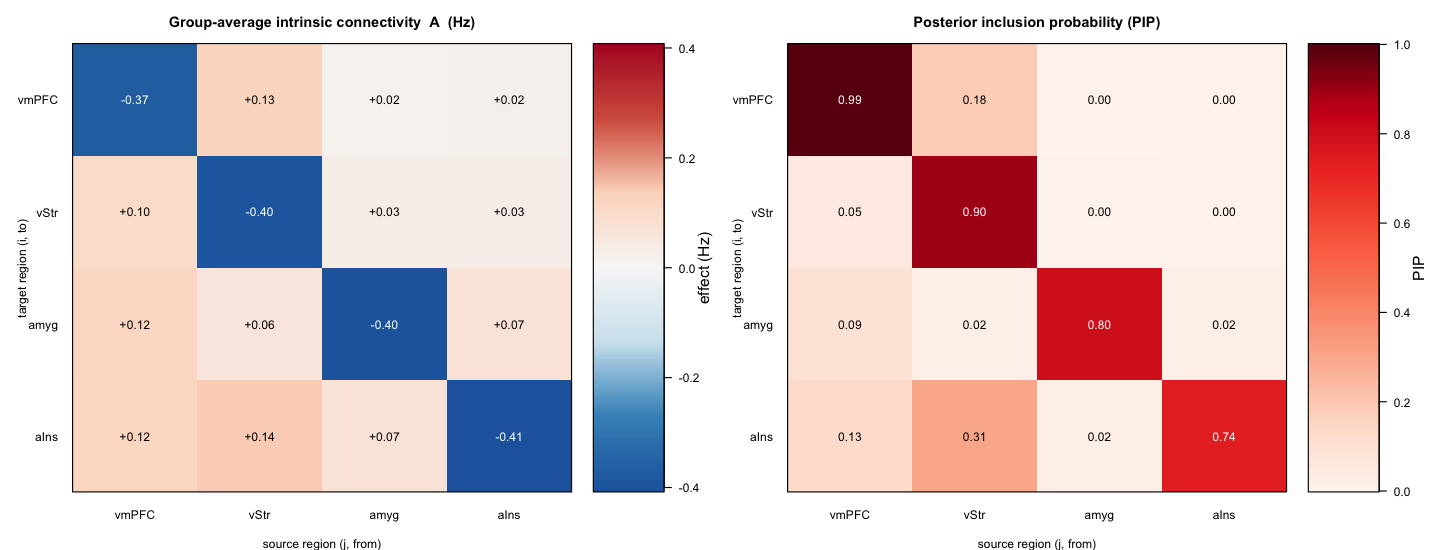}
\caption{Group-average intrinsic connectivity from the \emph{Proposed} model ($\nu=3$). 
\emph{Left:} the intrinsic coupling matrix $A$ in Hz. {Off-diagonal elements are displayed on their native Hz scale, while diagonal elements were transformed from SPM's internal log-scaling parameters using $a_{ii}^{\mathrm{Hz}}=-\tfrac{1}{2}\exp(\widetilde a_{ii})$.} Element $(i,j)$ is the directed influence of source region $j$ (column, ``from'') on target region $i$ (row, ``to''), with warm cells excitatory and cool cells inhibitory.
\emph{Right:} {the corresponding posterior inclusion probabilities, computed on the fitted internal parameter scale} (PIP, from $0$ to $1$; darker cells indicate stronger {inclusion support}). Regions are vmPFC, vStr, amygdala, and aIns. {Self-inhibition dominates. All four diagonal self-inhibition scaling parameters had PIP $>0.5$, with PIPs of $0.99$, $0.90$, $0.80$, and $0.74$ for vmPFC, vStr, amygdala, and aIns, respectively. These probabilities quantify evidence for departure from the SPM default of $-0.5$ Hz, not evidence for the presence or absence of the self-connections.}}
\label{fig:realA}
\end{figure}

With the empirical specification fixed, the intercept-only fit provided the first check of how the robust and sparse components behaved together. {The estimated scales were clearly ordered, with $\widehat{\tau}_0=0.070<\widehat{\tau}_1=0.345$. The residual scale was fixed at $\sigma^2=1$ for scale identification. After transforming the internal diagonal parameters to physical units, the estimated self-inhibition strengths ranged from $-0.41$ to $-0.37$ Hz (Figure~\ref{fig:realA}). The four corresponding diagonal self-inhibition scaling parameters, $A_{11}$, $A_{22}$, $A_{33}$, and $A_{44}$, were the only parameters with PIP $>0.5$, with PIPs of $0.995$, $0.896$, $0.802$, and $0.740$, respectively. Together with their negative internal estimates, these probabilities indicate support for self-inhibition weaker than the SPM default of $-0.5$ Hz. They quantify evidence for departure from the default rather than evidence for the presence or absence of the self-connections. The remaining intrinsic and modulatory parameters did not exceed the PIP threshold of $0.5$.} {All three generalized-EM starts converged to the same objective, and the integrated-PIP calculation had a maximum Monte Carlo standard error of $0.003$ and a maximum between-chain range of $0.012$.}

Table~\ref{tab:realparams} provides the complete group-average estimates underlying this summary and the method comparison developed below. Presenting all $22$ parameters makes the extent of shrinkage visible without treating weakly supported estimates as discoveries.

\begin{table}[tbp]
\centering
\small
\caption{Group-average (intercept) estimates of all $22$ DCM parameters under standard \emph{SPM-PEB} and the \emph{Proposed} (robust $+$ sparse) model. $A(i,j)$ denotes the intrinsic connection from region $j$ to region $i$, and $B(i,j,k)$ denotes input $k$ modulating that connection (region indices $1$--$4$ are vmPFC, vStr, amygdala, and aIns). For diagonal $A(i,i)$ terms, estimates and approximate SDs are reported after transformation to physical self-inhibition, $a_{ii}^{\mathrm{Hz}}=-\tfrac{1}{2}\exp(\widetilde a_{ii})$, with SDs obtained using the delta method. Their \emph{SPM-PEB} evidence indicators and \emph{Proposed}-model PIPs were computed on the internal log-scaling scale, where zero corresponds to the SPM default of $-0.5$ Hz. Off-diagonal $A$ and all $B$ parameters are reported in Hz. \emph{SPM-PEB} reports a posterior mean $\pm$ SD, with \textsuperscript{$\ast$} marking $|\mathrm{E}p|>3\,\mathrm{SD}$. The \emph{Proposed} model reports an estimate $\pm$ conditional Laplace SD and a PIP, with bold PIPs exceeding $0.5$.}\label{tab:realparams}
\begin{tabular}{l r@{\,$\pm$\,}l r@{\,$\pm$\,}l c}
\toprule
Parameter & \multicolumn{2}{c}{\emph{SPM-PEB} ($\mathrm{E}p\pm$SD)} & \multicolumn{2}{c}{\emph{Proposed} (estimate $\pm$ SD)} & \emph{Proposed} PIP\\
\midrule
A(1,1) & -0.314\textsuperscript{$\ast$} & 0.010 & -0.369 & 0.010 & \textbf{0.995}\\
A(2,1) & +0.101\textsuperscript{$\ast$} & 0.029 & +0.098 & 0.016 & 0.052\\
A(3,1) & +0.121\textsuperscript{$\ast$} & 0.029 & +0.117 & 0.011 & 0.091\\
A(4,1) & +0.135\textsuperscript{$\ast$} & 0.029 & +0.121 & 0.018 & 0.129\\
A(1,2) & +0.182\textsuperscript{$\ast$} & 0.032 & +0.128 & 0.023 & 0.184\\
A(2,2) & -0.348\textsuperscript{$\ast$} & 0.012 & -0.395 & 0.009 & \textbf{0.896}\\
A(3,2) & +0.067 & 0.031 & +0.059 & 0.023 & 0.016\\
A(4,2) & +0.148\textsuperscript{$\ast$} & 0.031 & +0.141 & 0.029 & 0.306\\
A(1,3) & -0.029 & 0.032 & +0.016 & 0.022 & 0.002\\
A(2,3) & +0.003 & 0.031 & +0.032 & 0.023 & 0.005\\
A(3,3) & -0.325\textsuperscript{$\ast$} & 0.012 & -0.401 & 0.012 & \textbf{0.802}\\
A(4,3) & +0.031 & 0.031 & +0.071 & 0.019 & 0.021\\
A(1,4) & -0.015 & 0.032 & +0.022 & 0.021 & 0.003\\
A(2,4) & +0.006 & 0.031 & +0.026 & 0.023 & 0.004\\
A(3,4) & +0.008 & 0.031 & +0.071 & 0.017 & 0.018\\
A(4,4) & -0.361\textsuperscript{$\ast$} & 0.013 & -0.406 & 0.011 & \textbf{0.740}\\
\midrule
B(2,1,1) & +0.010 & 0.227 & +0.017 & 0.036 & 0.007\\
B(1,2,1) & +0.144 & 0.232 & +0.010 & 0.061 & 0.080\\
B(1,3,2) & +0.040 & 0.274 & -0.002 & 0.059 & 0.048\\
B(4,3,2) & +0.041 & 0.251 & +0.015 & 0.046 & 0.013\\
B(1,4,2) & -0.398 & 0.269 & -0.015 & 0.056 & 0.038\\
B(3,4,2) & +0.082 & 0.249 & +0.016 & 0.049 & 0.020\\
\bottomrule
\end{tabular}
\end{table}

Because the coefficient estimates reflect both nonlocal shrinkage and adaptive Student-$t$ weighting, we next examined the weights directly before interpreting differences from \emph{SPM-PEB}.

\subsection{Subject-level robustness weights}

\begin{figure}[tbp]
\centering
\includegraphics[width=\linewidth]{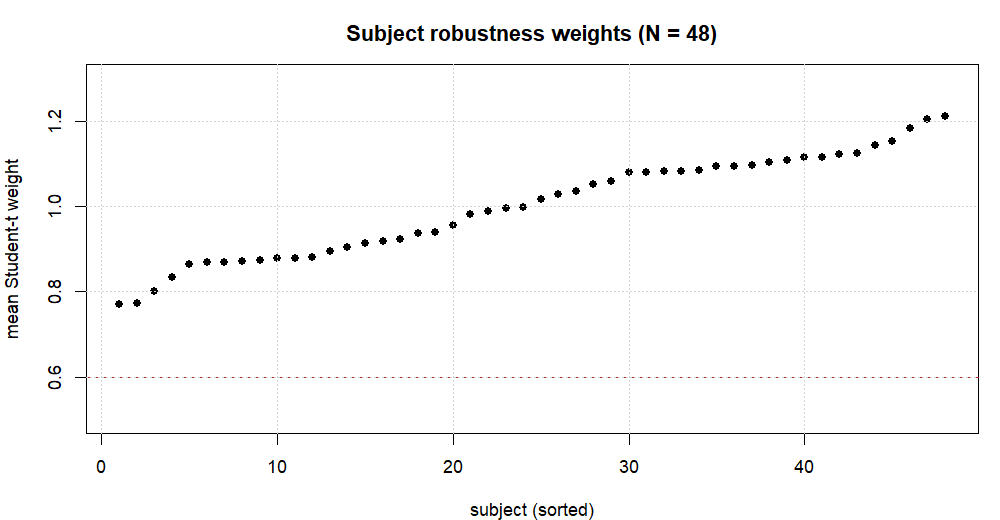}
\caption{Subject-level Student-$t$ robustness weights, sorted in increasing order. Each point is one participant's mean weight across the $22$ whitened {residual coordinates}; larger values indicate greater compatibility with the fitted group model, whereas values below $1$ indicate down-weighting. The dashed line at $0.6$ is a descriptive reference for heavy down-weighting, not an exclusion threshold. All $48$ means lie above this line.}
\label{fig:weights}
\end{figure}

The weight diagnostics indicated mild heterogeneity rather than a small set of severely discordant participants. Figure~\ref{fig:weights} shows, for each participant, the mean Student-$t$ weight across the $22$ whitened {residual coordinates}, ordered from smallest to largest. The most down-weighted participants had means near {$0.75$}, and all $48$ means exceeded $0.6$. The likelihood therefore reduced the influence of atypical {whitened residual coordinates} without imposing any additional exclusion within the analyzed sample. Having established the extent of down-weighting, we then assessed whether the robust and sparse fit materially altered the group-level conclusions relative to \emph{SPM-PEB}.

\subsection{Agreement with standard \emph{SPM-PEB}}
\begin{figure}[tbp]
\centering
\includegraphics[width=\linewidth]{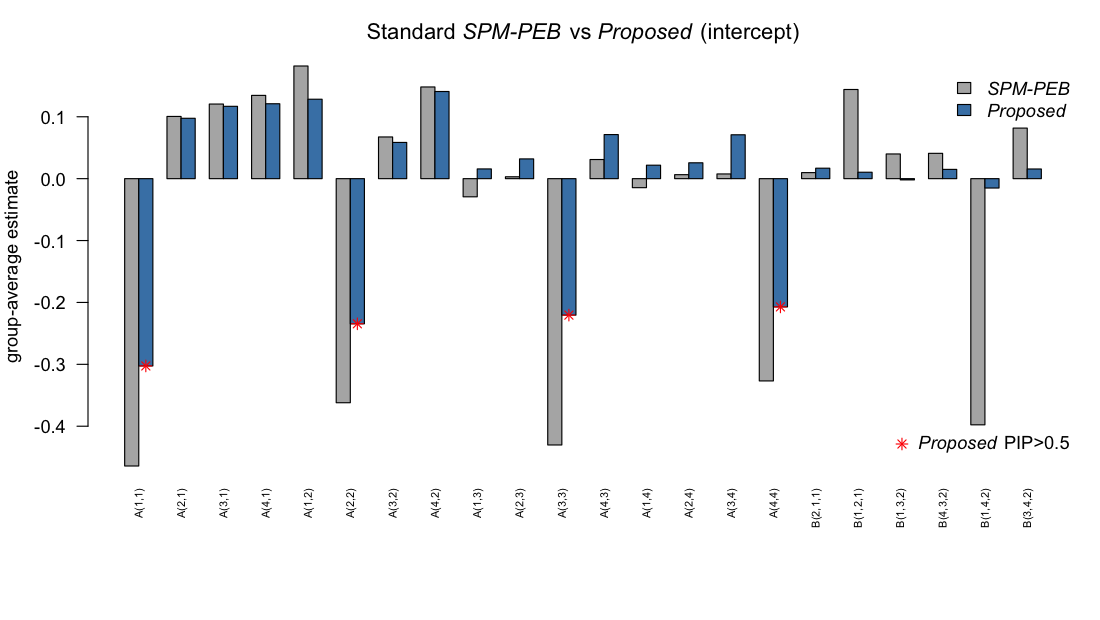}\\
\includegraphics[width=\linewidth]{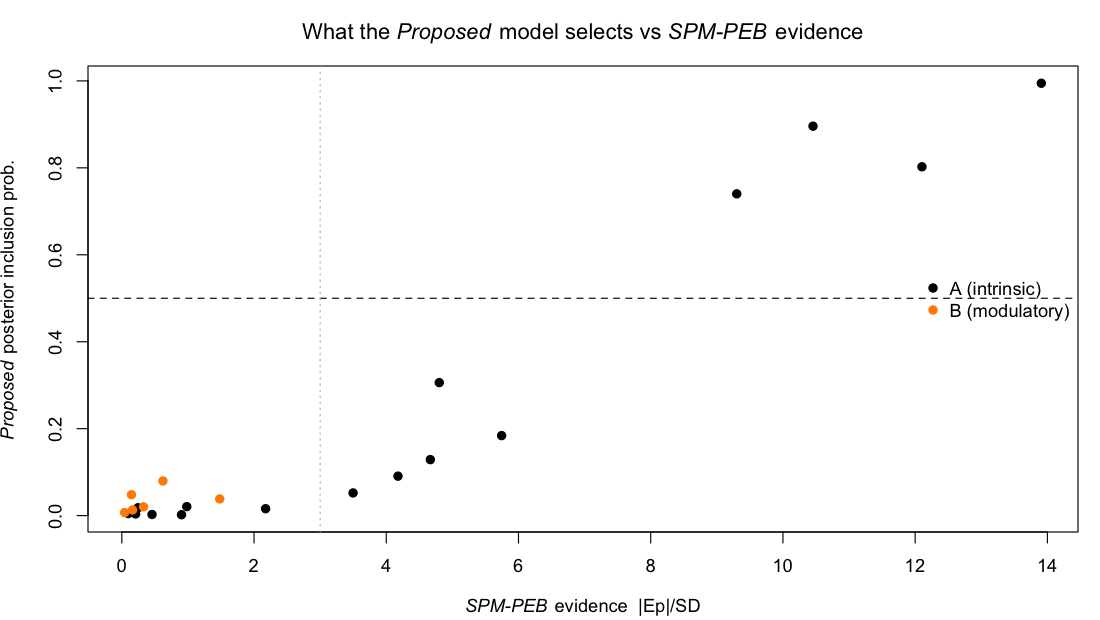}
\caption{Standard \emph{SPM-PEB} versus the \emph{Proposed} model for the NARPS group-average (intercept). \emph{Top:} group-level estimates for intrinsic $A(i,j)$ and modulatory $B(i,j,k)$ parameters; {for diagonal $A$ parameters, the plotted estimates and sign comparisons use SPM's internal log-scaling scale; off-diagonal $A$ and all $B$ parameters are in Hz;} region indices $1$--$4$ are vmPFC, vStr, amygdala, and aIns, and red stars mark parameters included by the \emph{Proposed} model at PIP $>0.5$. \emph{Bottom:} \emph{Proposed}-model PIP versus \emph{SPM-PEB} evidence $|\mathrm{E}p|/\mathrm{SD}$, colored by parameter type; dashed lines mark PIP $=0.5$ and evidence $=3$. The methods agree in direction for {$19$} of $22$ parameters and for every effect meeting either support criterion, while the \emph{Proposed} model selects only the strongest effects and shrinks the noisier modulatory parameters.}
\label{fig:cmp}
\end{figure}

The \emph{Proposed}-model estimates agreed in sign with standard \emph{SPM-PEB} for {$19$} of the $22$ group-level parameters (one-sided exact binomial {$p=4.28\times10^{-4}$}). {For diagonal $A$ parameters, sign agreement was evaluated on SPM's internal log-scaling scale and therefore indicates agreement in the direction of departure from the $-0.5$-Hz default, not agreement about the physical sign of a self-connection, which is constrained to be negative.} Because the empirical comparison was based on the posterior means and standard deviations returned by the \emph{SPM-PEB} fit, SPM support was summarized using $|\mathrm{E}p|/\mathrm{SD}>3$. This descriptive criterion is not calibrated to the \emph{Proposed}-model PIP threshold and is used only to contextualize directional agreement, not to compare selection probabilities across methods. Among the nine effects supported under either method's stated criterion---\emph{Proposed}-model PIP $>0.5$ or SPM evidence $|\mathrm{E}p|/\mathrm{SD}>3$---the methods agreed in sign for all nine ($9/9$; one-sided exact binomial {$p=1.95\times10^{-3}$}; Figure~\ref{fig:cmp}). The {three} sign disagreements involved estimates statistically indistinguishable from zero under both analyses. Differences were therefore concentrated in magnitude rather than direction, with a median \emph{Proposed}-to-PEB absolute-magnitude ratio of {$0.13$} for the noisier modulatory ($B$) parameters and {$0.92$} for the intrinsic ($A$) parameters. Thus, the \emph{Proposed} model preserved the direction of the supported PEB effects while applying stronger shrinkage to weak modulatory estimates.

%The simulation study uses BMR inclusion probabilities as the \emph{SPM-PEB} selection scores, whereas this section uses $|\mathrm{E}p|/\mathrm{SD}>3$. For consistency, report the real-data BMR inclusion probabilities if available; otherwise explain why the posterior mean-to-SD criterion is used here and emphasize that it is not calibrated identically to the \emph{Proposed}-model PIP.

\subsection{Gender and age covariate effects}
{Finally, we moved beyond the intercept model by jointly adding gender and standardized age as second-level covariates while retaining the same fitting settings. Table~\ref{tab:realcov} reports both covariate effects for all $22$ parameters. Most estimates were small under both methods, and the \emph{Proposed} model was markedly sparser. The gender effect on $A(3,2)$ was the only gender coefficient with PIP $>0.5$ (PIP $=0.83$), and the age effect on $A(1,1)$ was the only age coefficient exceeding this threshold (PIP $=0.58$). Two modulatory effects were immediately below the threshold: the gender effect on $B(3,4,2)$ and the age effect on $B(1,2,1)$ each had PIP $=0.48$. The two selected effects were positive under both \emph{SPM-PEB} and the \emph{Proposed} model, whereas sign disagreements elsewhere were confined to weakly supported estimates. The covariate analysis is therefore best viewed as a secondary demonstration of sparse adjustment rather than evidence for a gender- or age-dependent valuation mechanism.}

\begin{table}[tbp]
\centering
\small
\caption{Gender and age covariate effects on all $22$ NARPS DCM parameters under standard \emph{SPM-PEB} and the \emph{Proposed} (robust $+$ sparse) model. Gender is coded as M$-$F and age is measured per standard deviation. \emph{SPM-PEB} reports a posterior mean $\pm$ SD. Covariate effects on diagonal $A$ parameters are changes on SPM's internal log-scaling scale; their physical effect on self-inhibition depends on the intercept and covariate value. The \emph{Proposed} model reports a coefficient estimate $\pm$ conditional Laplace SD and a PIP, with bold PIPs exceeding $0.5$. Parameter naming follows Table~\ref{tab:realparams}.}
\label{tab:realcov}
\setlength{\tabcolsep}{2.5pt}
\begin{tabular}{l ccc ccc}
\toprule
 & \multicolumn{3}{c}{Gender (M$-$F)} & \multicolumn{3}{c}{Age (per SD)}\\
\cmidrule(lr){2-4}\cmidrule(lr){5-7}
Parameter & PEB\,$\pm$\,SD & \emph{Proposed} $\widehat{\beta}\pm$SD & PIP & PEB\,$\pm$\,SD & \emph{Proposed} $\widehat{\beta}\pm$SD & PIP\\
\midrule
A(1,1) & $-0.002\!\pm\!0.060$ & $+0.004\!\pm\!0.024$ & 0.018 & $+0.195\!\pm\!0.033$ & $+0.051\!\pm\!0.019$ & \textbf{0.577}\\
A(2,1) & $-0.048\!\pm\!0.052$ & $+0.008\!\pm\!0.021$ & 0.018 & $-0.083\!\pm\!0.028$ & $-0.042\!\pm\!0.015$ & 0.081\\
A(3,1) & $+0.011\!\pm\!0.052$ & $+0.002\!\pm\!0.017$ & 0.003 & $+0.007\!\pm\!0.028$ & $+0.008\!\pm\!0.011$ & 0.001\\
A(4,1) & $-0.014\!\pm\!0.052$ & $-0.013\!\pm\!0.022$ & 0.020 & $-0.041\!\pm\!0.028$ & $-0.020\!\pm\!0.015$ & 0.005\\
A(1,2) & $-0.006\!\pm\!0.057$ & $-0.017\!\pm\!0.023$ & 0.082 & $+0.014\!\pm\!0.031$ & $-0.014\!\pm\!0.017$ & 0.004\\
A(2,2) & $-0.080\!\pm\!0.063$ & $-0.002\!\pm\!0.023$ & 0.009 & $-0.086\!\pm\!0.033$ & $-0.012\!\pm\!0.018$ & 0.006\\
A(3,2) & $+0.115\!\pm\!0.056$ & $+0.041\!\pm\!0.023$ & \textbf{0.832} & $+0.051\!\pm\!0.030$ & $+0.025\!\pm\!0.017$ & 0.009\\
A(4,2) & $+0.114\!\pm\!0.056$ & $+0.033\!\pm\!0.024$ & 0.208 & $+0.039\!\pm\!0.030$ & $+0.007\!\pm\!0.020$ & 0.003\\
A(1,3) & $+0.136\!\pm\!0.059$ & $+0.008\!\pm\!0.022$ & 0.012 & $+0.029\!\pm\!0.032$ & $+0.005\!\pm\!0.017$ & 0.002\\
A(2,3) & $+0.060\!\pm\!0.057$ & $+0.022\!\pm\!0.022$ & 0.081 & $+0.031\!\pm\!0.031$ & $+0.009\!\pm\!0.018$ & 0.004\\
A(3,3) & $+0.099\!\pm\!0.064$ & $+0.012\!\pm\!0.024$ & 0.079 & $+0.126\!\pm\!0.035$ & $+0.030\!\pm\!0.019$ & 0.077\\
A(4,3) & $+0.042\!\pm\!0.057$ & $+0.030\!\pm\!0.021$ & 0.139 & $+0.006\!\pm\!0.031$ & $-0.003\!\pm\!0.016$ & 0.001\\
A(1,4) & $-0.117\!\pm\!0.058$ & $-0.005\!\pm\!0.022$ & 0.007 & $+0.007\!\pm\!0.031$ & $-0.004\!\pm\!0.016$ & 0.001\\
A(2,4) & $+0.055\!\pm\!0.056$ & $+0.019\!\pm\!0.022$ & 0.057 & $+0.061\!\pm\!0.030$ & $+0.031\!\pm\!0.018$ & 0.030\\
A(3,4) & $-0.055\!\pm\!0.056$ & $+0.004\!\pm\!0.020$ & 0.007 & $+0.012\!\pm\!0.031$ & $+0.001\!\pm\!0.016$ & 0.002\\
A(4,4) & $+0.109\!\pm\!0.063$ & $+0.014\!\pm\!0.024$ & 0.079 & $-0.018\!\pm\!0.035$ & $+0.006\!\pm\!0.019$ & 0.003\\
\midrule
B(2,1,1) & $+0.010\!\pm\!0.411$ & $+0.002\!\pm\!0.025$ & 0.033 & $+0.031\!\pm\!0.217$ & $+0.000\!\pm\!0.022$ & 0.006\\
B(1,2,1) & $+0.050\!\pm\!0.421$ & $+0.003\!\pm\!0.027$ & 0.357 & $+0.172\!\pm\!0.222$ & $+0.275\!\pm\!0.091$ & 0.485\\
B(1,3,2) & $-0.280\!\pm\!0.493$ & $-0.004\!\pm\!0.027$ & 0.390 & $-0.201\!\pm\!0.262$ & $-0.004\!\pm\!0.026$ & 0.156\\
B(4,3,2) & $+0.017\!\pm\!0.453$ & $-0.001\!\pm\!0.026$ & 0.070 & $+0.056\!\pm\!0.234$ & $-0.004\!\pm\!0.024$ & 0.017\\
B(1,4,2) & $-0.569\!\pm\!0.484$ & $-0.226\!\pm\!0.096$ & 0.160 & $+0.178\!\pm\!0.257$ & $+0.001\!\pm\!0.026$ & 0.062\\
B(3,4,2) & $-0.211\!\pm\!0.449$ & $-0.386\!\pm\!0.098$ & 0.484 & $+0.012\!\pm\!0.236$ & $+0.004\!\pm\!0.024$ & 0.021\\
\bottomrule
\end{tabular}
\end{table}

\subsection{Sensitivity to task version}
{Because participants completed either the equal-indifference or equal-range version of the mixed-gambles task, we repeated the group analysis after adding a centered indicator for task version. The primary and task-adjusted group-average estimates were highly similar ($r=0.988$), with a median absolute difference of $0.013$ and a maximum difference of $0.041$. The inclusion probabilities were more sensitive to this adjustment. Nine intrinsic effects had adjusted-intercept PIP $>0.5$, compared with four diagonal self-inhibition scaling parameters with PIP $>0.5$ in the primary intercept-only analysis. Six equal-range versus equal-indifference effects exceeded the PIP threshold of $0.5$: $A_{21}$, $A_{41}$, $A_{32}$, $A_{42}$, $A_{24}$, and $B_{12}^{(1)}$. Their signs agreed with the corresponding \emph{SPM-PEB} estimates, although none of the SPM task-version effects satisfied $|\mathrm{E}p|/\mathrm{SD}>3$. Thus, the group-average coefficient estimates were stable after adjustment for task version, but the empirical-Bayes inclusion probabilities were more sensitive to the second-level design. Given the modest sample size and exploratory nature of this analysis, these task-version effects should not be interpreted as established differences in effective connectivity. Complete results are reported in Supplementary Table~S19.} %\ref{tab:supp-task-version}.

\

Overall, the application connects the two methodological components observed separately in simulation, with Student-$t$ weighting moderating mild heterogeneity without hard exclusion and the nonlocal prior concentrating inclusion support on a small number of effects while strongly shrinking weak modulatory and covariate estimates. The close directional agreement with supported \emph{SPM-PEB} effects provides a useful reference point, but the single modest-sized data set does not establish universal superiority, demographic modulation, or a new substantive valuation finding.

%%%
\section{Discussion}\label{sec:discussion}
We developed a robust and sparse extension of group DCM that combines two inferential functions often handled separately. Student-$t$ weighting limits the leverage of atypical first-level posterior summaries, whereas the nonlocal spike-and-slab prior distinguishes negligible coefficients from effects separated from zero. Both components operate after covariance-aware pre-whitening and therefore retain the subject-level posterior uncertainty, Kronecker-expanded design, and between-subject covariance structure of conventional PEB. The resulting output augments group-effect estimates with {whitened-coordinate} robustness weights and coefficient-level PIPs, providing diagnostics that can be inspected alongside the scientific contrasts without automatic observation removal.

The simulations clarify that robustness and sparsity make complementary but noninterchangeable contributions. {On uncontaminated data, the five methods often performed similarly, indicating little efficiency cost from the robust formulation under the examined conditions. As contamination increased, the three Student-$t$ estimators retained substantially higher PR-AUC and lower RMSE than \emph{Sparse-only} and \emph{SPM-PEB}. Across the principal comparison, the three robust estimators had mean PR-AUCs between $0.954$ and $0.956$. The \emph{Proposed} estimator had the lowest average RMSE, $0.075$, and the highest MCC at the common $0.5$ threshold, $0.697$. \emph{Robust-only} achieved similar threshold-free discrimination but had MCC $0.488$ because its normal-approximation score selected many null coefficients. These results identify Student-$t$ weighting as the primary protection against atypical observations and sparse mixture regularization as the main contributor to conservative fixed-threshold selection.}

{The value of sparse mixture regularization was clearest when many candidate effects were null. In the $p=40$ experiment, \emph{Proposed} and \emph{Local-slab} retained perfect PR-AUC under contamination across the examined active fractions, while the Gaussian-likelihood methods deteriorated. The local Gaussian slab sometimes produced higher MCC than the pMOM slab in this strong-signal setting, so the nonlocal slab was not uniformly superior. In the covariate experiment, \emph{Proposed} retained the highest PR-AUC under contamination, while \emph{Local-slab} generally had slightly lower RMSE and fewer false selections. \emph{Robust-only} may therefore be attractive when coefficient ranking is primary and broad sparsity is not expected. A sparse mixture model is preferable when fixed-threshold false selections matter, and the combined robust--sparse formulation is most useful when both contamination and many null effects are plausible.} The low conditional $\mathrm{FDP}_{>0.95}$ of \emph{SPM-PEB} in the headline summary also illustrates that the purity of only the most strongly supported selections is distinct from global ranking and estimation performance, particularly because this conditional measure excludes replicates with no score above $0.95$.

The sensitivity analysis further informs the choice of Student-$t$ degrees of freedom. The tuned simulation benchmark selected $\nu=2$ and $\pi={0.6}$ using data generated independently of the evaluation replicates. Fits with $\nu=2$ and $\nu=3$ were nearly indistinguishable {in the clean and baseline conditions} and separated mainly under the most severe contamination, where $\nu=2$ performed better. {We used $\nu=3$ for the empirical analysis because it retains a finite variance while sacrificing little performance under mild-to-moderate contamination, and we fixed $\pi=0.5$ as a neutral prior inclusion probability rather than carrying the severe-contamination benchmark value directly into the application.} More generally, $\nu$ and $\pi$ should be prespecified, tuned on data independent of the primary evaluation, or examined in a transparent sensitivity analysis. PIPs should likewise be interpreted jointly with coefficient estimates and fitted prior scales rather than as universal evidence measures. The \emph{Proposed}-model PIPs, SPM BMR probabilities, local-mixture PIPs, and normal-approximation confidence scores used in simulation are not identically calibrated.

{The NARPS application was consistent with the more moderate part of the simulation regime. The subject-level mean weights indicated mild heterogeneity, with the most down-weighted means near $0.75$ and none below the descriptive reference of $0.6$. Thus, the robust likelihood moderated influence without identifying a basis for hard participant exclusion. The fitted prior scales were ordered as intended, $\widehat{\tau}_0=0.070<\widehat{\tau}_1=0.345$. Inclusion support concentrated on all four diagonal self-inhibition scaling parameters, with PIPs of $0.995$, $0.896$, $0.802$, and $0.740$. Together with their negative internal estimates, these probabilities indicate self-inhibition weaker than the SPM default of $-0.5$ Hz. The \emph{Proposed} and standard \emph{SPM-PEB} estimates agreed in sign for 19 of 22 group-average parameters and for all nine effects supported under either method's stated criterion. Their main difference was shrinkage magnitude, which was substantially stronger for the noisier modulatory parameters than for the intrinsic parameters. Adjustment for task version produced highly similar group-average estimates but changed several inclusion probabilities, demonstrating that PIPs can be more sensitive than coefficient estimates to the second-level design.}

{The first-level information checks also showed that the two task versions differed materially in the number of loss-dominant trials and in the resulting uncertainty of the loss-modulatory parameters. Equal-indifference participants contributed only one loss-dominant trial across the four runs, and their within-subject PEB posterior-to-prior variance ratios for loss modulation were correspondingly larger than those of equal-range participants. Although adjustment for task version left the group-average coefficient estimates largely unchanged, the loss-related and task-version PIPs should therefore be interpreted cautiously because their first-level information content was not uniform across participants.}

{The covariate results require similar caution. The gender effect on $A_{32}$ and the age effect on $A_{11}$ were the only covariate coefficients with PIP $>0.5$, with PIPs of $0.83$ and $0.58$, respectively. Both estimates were positive under \emph{SPM-PEB} and the \emph{Proposed} model. The gender effect on $B_{34}^{(2)}$ and the age effect on $B_{12}^{(1)}$ were immediately below the threshold, with PIPs of approximately $0.48$. The two methods used support measures with different calibrations, the analyzed sample contained only 48 participants, and it was restricted for computational reasons rather than designed to estimate demographic heterogeneity. The application also used a fixed set of anatomical regions and a prespecified DCM architecture. It therefore serves as a demonstration of sparse covariate adjustment, not as evidence for general gender- or age-dependent mechanisms of valuation.}

Several limitations define the scope of the present evidence. The simulations generated first-level posterior summaries from a controlled hierarchical model rather than repeatedly simulating fMRI time series and reinverting complete DCMs. They therefore test the proposed group-level estimator directly but do not capture every form of first-level model misspecification, hemodynamic uncertainty, or preprocessing artifact. The contamination shifts and geometries were deliberately stylized, and performance may differ for weaker, clustered, or covariate-dependent departures. The {whitened-coordinate} Student-$t$ formulation assumes that whitening has adequately represented the residual covariance. Inaccurate first-level covariances or between-subject covariance bases could leave residual dependence. {The whitening analysis showed similar aggregate performance across specifications but material changes in some coefficient estimates and PIPs, confirming that the whitening factor and parameter ordering are part of the model specification.} {In addition, the generalized EM/ECM procedure estimates an empirical-Bayes mode, and the post-fit parallel-tempering calculation integrates over the coefficients conditional on the fitted hyperparameters. It therefore does not propagate uncertainty in the prior scales or variance components.} The nonlocal objective can contain separated modes, making multiple initializations and sensitivity analyses important. {The covariance-component validation showed small average bias, although five of 50 estimates reached the lower bound, indicating finite-sample boundary behavior for a small variance component.} Finally, the simulation comparison used an R port of \emph{SPM-PEB}, and the empirical analysis used a single modest-sized subset without external replication.

These limitations suggest several extensions. {A fully Bayesian sampler or variational approximation that jointly treats the coefficients, inclusion indicators, prior scales, and between-subject covariance components could propagate uncertainty beyond the current conditional PIPs.} Subject-block or multivariate Student-$t$ errors could complement the {whitened-coordinate} weights when entire scans exhibit coordinated deviations. Structured nonlocal priors could encode reciprocity, network hierarchy, anatomical grouping, or shared support across experimental conditions. Further evaluation should include end-to-end DCM simulations, additional robust mixed-model and sparse-PEB comparators, calibration of interval estimates and selection scores, and independent clinical and population neuroimaging cohorts. {Planned development of an R package, with documented inputs, diagnostics, and reproducible examples, will be important for its use as a neuroinformatics tool rather than only as a methodological construction.}

%%%
\section{Conclusion}\label{sec:conclusion}
We introduced a robust and sparse group-DCM framework that preserves the first-level uncertainty and hierarchical design of PEB while adding Student-$t$ influence control and nonlocal coefficient selection. {The simulations showed that Student-$t$ weighting protects estimation and score ranking under contamination, while sparse mixture regularization reduces false selections at fixed thresholds. The pMOM slab did not uniformly outperform the local Gaussian slab.} Their combination was most effective when contamination and sparsity were both present, rather than uniformly dominating every comparator. {In the NARPS application, the method moderated mild heterogeneity without excluding participants, concentrated support on departures of all four diagonal self-inhibition scaling parameters from the SPM default, and preserved the direction of the effects supported by standard \emph{SPM-PEB} while shrinking weak modulatory estimates more strongly. Task-version adjustment left the group-average estimates stable but showed greater sensitivity in the inclusion probabilities.}

The fitted weights and PIPs are intended as interpretable diagnostics, not automatic rules for deleting observations or declaring neuroscientific discoveries. {Their use should be accompanied by inspection of data quality and model specification, sensitivity analyses for prior settings, whitening, and second-level design, and validation in independent samples.} With further work on uncertainty propagation, multivariate robustness, structured priors, and { development of an R package}, the proposed framework provides a practical foundation for robust and parsimonious group effective-connectivity analysis.

\

\section*{Data availability}
The empirical analysis used the publicly available NARPS dataset \cite{botvinik2019narps}, archived in OpenNeuro as dataset ds001734, version 1.0.4 (\url{https://doi.org/10.18112/openneuro.ds001734.v1.0.4}). The de-identified participant labels defining the analyzed sample and the derived subject-level DCM summaries are available in the accompanying GitHub repository at \url{https://github.com/Kay202/DCM-Robust-and-Sparse}.

\bibliographystyle{jasa}
\bibliography{mybib}  %%% Uncomment this line and comment out the ``thebibliography'' section below to use the external .bib file (using bibtex) .

\begin{thebibliography}{22}
\newcommand{\enquote}[1]{``#1''}
\expandafter\ifx\csname natexlab\endcsname\relax\def\natexlab#1{#1}\fi

\bibitem[\protect\citename{Alak{\"o}rkk{\"o} et~al.,
  }2017]{alakorkko2017effects}
Alak{\"o}rkk{\"o}, T., Saarim{\"a}ki, H., Glerean, E., Saram{\"a}ki, J., and
  Korhonen, O. (2017).
\newblock \enquote{Effects of spatial smoothing on functional brain networks.}
\newblock {\em European Journal of Neuroscience\/}, 46, 9, 2471--2480.

\bibitem[\protect\citename{Botvinik-Nezer et~al., }2020]{botvinik2019narps}
Botvinik-Nezer, R., Holzmeister, F., Camerer, C.~F., Dreber, A., Huber, J.,
  Johannesson, M., Kirchler, M., Iwanir, R., Mumford, J.~A., Adcock, R.~A.,
  et~al. (2020).
\newblock \enquote{Variability in the analysis of a single neuroimaging dataset
  by many teams.}
\newblock {\em Nature\/}, 582, 7810, 84--88.

\bibitem[\protect\citename{Botvinik-Nezer et~al., }2019]{botvinik2019fmri}
Botvinik-Nezer, R., Iwanir, R., Holzmeister, F., Huber, J., Johannesson, M.,
  Kirchler, M., Dreber, A., Camerer, C.~F., Poldrack, R.~A., and Schonberg, T.
  (2019).
\newblock \enquote{fMRI data of mixed gambles from the Neuroimaging Analysis
  Replication and Prediction Study.}
\newblock {\em Scientific data\/}, 6, 1, 106.

\bibitem[\protect\citename{Bowman et~al., }2008]{bowman2008bayesian}
Bowman, F.~D., Caffo, B., Bassett, S.~S., and Kilts, C. (2008).
\newblock \enquote{A Bayesian hierarchical framework for spatial modeling of
  fMRI data.}
\newblock {\em NeuroImage\/}, 39, 1, 146--156.

\bibitem[\protect\citename{Cabezas et~al., }2024]{cabezas2024robust}
Cabezas, A., Battiston, M., and Nemeth, C. (2024).
\newblock \enquote{Robust Bayesian Nonparametric Variable Selection for Linear
  Regression.}
\newblock {\em Stat\/}, 13, 2, e696.

\bibitem[\protect\citename{De and Ghosh, }2026]{de2026robust}
De, S. and Ghosh, J. (2026).
\newblock \enquote{Robust Bayesian Model Averaging for Linear Regression Models
  with Heavy-Tailed Errors.}
\newblock {\em Journal of Applied Statistics\/}, 53, 2, 304--330.

\bibitem[\protect\citename{Friston et~al., }2003]{friston2003dcm}
Friston, K.~J., Harrison, L., and Penny, W. (2003).
\newblock \enquote{Dynamic causal modelling.}
\newblock {\em Neuroimage\/}, 19, 4, 1273--1302.

\bibitem[\protect\citename{Friston et~al., }2016]{friston2016peb}
Friston, K.~J., Litvak, V., Oswal, A., Razi, A., Stephan, K.~E., Van~Wijk,
  B.~C., Ziegler, G., and Zeidman, P. (2016).
\newblock \enquote{Bayesian model reduction and empirical Bayes for group (DCM)
  studies.}
\newblock {\em Neuroimage\/}, 128, 413--431.

\bibitem[\protect\citename{Johnson and Rossell, }2010]{johnson2010nonlocal}
Johnson, V.~E. and Rossell, D. (2010).
\newblock \enquote{On the use of non-local prior densities in Bayesian
  hypothesis tests.}
\newblock {\em Journal of the Royal Statistical Society Series B: Statistical
  Methodology\/}, 72, 2, 143--170.

\bibitem[\protect\citename{Maruyama and Strawderman, }2014]{maruyama2014robust}
Maruyama, Y. and Strawderman, W.~E. (2014).
\newblock \enquote{Robust Bayesian Variable Selection in Linear Models with
  Spherically Symmetric Errors.}
\newblock {\em Biometrika\/}, 101, 4, 992--998.

\bibitem[\protect\citename{Penny et~al., }2010]{penny2010families}
Penny, W.~D., Stephan, K.~E., Daunizeau, J., Rosa, M.~J., Friston, K.~J.,
  Schofield, T.~M., and Leff, A.~P. (2010).
\newblock \enquote{Comparing families of dynamic causal models.}
\newblock {\em PLoS computational biology\/}, 6, 3, e1000709.

\bibitem[\protect\citename{Rossell and Rubio, }2018]{rossell2018tractable}
Rossell, D. and Rubio, F.~J. (2018).
\newblock \enquote{Tractable Bayesian Variable Selection: Beyond Normality.}
\newblock {\em Journal of the American Statistical Association\/}, 113, 524,
  1742--1758.

\bibitem[\protect\citename{Rossell and Telesca, }2017]{rossell2017nonlocal}
Rossell, D. and Telesca, D. (2017).
\newblock \enquote{Nonlocal priors for high-dimensional estimation.}
\newblock {\em Journal of the American Statistical Association\/}, 112, 517,
  254--265.

\bibitem[\protect\citename{Sanyal, }2025]{sanyal2025nonlocal}
Sanyal, N. (2025).
\newblock \enquote{Nonlocal Prior Mixture-Based Bayesian Wavelet Regression
  with Application to Noisy Imaging and Audio Data.}
\newblock {\em Mathematics\/}, 13, 16, 2642.

\bibitem[\protect\citename{Sanyal and Ferreira, }2012]{sanyal2012fmri}
Sanyal, N. and Ferreira, M.~A. (2012).
\newblock \enquote{Bayesian hierarchical multi-subject multiscale analysis of
  functional MRI data.}
\newblock {\em NeuroImage\/}, 63, 3, 1519--1531.

\bibitem[\protect\citename{Sanyal and Ferreira, }2017]{sanyal2017wavelet}
--- (2017).
\newblock \enquote{Bayesian wavelet analysis using nonlocal priors with an
  application to FMRI analysis.}
\newblock {\em Sankhya B\/}, 79, 2, 361--388.

\bibitem[\protect\citename{Stephan et~al., }2010]{stephan2010rules}
Stephan, K.~E., Penny, W.~D., Moran, R.~J., den Ouden, H.~E., Daunizeau, J.,
  and Friston, K.~J. (2010).
\newblock \enquote{Ten simple rules for dynamic causal modeling.}
\newblock {\em Neuroimage\/}, 49, 4, 3099--3109.

\bibitem[\protect\citename{Stephan et~al., }2007]{stephan2007hemodynamic}
Stephan, K.~E., Weiskopf, N., Drysdale, P.~M., Robinson, P.~A., and Friston,
  K.~J. (2007).
\newblock \enquote{Comparing hemodynamic models with DCM.}
\newblock {\em Neuroimage\/}, 38, 3, 387--401.

\bibitem[\protect\citename{Tom et~al., }2007]{tom2007}
Tom, S.~M., Fox, C.~R., Trepel, C., and Poldrack, R.~A. (2007).
\newblock \enquote{The Neural Basis of Loss Aversion in Decision-Making Under
  Risk.}
\newblock {\em Science\/}, 315, 5811, 515--518.

\bibitem[\protect\citename{Woolrich et~al., }2004]{woolrich2004multilevel}
Woolrich, M.~W., Behrens, T.~E., Beckmann, C.~F., Jenkinson, M., and Smith,
  S.~M. (2004).
\newblock \enquote{Multilevel linear modelling for FMRI group analysis using
  Bayesian inference.}
\newblock {\em Neuroimage\/}, 21, 4, 1732--1747.

\bibitem[\protect\citename{Zeidman et~al.,
  }2019{\natexlab{a}}]{zeidman2019part1}
Zeidman, P., Jafarian, A., Corbin, N., Seghier, M.~L., Razi, A., Price, C.~J.,
  and Friston, K.~J. (2019{\natexlab{a}}).
\newblock \enquote{A guide to group effective connectivity analysis, part 1:
  First level analysis with DCM for fMRI.}
\newblock {\em Neuroimage\/}, 200, 174--190.

\bibitem[\protect\citename{Zeidman et~al.,
  }2019{\natexlab{b}}]{zeidman2019part2}
Zeidman, P., Jafarian, A., Seghier, M.~L., Litvak, V., Cagnan, H., Price,
  C.~J., and Friston, K.~J. (2019{\natexlab{b}}).
\newblock \enquote{A guide to group effective connectivity analysis, part 2:
  Second level analysis with PEB.}
\newblock {\em Neuroimage\/}, 200, 12--25.

\end{thebibliography}

\clearpage

\renewcommand{\thesection}{S\arabic{section}}
\renewcommand{\thesubsection}{\thesection.\arabic{subsection}}
\renewcommand{\thetable}{S\arabic{table}}
\renewcommand{\thefigure}{S\arabic{figure}}

\begin{center}
{\bfseries \huge \noindent Supplementary Materials\\[10pt] \par
\Large ``Robust and Sparse Group Dynamic Causal Modeling via Student-$t$ Parametric Empirical Bayes and Nonlocal Priors'' \\[5pt] by Godfred Arhin and Nilotpal Sanyal}
\end{center}

\setcounter{section}{0}
\addtocontents{toc}{\protect\setcounter{tocdepth}{3}} 

\tableofcontents

\section{Overview}
\label{supp:overview}

This document provides detailed simulation evidence complementing the principal findings in the main manuscript. It reports tuning, numerical diagnostics, matched-error-rate comparisons, condition-specific results, contamination breadth, higher-dimensional sparsity, degrees-of-freedom and tuning sensitivity, whitening sensitivity, and algorithmic validation. Every simulation summary is based on 50 independently generated evaluation replicates unless stated otherwise. It also reports the empirical NARPS covariance sensitivity analysis and the task-version sensitivity analysis for the real-data application.

\section{Additional simulation results}
\label{supp:results}

\subsection{Tuning and numerical diagnostics}
\label{supp:tuning-diagnostics}

The prespecified tuning rule selected $(\nu,\pi)=(2,0.6)$ for the proposed estimator, with $(2,0.7)$ nearly tied (Table~\ref{tab:supp-tuning}). Separate tuning selected $\nu=2$ for the robust-only estimator and $\pi=0.5$ for the sparse-only estimator. The primary ablation comparison used the proposed estimator's $\nu$ and $\pi$ for all components shared with that model, while the arm-specific sparse-only choice was examined separately.

All final principal-grid fits satisfied their estimator-specific convergence rules. Table~\ref{tab:supp-diagnostics} reports optimization and posterior-sampling diagnostics. Integrated-PIP MCSEs remained below $0.036$ in the principal grid. Between-chain PIP ranges were larger for some fits, particularly for the sparse-only estimator, and should be considered when interpreting individual inclusion probabilities. Table~\ref{tab:supp-matched-fpr} complements the native threshold analysis by comparing the methods at approximately common cross-fitted false-positive rates.

\begin{table}[htbp]
\centering
\small
\setlength{\tabcolsep}{4pt}
\caption{Selected tuning settings and the nearly tied runner-up for the proposed estimator. Each candidate was averaged over 20 tuning datasets. The composite score was normalized within estimator family, so scores should not be compared across families.}
\label{tab:supp-tuning}
\begin{tabular}{lrrrrr}
\toprule
Estimator &  $\nu$  &  $\pi$  & PR-AUC & RMSE & Composite\\
\midrule
\emph{Proposed} & 2 & 0.6 & 0.955 & 0.071 & 0.969\\
\emph{Proposed} runner-up & 2 & 0.7 & 0.955 & 0.071 & 0.968\\
\emph{Robust-only} & 2 & -- & 0.946 & 0.079 & 1.000\\
\emph{Sparse-only} & 1000000 & 0.5 & 0.871 & 0.144 & 0.880\\
\bottomrule
\end{tabular}
\end{table}

\begin{table}[htbp]
\centering
\small
\setlength{\tabcolsep}{4pt}
\caption{Numerical diagnostics across the 1000 fits per estimator in the principal grid. PIP diagnostics apply only to estimators using posterior sampling.}
\label{tab:supp-diagnostics}
\resizebox{\textwidth}{!}{%
\begin{tabular}{lrrrrrr}
\toprule
Estimator & Fits & Convergence & Objective monotonicity & Max. PIP MCSE & Median chain range & Max. chain range\\
\midrule
\emph{Proposed} & 1000 & 1.000 & 1.000 & 0.036 & 0.058 & 0.309\\
\emph{Robust-only} & 1000 & 1.000 & 1.000 & -- & -- & --\\
\emph{Sparse-only} & 1000 & 1.000 & 1.000 & 0.033 & 0.109 & 0.349\\
\emph{Local-slab} & 1000 & 1.000 & 1.000 & 0.020 & 0.051 & 0.157\\
\emph{SPM-PEB} & 1000 & 1.000 & -- & -- & -- & --\\
\bottomrule
\end{tabular}
}
\end{table}

\begin{table}[htbp]
\centering
\small
\setlength{\tabcolsep}{4pt}
\caption{Cross-fitted selection performance at approximately matched empirical false-positive rates. Results pool the baseline and nine contamination--sample-size conditions, totaling 8000 coefficient evaluations and 500 replicate--condition units.}
\label{tab:supp-matched-fpr}
\resizebox{\textwidth}{!}{%
\begin{tabular}{rlrrrrr}
\toprule
Target FPR & Estimator & Observed FPR & Power & FDR & Mean selected & Discovery frequency\\
\midrule
0.05 & \emph{Proposed} & 0.050 & 0.773 & 0.044 & 7.568 & 1.000\\
0.05 & \emph{Robust-only} & 0.052 & 0.766 & 0.046 & 7.516 & 1.000\\
0.05 & \emph{Sparse-only} & 0.050 & 0.389 & 0.084 & 3.972 & 0.976\\
0.05 & \emph{Local-slab} & 0.052 & 0.790 & 0.044 & 7.734 & 1.000\\
0.05 & \emph{SPM-PEB} & 0.051 & 0.382 & 0.086 & 3.914 & 0.972\\
0.10 & \emph{Proposed} & 0.100 & 0.821 & 0.079 & 8.346 & 1.000\\
0.10 & \emph{Robust-only} & 0.104 & 0.805 & 0.084 & 8.224 & 1.000\\
0.10 & \emph{Sparse-only} & 0.101 & 0.455 & 0.136 & 4.932 & 0.998\\
0.10 & \emph{Local-slab} & 0.102 & 0.824 & 0.081 & 8.392 & 1.000\\
0.10 & \emph{SPM-PEB} & 0.101 & 0.460 & 0.135 & 4.976 & 0.996\\
\bottomrule
\end{tabular}
}
\end{table}

\subsection{Complete principal-grid results}
\label{supp:full-grid}

Tables~\ref{tab:supp-prauc-full}--\ref{tab:supp-mcc-full} report PR-AUC, coefficient RMSE, and MCC for all 20 principal-grid conditions. The global-null condition has no positive class, so PR-AUC is undefined and MCC is uninformative. Table~\ref{tab:supp-global-null} therefore reports its RMSE, false-positive rate, and mean number selected.

\begin{table}[htbp]
\centering
\scriptsize
\setlength{\tabcolsep}{4pt}
\caption{PR-AUC for every principal-grid condition. Entries are means with Monte Carlo standard errors in parentheses from 50 replicates.}
\label{tab:supp-prauc-full}
\resizebox{\textwidth}{!}{%
\begin{tabular}{lrrrrr}
\toprule
Condition & \emph{Proposed} & \emph{Robust-only} & \emph{Sparse-only} & \emph{Local-slab} & \emph{SPM-PEB}\\
\midrule
Baseline & 0.963 (0.003) & 0.962 (0.003) & 0.803 (0.014) & 0.962 (0.003) & 0.792 (0.014)\\
$\phi=0$, $N=24$ & 0.940 (0.005) & 0.939 (0.005) & 0.946 (0.005) & 0.940 (0.005) & 0.946 (0.005)\\
$\phi=0$, $N=48$ & 0.967 (0.004) & 0.968 (0.003) & 0.972 (0.003) & 0.967 (0.003) & 0.973 (0.003)\\
$\phi=0$, $N=80$ & 0.977 (0.003) & 0.977 (0.003) & 0.980 (0.003) & 0.975 (0.003) & 0.979 (0.003)\\
$\phi=0.10$, $N=24$ & 0.935 (0.006) & 0.931 (0.006) & 0.764 (0.014) & 0.935 (0.006) & 0.729 (0.015)\\
$\phi=0.10$, $N=48$ & 0.960 (0.004) & 0.960 (0.004) & 0.775 (0.012) & 0.960 (0.004) & 0.768 (0.013)\\
$\phi=0.10$, $N=80$ & 0.975 (0.003) & 0.975 (0.003) & 0.842 (0.012) & 0.975 (0.003) & 0.841 (0.011)\\
$\phi=0.20$, $N=24$ & 0.917 (0.006) & 0.906 (0.007) & 0.686 (0.015) & 0.920 (0.006) & 0.683 (0.015)\\
$\phi=0.20$, $N=48$ & 0.953 (0.004) & 0.953 (0.004) & 0.738 (0.016) & 0.954 (0.004) & 0.740 (0.015)\\
$\phi=0.20$, $N=80$ & 0.972 (0.003) & 0.973 (0.003) & 0.768 (0.013) & 0.972 (0.003) & 0.770 (0.013)\\
$\kappa=0$ (global null) & -- & -- & -- & -- & --\\
$\kappa=0.5$ & 0.894 (0.008) & 0.893 (0.008) & 0.703 (0.015) & 0.895 (0.008) & 0.696 (0.016)\\
$\kappa=1.5$ & 0.980 (0.003) & 0.981 (0.003) & 0.871 (0.010) & 0.981 (0.002) & 0.863 (0.011)\\
$\kappa=2$ & 0.992 (0.002) & 0.991 (0.002) & 0.908 (0.008) & 0.991 (0.002) & 0.901 (0.009)\\
Network $p=12$ & 0.974 (0.003) & 0.976 (0.003) & 0.887 (0.009) & 0.975 (0.003) & 0.876 (0.011)\\
Network $p=22$ & 0.947 (0.004) & 0.946 (0.004) & 0.763 (0.014) & 0.948 (0.004) & 0.745 (0.015)\\
Small diagonal covariance & 0.986 (0.002) & 0.988 (0.002) & 0.809 (0.013) & 0.986 (0.002) & 0.794 (0.014)\\
Large diagonal covariance & 0.923 (0.006) & 0.923 (0.006) & 0.794 (0.014) & 0.924 (0.006) & 0.789 (0.015)\\
Correlated covariance & 0.956 (0.004) & 0.939 (0.005) & 0.816 (0.013) & 0.952 (0.005) & 0.779 (0.014)\\
Covariate design & 0.765 (0.025) & 0.708 (0.026) & 0.326 (0.028) & 0.700 (0.026) & 0.270 (0.025)\\
\bottomrule
\end{tabular}
}
\end{table}

\begin{table}[htbp]
\centering
\scriptsize
\setlength{\tabcolsep}{4pt}
\caption{coefficient RMSE for every principal-grid condition. Entries are means with Monte Carlo standard errors in parentheses from 50 replicates.}
\label{tab:supp-rmse-full}
\resizebox{\textwidth}{!}{%
\begin{tabular}{lrrrrr}
\toprule
Condition & \emph{Proposed} & \emph{Robust-only} & \emph{Sparse-only} & \emph{Local-slab} & \emph{SPM-PEB}\\
\midrule
Baseline & 0.067 (0.002) & 0.073 (0.002) & 0.201 (0.006) & 0.071 (0.002) & 0.290 (0.008)\\
$\phi=0$, $N=24$ & 0.095 (0.003) & 0.103 (0.002) & 0.086 (0.003) & 0.102 (0.003) & 0.131 (0.003)\\
$\phi=0$, $N=48$ & 0.063 (0.002) & 0.069 (0.002) & 0.058 (0.002) & 0.067 (0.002) & 0.103 (0.002)\\
$\phi=0$, $N=80$ & 0.052 (0.002) & 0.058 (0.001) & 0.047 (0.001) & 0.054 (0.002) & 0.090 (0.002)\\
$\phi=0.10$, $N=24$ & 0.100 (0.003) & 0.109 (0.003) & 0.248 (0.007) & 0.110 (0.003) & 0.421 (0.011)\\
$\phi=0.10$, $N=48$ & 0.068 (0.002) & 0.074 (0.002) & 0.219 (0.006) & 0.072 (0.002) & 0.320 (0.008)\\
$\phi=0.10$, $N=80$ & 0.055 (0.002) & 0.061 (0.001) & 0.180 (0.005) & 0.058 (0.002) & 0.252 (0.006)\\
$\phi=0.20$, $N=24$ & 0.111 (0.004) & 0.124 (0.005) & 0.300 (0.008) & 0.125 (0.004) & 0.523 (0.014)\\
$\phi=0.20$, $N=48$ & 0.078 (0.003) & 0.081 (0.002) & 0.258 (0.007) & 0.087 (0.003) & 0.402 (0.011)\\
$\phi=0.20$, $N=80$ & 0.062 (0.002) & 0.065 (0.002) & 0.235 (0.006) & 0.068 (0.002) & 0.347 (0.008)\\
$\kappa=0$ (global null) & 0.023 (0.003) & 0.073 (0.002) & 0.176 (0.007) & 0.030 (0.002) & 0.288 (0.008)\\
$\kappa=0.5$ & 0.074 (0.002) & 0.073 (0.002) & 0.188 (0.006) & 0.070 (0.002) & 0.288 (0.008)\\
$\kappa=1.5$ & 0.062 (0.002) & 0.073 (0.002) & 0.211 (0.007) & 0.065 (0.002) & 0.293 (0.008)\\
$\kappa=2$ & 0.064 (0.002) & 0.073 (0.002) & 0.219 (0.007) & 0.065 (0.002) & 0.298 (0.008)\\
Network $p=12$ & 0.079 (0.003) & 0.076 (0.002) & 0.211 (0.005) & 0.085 (0.003) & 0.302 (0.009)\\
Network $p=22$ & 0.066 (0.002) & 0.078 (0.001) & 0.186 (0.005) & 0.071 (0.002) & 0.278 (0.007)\\
Small diagonal covariance & 0.037 (0.001) & 0.042 (0.001) & 0.200 (0.006) & 0.038 (0.001) & 0.245 (0.007)\\
Large diagonal covariance & 0.110 (0.004) & 0.124 (0.004) & 0.208 (0.006) & 0.120 (0.004) & 0.514 (0.013)\\
Correlated covariance & 0.080 (0.002) & 0.087 (0.002) & 0.201 (0.005) & 0.092 (0.003) & 0.348 (0.012)\\
Covariate design & 0.085 (0.002) & 0.127 (0.002) & 0.178 (0.003) & 0.082 (0.002) & 0.421 (0.008)\\
\bottomrule
\end{tabular}
}
\end{table}

\begin{table}[htbp]
\centering
\scriptsize
\setlength{\tabcolsep}{4pt}
\caption{MCC at the score threshold 0.5 for every principal-grid condition. Entries are means with Monte Carlo standard errors in parentheses from 50 replicates.}
\label{tab:supp-mcc-full}
\resizebox{\textwidth}{!}{%
\begin{tabular}{lrrrrr}
\toprule
Condition & \emph{Proposed} & \emph{Robust-only} & \emph{Sparse-only} & \emph{Local-slab} & \emph{SPM-PEB}\\
\midrule
Baseline & 0.730 (0.012) & 0.517 (0.028) & 0.260 (0.033) & 0.712 (0.010) & 0.240 (0.031)\\
$\phi=0$, $N=24$ & 0.660 (0.014) & 0.386 (0.032) & 0.701 (0.013) & 0.605 (0.014) & 0.712 (0.015)\\
$\phi=0$, $N=48$ & 0.734 (0.012) & 0.502 (0.025) & 0.748 (0.010) & 0.719 (0.009) & 0.776 (0.010)\\
$\phi=0$, $N=80$ & 0.767 (0.009) & 0.483 (0.025) & 0.772 (0.009) & 0.754 (0.009) & 0.835 (0.012)\\
$\phi=0.10$, $N=24$ & 0.618 (0.018) & 0.388 (0.030) & 0.232 (0.038) & 0.578 (0.015) & 0.187 (0.036)\\
$\phi=0.10$, $N=48$ & 0.728 (0.012) & 0.519 (0.026) & 0.210 (0.033) & 0.708 (0.011) & 0.223 (0.029)\\
$\phi=0.10$, $N=80$ & 0.759 (0.009) & 0.520 (0.023) & 0.396 (0.026) & 0.739 (0.010) & 0.381 (0.024)\\
$\phi=0.20$, $N=24$ & 0.574 (0.022) & 0.434 (0.028) & 0.020 (0.030) & 0.486 (0.018) & 0.074 (0.040)\\
$\phi=0.20$, $N=48$ & 0.663 (0.016) & 0.555 (0.023) & 0.158 (0.040) & 0.626 (0.013) & 0.189 (0.032)\\
$\phi=0.20$, $N=80$ & 0.736 (0.011) & 0.574 (0.022) & 0.229 (0.037) & 0.703 (0.010) & 0.243 (0.030)\\
$\kappa=0$ (global null) & -- & -- & -- & -- & --\\
$\kappa=0.5$ & 0.422 (0.018) & 0.400 (0.027) & 0.102 (0.034) & 0.376 (0.014) & 0.062 (0.034)\\
$\kappa=1.5$ & 0.800 (0.010) & 0.562 (0.026) & 0.415 (0.034) & 0.794 (0.008) & 0.416 (0.029)\\
$\kappa=2$ & 0.836 (0.010) & 0.580 (0.023) & 0.539 (0.029) & 0.839 (0.010) & 0.531 (0.027)\\
Network $p=12$ & 0.609 (0.015) & 0.405 (0.050) & 0.228 (0.035) & 0.586 (0.014) & 0.217 (0.032)\\
Network $p=22$ & 0.753 (0.011) & 0.435 (0.023) & 0.338 (0.032) & 0.734 (0.011) & 0.345 (0.026)\\
Small diagonal covariance & 0.781 (0.006) & 0.495 (0.022) & 0.266 (0.033) & 0.775 (0.005) & 0.277 (0.028)\\
Large diagonal covariance & 0.596 (0.018) & 0.426 (0.025) & 0.243 (0.035) & 0.542 (0.017) & 0.235 (0.029)\\
Correlated covariance & 0.681 (0.013) & 0.462 (0.029) & 0.306 (0.026) & 0.614 (0.013) & 0.228 (0.033)\\
Covariate design & 0.532 (0.026) & 0.219 (0.012) & 0.122 (0.022) & 0.593 (0.029) & 0.080 (0.030)\\
\bottomrule
\end{tabular}
}
\end{table}

\begin{table}[htbp]
\centering
\small
\setlength{\tabcolsep}{4pt}
\caption{Global-null performance. Ranking measures and MCC are undefined or uninformative when no coefficient is active, so the table reports estimation error and false selections.}
\label{tab:supp-global-null}
\begin{tabular}{lrrr}
\toprule
Estimator & RMSE & False-positive rate & Number selected\\
\midrule
\emph{Proposed} & 0.023 (0.003) & 0.009 (0.004) & 0.140 (0.057)\\
\emph{Robust-only} & 0.073 (0.002) & 0.525 (0.018) & 8.400 (0.296)\\
\emph{Sparse-only} & 0.176 (0.007) & 0.215 (0.018) & 3.440 (0.292)\\
\emph{Local-slab} & 0.030 (0.002) & 0.004 (0.003) & 0.060 (0.044)\\
\emph{SPM-PEB} & 0.288 (0.008) & 0.141 (0.014) & 2.260 (0.217)\\
\bottomrule
\end{tabular}
\end{table}

\subsection{Contamination breadth and covariate robustness}
\label{supp:breadth}

Table~\ref{tab:supp-magnitude} gives exact results for the contamination-magnitude analysis shown graphically in the main manuscript. The Student-$t$ methods remained stable as the magnitude increased, while the Gaussian-likelihood methods deteriorated. The same robust methods were also stable across cell-wise, whole-subject, and structured-subject contamination geometries (Table~\ref{tab:supp-geometry}).

The covariate experiments were more difficult because evaluation covered 32 null and three active covariate coefficients in each replicate. Under contamination, the proposed estimator retained higher PR-AUC than the other methods, while the local-slab estimator generally made fewer false selections and achieved slightly lower RMSE (Table~\ref{tab:supp-covariate-contamination}). Restricting contamination to simulated male subjects produced the same qualitative ordering (Table~\ref{tab:supp-subgroup}).

\begin{table}[htbp]
\centering
\scriptsize
\setlength{\tabcolsep}{4pt}
\caption{Performance across contamination magnitudes. Entries are means with Monte Carlo standard errors in parentheses from 50 replicates.}
\label{tab:supp-magnitude}
\begin{tabular}{llrrr}
\toprule
Condition & Estimator & PR-AUC & RMSE & MCC\\
\midrule
Clean & \emph{Proposed} & 0.967 (0.004) & 0.063 (0.002) & 0.738 (0.012)\\
Clean & \emph{Robust-only} & 0.968 (0.003) & 0.069 (0.002) & 0.502 (0.025)\\
Clean & \emph{Sparse-only} & 0.972 (0.003) & 0.058 (0.002) & 0.742 (0.012)\\
Clean & \emph{Local-slab} & 0.967 (0.004) & 0.066 (0.002) & 0.720 (0.008)\\
Clean & \emph{SPM-PEB} & 0.973 (0.003) & 0.103 (0.002) & 0.776 (0.010)\\
$\phi=0.10$, $\delta=2$ & \emph{Proposed} & 0.957 (0.004) & 0.072 (0.002) & 0.725 (0.014)\\
$\phi=0.10$, $\delta=2$ & \emph{Robust-only} & 0.956 (0.004) & 0.079 (0.002) & 0.440 (0.032)\\
$\phi=0.10$, $\delta=2$ & \emph{Sparse-only} & 0.930 (0.006) & 0.103 (0.003) & 0.587 (0.019)\\
$\phi=0.10$, $\delta=2$ & \emph{Local-slab} & 0.957 (0.004) & 0.075 (0.002) & 0.691 (0.013)\\
$\phi=0.10$, $\delta=2$ & \emph{SPM-PEB} & 0.928 (0.006) & 0.148 (0.003) & 0.591 (0.019)\\
$\phi=0.10$, $\delta=3$ & \emph{Proposed} & 0.960 (0.004) & 0.069 (0.002) & 0.723 (0.014)\\
$\phi=0.10$, $\delta=3$ & \emph{Robust-only} & 0.959 (0.004) & 0.076 (0.002) & 0.481 (0.030)\\
$\phi=0.10$, $\delta=3$ & \emph{Sparse-only} & 0.887 (0.009) & 0.135 (0.004) & 0.476 (0.027)\\
$\phi=0.10$, $\delta=3$ & \emph{Local-slab} & 0.959 (0.004) & 0.073 (0.002) & 0.700 (0.012)\\
$\phi=0.10$, $\delta=3$ & \emph{SPM-PEB} & 0.884 (0.009) & 0.190 (0.004) & 0.480 (0.026)\\
$\phi=0.10$, $\delta=6$ & \emph{Proposed} & 0.960 (0.004) & 0.068 (0.002) & 0.728 (0.012)\\
$\phi=0.10$, $\delta=6$ & \emph{Robust-only} & 0.960 (0.004) & 0.074 (0.002) & 0.519 (0.026)\\
$\phi=0.10$, $\delta=6$ & \emph{Sparse-only} & 0.779 (0.012) & 0.219 (0.005) & 0.222 (0.034)\\
$\phi=0.10$, $\delta=6$ & \emph{Local-slab} & 0.960 (0.004) & 0.072 (0.002) & 0.707 (0.011)\\
$\phi=0.10$, $\delta=6$ & \emph{SPM-PEB} & 0.768 (0.013) & 0.320 (0.008) & 0.223 (0.029)\\
\bottomrule
\end{tabular}
\end{table}

\begin{table}[htbp]
\centering
\small
\setlength{\tabcolsep}{4pt}
\caption{Robustness to contamination geometry at $15\%$ contamination. Entries are means with Monte Carlo standard errors in parentheses from 50 replicates.}
\label{tab:supp-geometry}
\resizebox{\textwidth}{!}{%
\begin{tabular}{lrrrrrr}
\toprule
Estimator & Cell PR-AUC & Cell RMSE & Subject PR-AUC & Subject RMSE & Structured PR-AUC & Structured RMSE\\
\midrule
\emph{Proposed} & 0.958 (0.004) & 0.074 (0.002) & 0.960 (0.004) & 0.073 (0.003) & 0.961 (0.004) & 0.074 (0.003)\\
\emph{Robust-only} & 0.957 (0.004) & 0.078 (0.002) & 0.961 (0.004) & 0.080 (0.002) & 0.963 (0.004) & 0.080 (0.002)\\
\emph{Sparse-only} & 0.734 (0.016) & 0.240 (0.007) & 0.746 (0.016) & 0.237 (0.006) & 0.832 (0.013) & 0.230 (0.020)\\
\emph{Local-slab} & 0.960 (0.004) & 0.080 (0.003) & 0.961 (0.004) & 0.079 (0.003) & 0.960 (0.004) & 0.080 (0.003)\\
\emph{SPM-PEB} & 0.727 (0.015) & 0.364 (0.010) & 0.752 (0.016) & 0.361 (0.008) & 0.831 (0.013) & 0.331 (0.030)\\
\bottomrule
\end{tabular}
}
\end{table}

\begin{table}[htbp]
\centering
\scriptsize
\setlength{\tabcolsep}{4pt}
\caption{Covariate-effect performance under cell-wise contamination. The intercept coefficients were excluded from evaluation. Entries are means with Monte Carlo standard errors in parentheses from 50 replicates.}
\label{tab:supp-covariate-contamination}
\begin{tabular}{llrrr}
\toprule
Condition & Estimator & PR-AUC & RMSE & MCC\\
\midrule
$\phi=0$ & \emph{Proposed} & 0.786 (0.026) & 0.080 (0.003) & 0.608 (0.032)\\
$\phi=0$ & \emph{Robust-only} & 0.770 (0.028) & 0.118 (0.003) & 0.219 (0.011)\\
$\phi=0$ & \emph{Sparse-only} & 0.853 (0.022) & 0.073 (0.003) & 0.677 (0.027)\\
$\phi=0$ & \emph{Local-slab} & 0.750 (0.027) & 0.073 (0.002) & 0.640 (0.033)\\
$\phi=0$ & \emph{SPM-PEB} & 0.844 (0.023) & 0.132 (0.003) & 0.755 (0.030)\\
$\phi=0.10$ & \emph{Proposed} & 0.794 (0.021) & 0.083 (0.002) & 0.567 (0.027)\\
$\phi=0.10$ & \emph{Robust-only} & 0.738 (0.028) & 0.127 (0.003) & 0.215 (0.013)\\
$\phi=0.10$ & \emph{Sparse-only} & 0.253 (0.027) & 0.184 (0.003) & 0.069 (0.025)\\
$\phi=0.10$ & \emph{Local-slab} & 0.740 (0.023) & 0.076 (0.002) & 0.582 (0.038)\\
$\phi=0.10$ & \emph{SPM-PEB} & 0.239 (0.024) & 0.448 (0.010) & 0.081 (0.030)\\
$\phi=0.20$ & \emph{Proposed} & 0.712 (0.027) & 0.085 (0.002) & 0.490 (0.025)\\
$\phi=0.20$ & \emph{Robust-only} & 0.660 (0.032) & 0.139 (0.003) & 0.246 (0.013)\\
$\phi=0.20$ & \emph{Sparse-only} & 0.207 (0.020) & 0.184 (0.004) & 0.074 (0.019)\\
$\phi=0.20$ & \emph{Local-slab} & 0.641 (0.028) & 0.081 (0.002) & 0.428 (0.044)\\
$\phi=0.20$ & \emph{SPM-PEB} & 0.212 (0.022) & 0.544 (0.011) & 0.002 (0.026)\\
\bottomrule
\end{tabular}
\end{table}

\begin{table}[htbp]
\centering
\small
\setlength{\tabcolsep}{4pt}
\caption{Covariate-effect performance before and after subgroup masking. In the masking condition, whole-subject contamination affected $30\%$ of simulated male subjects.}
\label{tab:supp-subgroup}
\resizebox{\textwidth}{!}{%
\begin{tabular}{lrrrrrr}
\toprule
Estimator & Clean PR-AUC & Clean RMSE & Clean MCC & Masked PR-AUC & Masked RMSE & Masked MCC\\
\midrule
\emph{Proposed} & 0.780 (0.025) & 0.080 (0.003) & 0.611 (0.033) & 0.736 (0.028) & 0.087 (0.002) & 0.522 (0.026)\\
\emph{Robust-only} & 0.770 (0.028) & 0.118 (0.003) & 0.218 (0.011) & 0.688 (0.031) & 0.131 (0.003) & 0.237 (0.015)\\
\emph{Sparse-only} & 0.851 (0.023) & 0.073 (0.003) & 0.674 (0.027) & 0.226 (0.021) & 0.207 (0.006) & 0.045 (0.022)\\
\emph{Local-slab} & 0.757 (0.026) & 0.073 (0.002) & 0.640 (0.033) & 0.696 (0.025) & 0.078 (0.002) & 0.555 (0.035)\\
\emph{SPM-PEB} & 0.844 (0.023) & 0.132 (0.003) & 0.755 (0.030) & 0.235 (0.024) & 0.525 (0.012) & 0.074 (0.032)\\
\bottomrule
\end{tabular}
}
\end{table}

\subsection{Higher-dimensional sparsity}
\label{supp:sparse}

Table~\ref{tab:supp-sparse} reports the complete $p=40$ experiment. With clean data and strong nonzero effects, all estimators ranked the active coefficients perfectly. The spike-and-slab estimators nevertheless produced much higher MCC than robust ridge because they selected far fewer null coefficients. Under contamination, the two robust spike-and-slab estimators retained high selection and estimation performance, while the Gaussian-likelihood methods deteriorated. The local Gaussian slab equaled or exceeded the pMOM slab on several fixed-threshold measures in this experiment, so these results do not establish uniform superiority of the nonlocal slab.

\begin{table}[htbp]
\centering
\scriptsize
\setlength{\tabcolsep}{4pt}
\caption{Higher-dimensional sparse-regime performance for $N=48$ and $p=40$. Entries are means with Monte Carlo standard errors in parentheses from 50 replicates.}
\label{tab:supp-sparse}
\resizebox{\textwidth}{!}{%
\begin{tabular}{rrlrrrr}
\toprule
Active fraction &  $\phi$  & Estimator & PR-AUC & RMSE & MCC & FPR\\
\midrule
0.05 & 0.00 & \emph{Proposed} & 1.000 (0.000) & 0.027 (0.001) & 0.937 (0.015) & 0.009 (0.003)\\
0.05 & 0.00 & \emph{Robust-only} & 1.000 (0.000) & 0.074 (0.001) & 0.192 (0.004) & 0.571 (0.011)\\
0.05 & 0.00 & \emph{Sparse-only} & 1.000 (0.000) & 0.025 (0.002) & 0.947 (0.015) & 0.008 (0.003)\\
0.05 & 0.00 & \emph{Local-slab} & 1.000 (0.000) & 0.036 (0.001) & 0.951 (0.013) & 0.007 (0.002)\\
0.05 & 0.00 & \emph{SPM-PEB} & 1.000 (0.000) & 0.088 (0.001) & 0.832 (0.019) & 0.027 (0.004)\\
0.05 & 0.10 & \emph{Proposed} & 1.000 (0.000) & 0.045 (0.001) & 0.784 (0.018) & 0.036 (0.004)\\
0.05 & 0.10 & \emph{Robust-only} & 1.000 (0.000) & 0.078 (0.001) & 0.190 (0.004) & 0.575 (0.010)\\
0.05 & 0.10 & \emph{Sparse-only} & 0.553 (0.042) & 0.185 (0.003) & 0.278 (0.015) & 0.307 (0.010)\\
0.05 & 0.10 & \emph{Local-slab} & 1.000 (0.000) & 0.037 (0.001) & 0.946 (0.014) & 0.008 (0.002)\\
0.05 & 0.10 & \emph{SPM-PEB} & 0.551 (0.042) & 0.311 (0.005) & 0.384 (0.032) & 0.124 (0.008)\\
0.10 & 0.00 & \emph{Proposed} & 1.000 (0.000) & 0.032 (0.001) & 0.959 (0.010) & 0.011 (0.003)\\
0.10 & 0.00 & \emph{Robust-only} & 1.000 (0.000) & 0.074 (0.001) & 0.267 (0.006) & 0.569 (0.011)\\
0.10 & 0.00 & \emph{Sparse-only} & 1.000 (0.000) & 0.030 (0.002) & 0.962 (0.011) & 0.010 (0.003)\\
0.10 & 0.00 & \emph{Local-slab} & 1.000 (0.000) & 0.039 (0.001) & 0.970 (0.008) & 0.007 (0.002)\\
0.10 & 0.00 & \emph{SPM-PEB} & 1.000 (0.000) & 0.094 (0.001) & 0.894 (0.013) & 0.028 (0.004)\\
0.10 & 0.10 & \emph{Proposed} & 1.000 (0.000) & 0.044 (0.001) & 0.885 (0.012) & 0.029 (0.003)\\
0.10 & 0.10 & \emph{Robust-only} & 1.000 (0.000) & 0.078 (0.001) & 0.266 (0.005) & 0.571 (0.010)\\
0.10 & 0.10 & \emph{Sparse-only} & 0.585 (0.029) & 0.191 (0.003) & 0.344 (0.016) & 0.314 (0.012)\\
0.10 & 0.10 & \emph{Local-slab} & 1.000 (0.000) & 0.041 (0.001) & 0.968 (0.008) & 0.008 (0.002)\\
0.10 & 0.10 & \emph{SPM-PEB} & 0.590 (0.028) & 0.311 (0.005) & 0.479 (0.025) & 0.123 (0.008)\\
0.25 & 0.00 & \emph{Proposed} & 1.000 (0.000) & 0.043 (0.001) & 0.975 (0.006) & 0.014 (0.004)\\
0.25 & 0.00 & \emph{Robust-only} & 1.000 (0.000) & 0.074 (0.001) & 0.403 (0.008) & 0.565 (0.012)\\
0.25 & 0.00 & \emph{Sparse-only} & 1.000 (0.000) & 0.039 (0.001) & 0.975 (0.006) & 0.014 (0.004)\\
0.25 & 0.00 & \emph{Local-slab} & 1.000 (0.000) & 0.047 (0.001) & 0.988 (0.004) & 0.007 (0.002)\\
0.25 & 0.00 & \emph{SPM-PEB} & 1.000 (0.000) & 0.106 (0.002) & 0.947 (0.007) & 0.029 (0.004)\\
0.25 & 0.10 & \emph{Proposed} & 1.000 (0.000) & 0.048 (0.001) & 0.956 (0.007) & 0.024 (0.004)\\
0.25 & 0.10 & \emph{Robust-only} & 1.000 (0.000) & 0.078 (0.001) & 0.402 (0.008) & 0.567 (0.011)\\
0.25 & 0.10 & \emph{Sparse-only} & 0.755 (0.013) & 0.203 (0.003) & 0.497 (0.017) & 0.290 (0.012)\\
0.25 & 0.10 & \emph{Local-slab} & 1.000 (0.000) & 0.049 (0.001) & 0.984 (0.004) & 0.009 (0.002)\\
0.25 & 0.10 & \emph{SPM-PEB} & 0.754 (0.013) & 0.313 (0.005) & 0.580 (0.019) & 0.121 (0.009)\\
\bottomrule
\end{tabular}
}
\end{table}

\subsection{Sensitivity and algorithmic validation}
\label{supp:sensitivity}

The $\nu=2$ and $\nu=3$ fits were nearly indistinguishable in the clean and baseline conditions, but $\nu=2$ performed better at $20\%$ contamination (Table~\ref{tab:supp-nu}). Using the separately tuned $\pi=0.5$ instead of the shared $\pi=0.6$ produced only small changes for the sparse-only estimator (Table~\ref{tab:supp-arm-tuning}).

The empirical-covariance analysis compared the stylized correlated covariance with first-level covariance matrices sampled with replacement from the corrected NARPS archive (Table~\ref{tab:supp-empirical-covariance}). All 500 fits converged. Across the posterior-sampling estimators, the largest PIP MCSE was $0.025$ and the largest between-chain PIP range was $0.196$. The proposed and local-slab estimators achieved near-perfect discrimination under the resampled NARPS matrices, with PR-AUCs of $0.998$, and their false-positive rates at the common $0.5$ threshold were zero. Robust-only achieved the lowest RMSE under the empirical matrices but had a false-positive rate of $0.551$, again showing that robust weighting alone does not provide sparse selection. Sparse-only changed little across covariance regimes. The qualitative advantage of combining robustness with a mixture prior was unchanged.

Aggregate performance was similar across whitening specifications, but coefficient-level estimates and PIPs could differ materially (Table~\ref{tab:supp-whitening}). This behavior is expected because independent element-wise Student-$t$ coordinates are not invariant to orthogonal rotations. The chosen precision-Cholesky factor and the DCM parameter ordering are therefore part of the fitted model specification.

The scalar calculation in Table~\ref{tab:supp-pip-validation} confirms that responsibility at the posterior mode can be entirely different from the integrated PIP. This supports the posterior-integration step used in the revised estimator. The covariance-component validation used the Gaussian-limit likelihood to match the Gaussian data-generating model. The mean estimate was $0.0192$ when the true value was $0.0225$, with bias $-0.0033$ and RMSE $0.0111$. Five of the 50 estimates reached the lower bound, indicating some finite-sample boundary behavior. All fits converged, both objective sequences were monotone in every replicate, and the finite-difference gradients were small (Table~\ref{tab:supp-objective-validation}). These results support the internal optimization calculations while showing that a small variance component can still be estimated at the boundary in individual samples.

\begin{table}[htbp]
\centering
\scriptsize
\setlength{\tabcolsep}{4pt}
\caption{Sensitivity to the Student-$t$ degrees of freedom. Each $\nu=2$ and $\nu=3$ comparison used the same simulated datasets. Entries are replicate means.}
\label{tab:supp-nu}
\resizebox{\textwidth}{!}{%
\begin{tabular}{llrrrrrr}
\toprule
Condition & Estimator & PR $\nu=2$ & PR $\nu=3$ & RMSE $\nu=2$ & RMSE $\nu=3$ & MCC $\nu=2$ & MCC $\nu=3$\\
\midrule
Baseline & \emph{Proposed} & 0.963 & 0.962 & 0.067 & 0.066 & 0.730 & 0.724\\
Baseline & \emph{Robust-only} & 0.962 & 0.962 & 0.073 & 0.071 & 0.517 & 0.542\\
Baseline & \emph{Local-slab} & 0.962 & 0.963 & 0.071 & 0.071 & 0.712 & 0.699\\
$\phi=0$, $N=48$ & \emph{Proposed} & 0.967 & 0.969 & 0.063 & 0.061 & 0.734 & 0.740\\
$\phi=0$, $N=48$ & \emph{Robust-only} & 0.968 & 0.970 & 0.069 & 0.067 & 0.502 & 0.513\\
$\phi=0$, $N=48$ & \emph{Local-slab} & 0.967 & 0.968 & 0.067 & 0.065 & 0.719 & 0.726\\
$\phi=0.20$, $N=48$ & \emph{Proposed} & 0.953 & 0.936 & 0.078 & 0.094 & 0.663 & 0.608\\
$\phi=0.20$, $N=48$ & \emph{Robust-only} & 0.953 & 0.941 & 0.081 & 0.099 & 0.555 & 0.581\\
$\phi=0.20$, $N=48$ & \emph{Local-slab} & 0.954 & 0.940 & 0.087 & 0.107 & 0.626 & 0.552\\
\bottomrule
\end{tabular}
}
\end{table}

\begin{table}[htbp]
\centering
\small
\setlength{\tabcolsep}{4pt}
\caption{Sensitivity of the sparse-only ablation to shared versus arm-specific tuning. The primary comparison used the proposed estimator's $\pi=0.6$, whereas arm-specific tuning selected $\pi=0.5$. Entries are replicate means.}
\label{tab:supp-arm-tuning}
\resizebox{\textwidth}{!}{%
\begin{tabular}{lrrrrrr}
\toprule
Condition & PR $\pi=.6$ & PR $\pi=.5$ & RMSE $\pi=.6$ & RMSE $\pi=.5$ & MCC $\pi=.6$ & MCC $\pi=.5$\\
\midrule
Baseline & 0.803 & 0.807 & 0.201 & 0.195 & 0.260 & 0.259\\
$\phi=0$, $N=48$ & 0.972 & 0.971 & 0.058 & 0.059 & 0.748 & 0.742\\
$\phi=0.20$, $N=48$ & 0.738 & 0.745 & 0.258 & 0.244 & 0.158 & 0.194\\
\bottomrule
\end{tabular}
}
\end{table}

\begin{table}[htbp]
\centering
\scriptsize
\setlength{\tabcolsep}{4pt}
\caption{Sensitivity to empirical first-level covariance matrices. The experiment used $N=48$, $p=22$, and $10\%$ cell-wise contamination of magnitude $3$. The stylized condition used covariance $0.20\bfR$, while the empirical condition sampled subject-level covariance matrices with replacement from the corrected NARPS archive. Entries are means with Monte Carlo standard errors in parentheses from 50 replicates.}
\label{tab:supp-empirical-covariance}
\resizebox{\textwidth}{!}{%
\begin{tabular}{llrrrr}
\toprule
Covariance & Estimator & PR-AUC & RMSE & MCC & FPR\\
\midrule
Stylized correlated & \emph{Proposed} & 0.943 (0.005) & 0.066 (0.002) & 0.763 (0.014) & 0.018 (0.006)\\
Stylized correlated & \emph{Robust-only} & 0.928 (0.006) & 0.084 (0.002) & 0.432 (0.024) & 0.533 (0.021)\\
Stylized correlated & \emph{Sparse-only} & 0.884 (0.010) & 0.125 (0.003) & 0.596 (0.021) & 0.094 (0.013)\\
Stylized correlated & \emph{Local-slab} & 0.941 (0.005) & 0.076 (0.003) & 0.706 (0.013) & 0.006 (0.004)\\
Stylized correlated & \emph{SPM-PEB} & 0.821 (0.012) & 0.241 (0.008) & 0.504 (0.027) & 0.147 (0.017)\\
Resampled NARPS & \emph{Proposed} & 0.998 (0.001) & 0.045 (0.002) & 0.799 (0.009) & 0.000 (0.000)\\
Resampled NARPS & \emph{Robust-only} & 0.956 (0.005) & 0.041 (0.001) & 0.472 (0.023) & 0.551 (0.022)\\
Resampled NARPS & \emph{Sparse-only} & 0.883 (0.008) & 0.121 (0.003) & 0.590 (0.021) & 0.084 (0.011)\\
Resampled NARPS & \emph{Local-slab} & 0.998 (0.001) & 0.048 (0.002) & 0.779 (0.010) & 0.000 (0.000)\\
Resampled NARPS & \emph{SPM-PEB} & 0.910 (0.006) & 0.177 (0.006) & 0.617 (0.021) & 0.039 (0.008)\\
\bottomrule
\end{tabular}
}
\end{table}

\begin{table}[htbp]
\centering
\small
\setlength{\tabcolsep}{4pt}
\caption{Whitening and parameter-order sensitivity for the proposed estimator. Difference columns give averages or maxima across 50 replicates relative to precision-Cholesky whitening in the original order.}
\label{tab:supp-whitening}
\resizebox{\textwidth}{!}{%
\begin{tabular}{lrrrrrr}
\toprule
Specification & PR-AUC & RMSE & Mean max. $|\Delta\widehat\beta|$ & Largest $|\Delta\widehat\beta|$ & Mean max. $|\Delta\mathrm{PIP}|$ & Largest $|\Delta\mathrm{PIP}|$\\
\midrule
Precision Cholesky & 0.958 & 0.077 & -- & -- & -- & --\\
Symmetric & 0.957 & 0.074 & 0.091 & 0.182 & 0.251 & 0.649\\
Reversed order & 0.954 & 0.079 & 0.137 & 0.222 & 0.341 & 0.672\\
\bottomrule
\end{tabular}
}
\end{table}

\begin{table}[htbp]
\centering
\small
\setlength{\tabcolsep}{4pt}
\caption{Algorithmic validation under a clean Gaussian single-component covariance design fitted with the Gaussian-limit likelihood.}
\label{tab:supp-objective-validation}
\begin{tabular}{lr}
\toprule
Quantity & Value\\
\midrule
Replicates & 50\\
True $\alpha$ & 0.0225\\
Mean $\widehat{\alpha}$ & 0.01922\\
Bias & -0.00328\\
RMSE & 0.01113\\
Estimates at lower bound & 5 of 50\\
Maximum $|\partial \ell/\partial\log\alpha|$ & 2.25e-05\\
Maximum coefficient gradient & 8.26e-04\\
Convergence rate & 1.000\\
Observed-objective monotonicity & 1.000\\
Expected-objective monotonicity & 1.000\\
\bottomrule
\end{tabular}
\end{table}

\begin{table}[htbp]
\centering
\small
\setlength{\tabcolsep}{4pt}
\caption{Scalar validation showing why the slab responsibility at a posterior mode cannot substitute for the integrated PIP.}
\label{tab:supp-pip-validation}
\begin{tabular}{lr}
\toprule
Quantity & Value\\
\midrule
Posterior mode & 0.0099\\
Responsibility at posterior mode & 0.00004\\
Integrated PIP & 0.7396\\
\bottomrule
\end{tabular}
\end{table}

\clearpage

%%%
\section{Additional real data results}

\subsection{Sensitivity to task version}
\label{sec:supp-task-version}

Participants completed one of two versions of the mixed-gambles task. Of the 48 participants, 25 completed the equal-indifference version and 23 completed the equal-range version. To examine whether this design difference affected the empirical findings, we repeated the group analysis after adding a centered indicator for task version to the second-level design. With this coding, the intercept represents the sample-average connectivity effect, while the task-version coefficient represents the equal-range minus equal-indifference contrast. All other model specifications and estimation settings were unchanged.

The generalized EM algorithm converged from all three initializations with monotone objective and surrogate-function sequences. The solution with the highest normalized log posterior was retained. At this solution, the estimated spike and slab scales were $\widehat{\tau}_0=0.0199$ and $\widehat{\tau}_1=0.2381$, respectively. Integrated PIPs were computed using parallel tempering. The cold-chain acceptance rates ranged from $0.408$ to $0.417$, the large-jump acceptance rates ranged from $0.156$ to $0.157$, and the minimum adjacent-temperature swap rate was $0.154$. The maximum PIP Monte Carlo standard error was $0.0119$, and the largest PIP range across independent chains was $0.051$. These diagnostics indicated adequate numerical and Monte Carlo stability.

The primary and task-adjusted intercept estimates were highly similar ($r=0.988$). Their median absolute difference was $0.013$, and the maximum absolute difference was $0.041$. The inclusion probabilities were more sensitive to the expanded design. Nine intrinsic effects had adjusted-intercept PIP $>0.5$, whereas four diagonal self-inhibition scaling parameters exceeded this threshold in the primary intercept-only analysis. Six task-version effects had PIP $>0.5$: $A_{21}$, $A_{41}$, $A_{32}$, $A_{42}$, $A_{24}$, and $B_{12}^{(1)}$. The signs of all six effects agreed with the corresponding \emph{SPM-PEB} estimates, although none of the SPM task-version effects satisfied $|\mathrm{E}p|/\mathrm{SD}>3$. Thus, adjustment for task version had little effect on the estimated group-average coefficients but had a greater influence on the empirical-Bayes inclusion probabilities. Given the modest sample size and the exploratory nature of this analysis, the task-version effects should not be interpreted as established differences in effective connectivity.

\begin{table}[htbp]
\centering
\caption{Sensitivity of the real-data results to adjustment for task version. The task-version coefficient represents the equal-range minus equal-indifference contrast. Estimates for diagonal $A$ parameters are reported on SPM's internal log-scaling scale, for which zero represents the default self-inhibition of $-0.5$ Hz. The primary PIP is from the intercept-only analysis. The adjusted estimate and PIP refer to the intercept in the model containing the centered task-version indicator. SPM task-version results are reported as posterior mean (posterior standard deviation).}
\label{tab:supp-task-version}
\small
\setlength{\tabcolsep}{4pt}
\renewcommand{\arraystretch}{1.05}
\begin{tabular}{l
                >{\centering\arraybackslash}p{1.7cm}
                >{\centering\arraybackslash}p{1.7cm}
                >{\centering\arraybackslash}p{1.7cm}
                >{\centering\arraybackslash}p{1.7cm}
                >{\centering\arraybackslash}p{1.7cm}
                >{\centering\arraybackslash}p{2.4cm}}
\hline
Parameter &
Primary intercept PIP &
Adjusted intercept estimate &
Adjusted intercept PIP &
Task-version estimate &
Task-version PIP &
SPM task-version estimate (SD) \\
\hline
$A_{11}$ & 0.995 & $-0.302$ & 1.000 & $-0.015$ & 0.366 & $-0.157$ (0.065) \\
$A_{21}$ & 0.052 & $ 0.112$ & 0.985 & $ 0.031$ & 0.924 & $ 0.118$ (0.054) \\
$A_{31}$ & 0.091 & $ 0.127$ & 1.000 & $ 0.021$ & 0.070 & $ 0.086$ (0.055) \\
$A_{41}$ & 0.129 & $ 0.140$ & 0.999 & $ 0.022$ & 0.645 & $ 0.086$ (0.055) \\
$A_{12}$ & 0.184 & $ 0.158$ & 0.997 & $ 0.004$ & 0.022 & $-0.036$ (0.061) \\
$A_{22}$ & 0.896 & $-0.237$ & 1.000 & $ 0.008$ & 0.045 & $ 0.129$ (0.068) \\
$A_{32}$ & 0.016 & $ 0.021$ & 0.065 & $-0.019$ & 0.810 & $-0.167$ (0.060) \\
$A_{42}$ & 0.306 & $ 0.177$ & 0.995 & $-0.016$ & 0.856 & $-0.178$ (0.060) \\
$A_{13}$ & 0.002 & $ 0.007$ & 0.003 & $ 0.009$ & 0.062 & $ 0.063$ (0.063) \\
$A_{23}$ & 0.005 & $ 0.012$ & 0.008 & $-0.010$ & 0.132 & $-0.068$ (0.061) \\
$A_{33}$ & 0.802 & $-0.230$ & 1.000 & $-0.001$ & 0.018 & $ 0.109$ (0.070) \\
$A_{43}$ & 0.021 & $ 0.030$ & 0.180 & $-0.012$ & 0.112 & $-0.101$ (0.061) \\
$A_{14}$ & 0.003 & $ 0.009$ & 0.003 & $ 0.009$ & 0.058 & $ 0.011$ (0.062) \\
$A_{24}$ & 0.004 & $ 0.011$ & 0.012 & $-0.016$ & 0.765 & $-0.119$ (0.060) \\
$A_{34}$ & 0.018 & $ 0.036$ & 0.210 & $-0.007$ & 0.015 & $-0.036$ (0.060) \\
$A_{44}$ & 0.740 & $-0.214$ & 1.000 & $-0.002$ & 0.021 & $-0.127$ (0.069) \\
$B_{21}^{(1)}$ & 0.007 & $ 0.003$ & 0.015 & $-0.002$ & 0.096 & $-0.086$ (0.428) \\
$B_{12}^{(1)}$ & 0.080 & $ 0.001$ & 0.122 & $ 0.420$ & 0.712 & $ 0.665$ (0.443) \\
$B_{13}^{(2)}$ & 0.048 & $-0.000$ & 0.098 & $ 0.001$ & 0.390 & $ 0.460$ (0.557) \\
$B_{43}^{(2)}$ & 0.013 & $ 0.005$ & 0.046 & $-0.194$ & 0.149 & $-0.329$ (0.492) \\
$B_{14}^{(2)}$ & 0.038 & $-0.002$ & 0.106 & $ 0.273$ & 0.296 & $ 0.727$ (0.539) \\
$B_{34}^{(2)}$ & 0.020 & $ 0.005$ & 0.077 & $ 0.001$ & 0.140 & $-0.032$ (0.481) \\
\hline
\end{tabular}
\end{table}

\subsection{First-level input information and diagnostics}
\label{sec:supp-first-level-diagnostics}

The two mixed-gambles task versions differed markedly in the information available for estimating loss modulation. We therefore examined the original event counts, the ranks of the DCM input and first-level GLM task matrices, run-level DCM fit diagnostics, and prior-to-posterior variance changes. Complete subject-by-run results are available in the public reproducibility repository. Tables~\ref{tab:supp-first-level-inputs}--\ref{tab:supp-variance-reduction} provide compact summaries.

\begin{table}[htbp]
\centering
\caption{Condition counts and input-information diagnostics by task version and run. Panel A reports the number of events per participant. All participants within a task version had the same event counts in each run. The zero-loss column gives the number of participants with no loss-dominant event in that run. Panel B reports median ranks with ranges in parentheses. A matrix was counted as full rank relative to its nonzero columns. EI denotes equal-indifference and ER denotes equal-range.}
\label{tab:supp-first-level-inputs}
\small
\setlength{\tabcolsep}{3pt}
\textit{Panel A: Condition counts}\par\smallskip
\begin{tabular}{llrrrrr}
\toprule
Task version & Run & $N$ & \shortstack{Gain-\\dominant} & \shortstack{Loss-\\dominant} & Intermediate & \shortstack{Zero\\loss}\\
\midrule
EI & 1 & 25 & 36 & 1  & 27 & 0\\
EI & 2 & 25 & 32 & 0  & 32 & 25\\
EI & 3 & 25 & 36 & 0  & 28 & 25\\
EI & 4 & 25 & 32 & 0  & 32 & 25\\
ER & 1 & 23 & 10 & 8  & 46 & 0\\
ER & 2 & 23 & 10 & 8  & 46 & 0\\
ER & 3 & 23 & 9  & 11 & 44 & 0\\
ER & 4 & 23 & 7  & 9  & 48 & 0\\
\bottomrule
\end{tabular}

\vspace{0.8em}
\textit{Panel B: Input and task-design ranks}\par\smallskip
\begin{tabular}{llrrrrr}
\toprule
Task version & Run & $N$ & \shortstack{DCM input\\rank} & \shortstack{DCM full\\rank} & \shortstack{GLM task\\rank} & \shortstack{GLM full\\rank}\\
\midrule
EI & 1 & 25 & 3 (3--3) & 25/25 & 23 (22--23) & 25/25\\
EI & 2 & 25 & 2 (2--2) & 25/25 & 22 (21--23) & 0/25\\
EI & 3 & 25 & 2 (2--2) & 25/25 & 22 (21--27) & 0/25\\
EI & 4 & 25 & 2 (2--2) & 25/25 & 21 (21--27) & 0/25\\
ER & 1 & 23 & 3 (3--3) & 23/23 & 22 (22--24) & 23/23\\
ER & 2 & 23 & 3 (3--3) & 23/23 & 22 (22--25) & 23/23\\
ER & 3 & 23 & 3 (3--3) & 23/23 & 23 (22--27) & 23/23\\
ER & 4 & 23 & 3 (3--3) & 23/23 & 23 (22--26) & 23/23\\
\bottomrule
\end{tabular}
\end{table}

The absent loss-dominant events explain the rank reduction in equal-indifference runs 2--4. The DCM input matrices remained full rank after identically zero inputs were excluded. The corresponding GLM task blocks were not full rank when all task columns, including empty-condition columns, were counted.

\begin{table}[htbp]
\centering
\caption{Run-level DCM fit diagnostics by task version. Free energy, percentage of variance explained, and residual root mean square (RMS) are reported as medians with interquartile ranges in parentheses. Covariance columns give the number of finite or positive-semidefinite posterior covariance matrices divided by the number of run-level DCMs.}
\label{tab:supp-first-level-fit}
\small
\setlength{\tabcolsep}{3pt}
\begin{tabular}{lrrrrrr}
\toprule
Task version & \shortstack{Run-level\\DCMs} & Free energy & \shortstack{Variance\\explained (\%)} & \shortstack{Residual\\RMS} & \shortstack{Finite\\covariance} & \shortstack{PSD\\covariance}\\
\midrule
Equal-indifference & 100 & $-1234.69$ ($-1509.91$, $-868.80$) & 2.79 (1.36, 6.23) & 0.200 (0.183, 0.216) & 100/100 & 100/100\\
Equal-range        & 92  & $-1481.03$ ($-1844.27$, $-947.78$) & 4.49 (1.39, 13.65) & 0.214 (0.185, 0.237) & 92/92 & 92/92\\
All                & 192 & $-1310.43$ ($-1717.00$, $-886.14$) & 3.36 (1.36, 8.87) & 0.205 (0.184, 0.224) & 192/192 & 192/192\\
\bottomrule
\end{tabular}
\end{table}

\begin{table}[htbp]
\centering
\caption{Prior-to-posterior variance changes for the six modulatory parameters. Entries are median posterior-to-prior variance ratios with interquartile ranges in parentheses. Ratios below one indicate reduced uncertainty relative to the prior. The first two parameters are gain-related and the final four are loss-related. EI denotes equal-indifference and ER denotes equal-range.}
\label{tab:supp-variance-reduction}
\small
\begin{tabular}{llrr}
\toprule
Analysis stage & Parameter & EI & ER\\
\midrule
Run-level DCM & $B(2,1,1)$ & 0.049 (0.018, 0.151) & 0.051 (0.008, 0.153)\\
Run-level DCM & $B(1,2,1)$ & 0.321 (0.138, 0.610) & 0.290 (0.104, 0.844)\\
Run-level DCM & $B(1,3,2)$ & 1.000 (1.000, 1.000) & 0.616 (0.422, 0.862)\\
Run-level DCM & $B(4,3,2)$ & 1.000 (1.000, 1.000) & 0.305 (0.090, 0.708)\\
Run-level DCM & $B(1,4,2)$ & 1.000 (1.000, 1.000) & 0.543 (0.253, 0.826)\\
Run-level DCM & $B(3,4,2)$ & 1.000 (1.000, 1.000) & 0.200 (0.049, 0.595)\\
\midrule
Within-subject PEB & $B(2,1,1)$ & 0.043 (0.031, 0.057) & 0.027 (0.021, 0.033)\\
Within-subject PEB & $B(1,2,1)$ & 0.100 (0.070, 0.169) & 0.066 (0.040, 0.125)\\
Within-subject PEB & $B(1,3,2)$ & 0.792 (0.747, 0.800) & 0.199 (0.104, 0.316)\\
Within-subject PEB & $B(4,3,2)$ & 0.632 (0.486, 0.787) & 0.074 (0.040, 0.145)\\
Within-subject PEB & $B(1,4,2)$ & 0.793 (0.725, 0.798) & 0.139 (0.083, 0.244)\\
Within-subject PEB & $B(3,4,2)$ & 0.703 (0.328, 0.762) & 0.064 (0.035, 0.095)\\
\bottomrule
\end{tabular}
\end{table}

At the run level, the loss-modulatory posterior variances remained essentially equal to their priors in the equal-indifference group because three of the four runs contained no loss-dominant trials. Within-subject PEB combined information across runs, but uncertainty remained substantially greater than in the equal-range group. Pooling the four loss-modulatory parameters across participants, the median within-subject PEB posterior-to-prior variance ratio was 0.760 for equal-indifference participants and 0.099 for equal-range participants.

\end{document}